\documentclass[11pt,a4paper]{article}
\pdfoutput=1
\usepackage{jheppub}
\usepackage{gensymb}
 \usepackage{multirow}
 \usepackage[table,xcdraw]{xcolor}
\DeclareSymbolFont{matha}{OML}{txmi}{m}{it}% txfonts
\DeclareMathSymbol{\varv}{\mathord}{matha}{118}
\usepackage{amsmath}
\usepackage{amssymb}
\usepackage{graphicx}
\usepackage{subfigure}
\usepackage{pstricks}
\usepackage{bm}
\usepackage{pbox}
\usepackage{placeins}
\usepackage{booktabs}
\usepackage[T1]{fontenc}
\usepackage{footnote}
\usepackage{pdfpages}
\usepackage{hhline}
\usepackage{multirow}
\usepackage{multicol}
\usepackage{enumitem}
\usepackage[toc,page]{appendix}
\allowdisplaybreaks
\usepackage{comment}
\usepackage{float}                                             
\usepackage{mathtools}                  
\usepackage{textcomp}
\usepackage{gensymb}

\usepackage{relsize}

\usepackage{slashed}

\usepackage[numbers]{natbib}
\usepackage{notoccite}

\usepackage{rotating}
\usepackage{tabularx}
\usepackage[labelfont=bf]{caption} % optional

\FloatBarrier
\usepackage[many]{tcolorbox}

\renewcommand{\theequation}{\thesection.\arabic{equation}}
\renewcommand{\thetable}{\Roman{table}}

\definecolor{MyDarkBlue}{rgb}{0.1, 0.1, 0.8} %defining the color 'MyDarkBlue'
\definecolor{SBlue}{rgb}{0.2, 0.4, 0.7} %defining the color 'MyDarkBlue'
\definecolor{MyLightBlue}{rgb}{0.22,0.51,0.9}
\definecolor{MyGreen}{rgb}{0.0, 0.5, 0.0}
\definecolor{BrickRed}{rgb}{0.8, 0.25, 0.33}
\RequirePackage{hyperref}
\hypersetup{colorlinks, citecolor=MyGreen,linkcolor=MyDarkBlue, urlcolor=BrickRed}

\begin{document}
%\preprint{OSU-HEP-20-14}

\title{
Left-Right Symmetric Model without Higgs Triplets
 }
 \author[a]{K.S. Babu,}
 \author[a]{Anil Thapa}
 
 \affiliation[a]{Department of Physics, Oklahoma State University, Stillwater, OK 74078, USA}

\emailAdd{babu@okstate.edu, thapaa@okstate.edu}

\preprint{OSU-HEP-20-15}

%%%%%%%%%%%%%%%%%%%%%%%%%%%%%%%%%%%%%%%%%%%%%%%%%%%%%%%%%%%%%%%%%%%%%%%%%%%%%%%%%%%%%

\abstract{
We develop a minimal left-right symmetric model based on the gauge group $SU(3)_C \otimes SU(2)_L \otimes SU(2)_R \otimes U(1)_{B-L}$ wherein the Higgs triplets conventionally employed for symmetry breaking are replaced by Higgs doublets.  Majorana masses for the right-handed neutrinos $(\nu_R$) are induced via two-loop diagrams involving a charged scalar field $\eta^+$.  This setup is shown to provide excellent fits to neutrino oscillation data via the seesaw mechanism for the entire range of the $W_R^\pm$ mass, from TeV to the GUT scale. When the $W_R^\pm$ mass is at the TeV scale, the $\nu_R$ masses turn out to be in the MeV range. We analyze constraints from low energy experiments, early universe cosmology and from supernova 1987a on such a scenario and show its consistency.  We also study collider implications of a relatively light $\eta^+$ scalar through its decay into multi-lepton final states and derive a lower limit of 390 GeV on its mass from the LHC, which can be improved to 555 GeV in its high luminosity run.
}
%%%%%%%%%%%%%%%%%%%%%%%%%%%%%%%%%%%%%%%%%%%%%%%%%%%%%%%%%%%%%%%%%%%%%%%%%%%%%%%%%%%%%%%%%

\maketitle
%\newpage
%%%%%%%%%%%%%%%%%%%%%%%%%%%%%%%%%%%%%%%%%%%%%%%%%%%%%%%%%%%%%%%%%%%%%%%%%%%%%%%%%%%%%%%%%%%%%%%%%%%%%%%%%%%%%%%%%%%%%%%%%%%%%%%%%%%%%%%%%%%%%%%%%%%%%%%
\section{Introduction}

Left-right symmetric models (LRSM) based on the gauge group $SU(3)_C \otimes SU(2)_L \otimes SU(2)_R \otimes U(1)_{B-L}$ \cite{Pati:1974yy, Mohapatra:1974gc,Mohapatra:1974hk, Senjanovic:1975rk,Senjanovic:1978ev,Mohapatra:1979ia,
Mohapatra:1980yp} are attractive extensions of the Standard Model on several grounds.  They explain Parity violation as a spontaneous phenomenon resulting from gauge symmetry breaking. They incorporate the right-handed neutrino $(\nu_R$) as an essential component of the right-handed lepton doublet, paving the way for neutrino mass generation by the seesaw mechanism \cite{Mohapatra:1979ia,Mohapatra:1980yp}.  The promotion of hypercharge $Y$ of the Standard Model into $(B-L)$ in LRSM may shed deeper insight into its origin from higher unification such as $SO(10)$.  And these models lead to a variety of interesting phenomena, if the left-right symmetry is realized near the TeV scale, that can be tested in ongoing and forthcoming low energy as well as in high energy collider experiments.

For consistent phenomenology the $SU(2)_L \otimes SU(2)_R \otimes U(1)_{B-L}$ gauge symmetry should break spontaneously down to $SU(2)_L \otimes U(1)_Y$ via the Higgs mechanism at a scale $v_R$ much larger than the electroweak symmetry breaking scale $v_L$.  In the early constructions of LRSM, before the advent of the seesaw mechanism to generate small neutrino masses \cite{Minkowski:1977sc,Yanagida:1979as,GellMann:1980vs,Glashow:1979nm,Mohapatra:1979ia,Mohapatra:1980yp}, a pair of Higgs doublets $\chi_L(2,1,1) + \chi_R(1,2,1)$ and a Higgs bidoublet $\Phi(2,2,0)$ were employed for this purpose \cite{Mohapatra:1974gc,Mohapatra:1974hk, Senjanovic:1975rk,Senjanovic:1978ev}.  (The quantum numbers here refer to  $SU(2)_L \otimes SU(2)_R \otimes U(1)_{B-L}$ transformations.) When the neutral component of $\chi_R$ develops a vacuum expectation value (VEV), $\left\langle \chi_R^0 \right\rangle = v_R/\sqrt{2}$, the gauge symmetry breaks down to $SU(2)_L \times U(1)_Y$, giving masses of order $v_R$ to the $W_R^\pm$ and the $Z_R$ gauge bosons.  The Higgs bidoublet $\Phi(2,2,0)$ is used to generate quark and lepton masses, including neutrino Dirac masses. The smallness of neutrino masses compared to the charged fermions masses remains unexplained in this scenario.

The discovery of the seesaw mechanism caused a major shift in the thinking on Higgs multiplets needed for symmetry breaking in LRSM. It was pointed out in Ref. \cite{Mohapatra:1979ia,Mohapatra:1980yp} that a pair of Higgs triplets $\Delta_L(1,3,2) + \Delta_R(3,1,2)$ can simultaneously generate $W_R^\pm$ and $Z_R$ gauge bosons masses and  Majorana masses for the $\nu_R$ fields, thus realizing the seesaw mechanism. After this observation, a Higgs sector consisting of $\{\Delta_L(1,3,2) + \Delta_R(3,1,2) + \Phi(2,2,0)\}$ has become standard in the discussion of LRSM models. One feature of this Higgs system distinct from the Higgs doublet scenario of early years is the appearance of a pair of doubly charged scalars $\delta^{\pm\pm}_L$ and $\delta^{\pm\pm}_R$ in the physical spectrum. The presence of the $\Delta_L(3,1,2)$ Higgs field, which is the Parity partner of the $\Delta_R(1,3,2)$ Higgs field used for $SU(2)_R \times U(1)_{B-L}$  symmetry breaking, provides a compelling motivation for type-II seesaw mechanism for small neutrino masses in this context \cite{Schechter:1980gr,Mohapatra:1979ia,Lazarides:1980nt}, which is in addition to contributions from the type-I seesaw mechanism \cite{Minkowski:1977sc,Yanagida:1979as,GellMann:1980vs,Glashow:1979nm,Mohapatra:1979ia,Mohapatra:1980yp}.
The phenomenology of this class of minimal left-right symmetric models has been well studied in the context of flavor physics  \cite{Beall:1981ze,Chang:1982dp,Branco:1982wp,Harari:1983gq,Ecker:1983dj,Gilman:1983ce,Ecker:1985vv,London:1989cf,Babu:1993hx,
Rizzo:1994aj,Barenboim:1996nd,Pospelov:1996fq,Ball:1999mb,Raidal:2002ph,Zhang:2007fn,Zhang:2007da,Blanke:2011ry,Chakrabortty:2012mh,Barry:2013xxa,Bertolini:2014sua,Senjanovic:2014pva,Senjanovic:2015yea,Das:2016vkr}, neutrino masses and cosmology \cite{Hirsch:1996qw,Joshipura:2001ya,Babu:2005bh,Nemevsek:2011aa,Nemevsek:2012cd,Das:2012ii,Nemevsek:2012iq,Barry:2013xxa,Chen:2013foz,Dev:2013oxa,Awasthi:2013ff,Dev:2014xea,Senjanovic:2018xtu,Heeck:2015qra}, Higgs boson physics \cite{Mohapatra:1980qe,Gunion:1989in,Deshpande:1990ip,Barenboim:2001vu,Zhang:2007da,Holthausen:2009uc}, as well as collider physics \cite{Holstein:1977qn,Mohapatra:1977be,Beg:1977ti,Barger:1978rj,Barger:1982sk,Keung:1983uu,Cvetic:1991kh,Maalampi:1992np,Tello:2010am,Maiezza:2010ic,Nemevsek:2011hz,Nemevsek:2012cd,Das:2012ii,Bambhaniya:2013wza,Patra:2015bga,Dev:2016dja,Lindner:2016lxq,Mitra:2016kov}.  

The purpose of this paper is to develop an alternate minimal version of LRSM which uses a Higgs system consisting of \{$\chi_L(1,2,1)+\chi_R(2,1,1) + \Phi(2,2,0)$ for gauge symmetry breaking and fermion mass generation, as was done in the early papers \cite{Mohapatra:1974gc,Mohapatra:1974hk, Senjanovic:1975rk,Senjanovic:1978ev}.  As for realizing the seesaw mechanism, a new singlet scalar $\eta^+(1,1,2)$ is introduced which has Yukawa couplings to the right-handed neutrino that violates lepton number.  Majorana masses for the $\nu_R$ fields are induced via two-loop diagrams involving the $\eta^+$ field.  This Higgs sector is arguably a little simpler than that of the  standard left-right model. The physical scalar spectrum in this scenario consists of four neutral scalars, two  pseudoscalars, and three charged scalars.  This is to be compared with the physical spectrum of standard left-right model which has one less charged scalar, but two doubly charged scalars.

The phenomenology of the model developed here is also distinct from that of the standard left-right model with respect to  neutrino physics, Higgs boson physics and collider signals. A careful analysis of this model shows that if the $W_R^\pm$ gauge boson has a mass near 5 TeV, two of the $\nu_R$ fields would have masses in the few MeV range, leading to interesting low energy phenomena. Such a scenario is constrained by early universe cosmology as well as by supernova 1987a energy loss in $\nu_R$. By analyzing these constraints we show the consistency of such a low mass $W_R^\pm$  scenario. As the mass of $W_R^\pm$ increases, so does the $\nu_R$ masses.  We show that the entire range of $W_R^\pm$ masses, from a few TeV to the GUT scale of $10^{16}$ GeV, is consistent within the model.

Left-right symmetric models involving this set of Higgs boson have been studied previously \cite{Babu:1988qv,FileviezPerez:2016erl}. In the early work of Ref. \cite{Babu:1988qv} the $\nu_R$ fields were found to be as light as the usual neutrinos.  In the recent work of Ref. \cite{FileviezPerez:2016erl} the $\nu_R$ fields were found to have masses of order 400 MeV or less.  These results were obtained  based on the evaluation of one-loop diagrams for $\nu_R$ Majorana masses, which are proportional to the charged lepton masses.  We observe here that there are more important two-loop diagrams for $\nu_R$ masses that do not rely on electroweak symmetry breaking parameters.  The $\nu_R$ mass arising from such diagrams scale linearly with $v_R$, suppressed by a two-loop factor.  This allows for the $\nu_R$ mass to be anywhere from MeV to $10^{14}$ GeV, depending on the scale $v_R$ where the $SU(2)_R$
gauge symmetry breaks.  In terms of effective operators, the $\nu_R$ mass receives two contributions:
\begin{equation}
    {\cal O}_1 = c_1\, \Psi_R \Psi_R (\chi_L^T \Phi \chi_R),~~~~~~{\cal O}_2 = c_2 \,\Psi_R \Psi_R (\chi_R \chi_R)~
\end{equation}
where $\Psi_R$ denotes the right-handed lepton doublet. 
The Wilson coefficients $c_1$ and $c_2$ are found to be of order
\begin{equation}
    c_1 \sim \frac{ (y_\tau^2 f \alpha_4)}{16 \pi^2}\left(\frac{1}{M^2}\right),~~~~~~ c_2 \sim \frac{(y_\tau^2 f \alpha_4)}{(16 \pi^2)^2} \left(\frac{\mu_4}{M^2}\right)~.
    \label{op2}
\end{equation}
Here $y_\tau$ is the tau-lepton Yukawa coupling, $f$ is the Yukawa coupling of the charged scalar $\eta^+$, $\alpha_4$ and $\mu_4$ are  scalar quartic and cubic couplings which together violate lepton number, and $M\sim v_R$ is the scale of new physics where these operators are induced. The operator ${\cal O}_1$ is realized through one-loop diagrams, while ${\cal O}_2$ is realized through two-loop diagrams.  In spite of the additional loop suppression, it is clear that as $v_R$ takes values much larger than $v_L$, contributions from ${\cal O}_2$ will dominate over ${\cal O}_1$.  Establishing this fact is an important result of the present paper.

As we shall show explicitly from a symmetry breaking analysis of the model presented in Sec. \ref{sec:scalar}, all scalar fields have masses of order $v_R$ or smaller, except for the $\eta^+$ field, which can have an arbitrarily large mass.  The mass parameter $M$ appearing in Eq. (\ref{op2}) is {\it not} the mass of $\eta^+$, but of the other scalar fields of the model which are of order $v_R$.  Thus the model allows for the $\eta^+$ field to be integrated out while still yielding the $\nu_R$ Majorana mass operators of Eq. (\ref{op2}). The $\nu_R$ mass depends only logarithmically on the $\eta^+$ mass. While this is a simplifying feature of the model, phenomenology would require a relative large $W_R^\pm$ mass of order 50 TeV or larger in this case.  We shall consider therefore the more general case of $\eta^+$ mass being of order the $W_R^\pm$ mass, in which case both masses can be as low as 5 TeV.

Our careful evaluation of the two-loop diagrams that generate Majorana masses for the $\nu_R$ fields confirms that these diagrams dominate over the one-loop diagrams for the entire range of $W_R^\pm$ mass.  We have analyzed the phenomenology of two specific scenarios, one where the $W_R^\pm$ gauge boson is light, with a mass in the few TeV range -- so that it is observable at the LHC, and one where it is much heavier.  In the former case the $\nu_R$ Majorana mass is in the few MeV range, which can potentially modify standard big bang cosmology, unless the $\nu_R$ decays before the onset of nucleosynthesis.  Satisfying this constraint requires that the $\eta^+$ scalar should have a mass of order a few TeV as well. We have re-evaluated the constraints on $W_R^\pm$ mass arising from the energy loss in $\nu_R$ from supernova 1987a. Including the full cross section for $\nu_R$ production, as well as certain interference terms that were previously ignored, we found the constraint on $W_R^\pm$ mass to be  $M_{W_R} > 4.6$ TeV, which is somewhat weaker than the limit of $M_{w_R} > 23$ TeV found in Ref. \cite{Barbieri:1988av}. 

 For intermediate value of $W_R^\pm$ mass, the $\eta^+$ scalar may be accessible to collider experiments which could lead to multi-lepton signals from the decay of $\eta^+ \eta^-$ pairs.  We have analyzed the current constraint from LHC, and obtained a limit $m_{\eta^+} \geq 390$ GeV, which may be increased to 555 GeV at the high luminosity run of the LHC.

The rest of the paper is organized as follows. The LRSM without Higgs triplets is outlined in Sec. \ref{sec:model}.   In Sec. \ref{sec:scalar} the scalar sector of this model is presented and analyzed. Here the masses of the Higgs field are laid out with a few simplifying assumptions. In Sec. \ref{sec:gauge} we summarize the the gauge boson masses and mixings in the  model. In Sec. \ref{sec:numass} we study the generation of right-handed neutrino mass via one-loop and two-loop diagrams. In Sec. \ref{sec:lowLR} we present fits to the neutrino oscillation data with a TeV scale $W_R$.  Here we summarize various experimental and cosmological limits on an MeV scale sterile neutrino and show that these constraints are satisfied in the model. In Sec. \ref{Sec:sn} we revisit the supernova constraints on $W_R$ mass valid when the $\nu_R$ mass is below 10 MeV. Sec. \ref{sec:nufit} provides fits to the neutrino mass matrix with the neutrino oscillation data. In Sec. \ref{sec:collider} we discuss the collider implication of this model by analyzing the production and decay of charged scalar singlet. We finally conclude in Sec. \ref{sec:summary}.

%%%%%%%%%%%%%%%%%5

\section{LR Symmetric Model without Higgs Triplets} \label{sec:model}

Here we present the basic ingredients of the minimal LRSM model without Higgs triplets.  The model is based on the gauge symmetry $SU(3)_C \otimes SU(2)_L \otimes SU(2)_R \otimes U(1)_{B-L}$ \cite{Pati:1974yy, Mohapatra:1974gc,Mohapatra:1974hk, Senjanovic:1975rk,Senjanovic:1978ev,Mohapatra:1979ia,
Mohapatra:1980yp} under which the fermion fields transform as left-handed doublets and right-handed doublets: 
%\vspace{-2mm}
\begin{align}
Q_L &= \begin{pmatrix}
 u_L \\
 d_L \\
\end{pmatrix} \sim (3,2,1,1/3), \hspace{10mm}
%%%%%%%%%%%%%%%%%%%%%%%%%%%
Q_R = \begin{pmatrix}
 u_R \\
 d_R \\
\end{pmatrix} \sim (3,1,2,1/3), \nonumber\\[5pt]
%%%%%%%%%%%%%%%%%%%%%%%%%%%%%%%5
\Psi_L &= \begin{pmatrix}
 \nu_L \\
 e_L \\
\end{pmatrix} \sim (1,2,1,-1), \hspace{12mm}
%%%%%%%%%%%%%%%%%%%%%%%%%%%%%%%%%%%%
\Psi_R = \begin{pmatrix}
 \nu_R \\
 e_R \\
\end{pmatrix}\sim (1,1,2,-1).
\label{eq:fermion}
\end{align}
%%%%%%%%%%%%%%
Here the generation index is suppressed, but should be assumed. Under Parity symmetry $Q_L \leftrightarrow Q_R$ and $\Psi_L \leftrightarrow \Psi_R$, which is possible due to the enhanced gauge symmetry. Note that the right-handed neutrino $\nu_R$ is required to complete the lepton multiplet, unlike in the Standard Model, leading to tiny neutrino masses via the seesaw mechanism.

The $SU(2)_R \otimes U(1)_{B-L}$ symmetry is broken spontaneously down to $U(1)_Y$ at a scale $v_R \gg v_L$ where $v_L$ denotes the electroweak symmetry breaking scale.  Furthermore, realistic fermion masses should be generated through couplings to the Higgs fields.  In the model developed here, these requirements are achieved by the choice of the following Higgs fields:
\begin{align}
\chi_L & = \begin{pmatrix}
\chi_L^+ \\
 \chi_L^0 \\
\end{pmatrix} \sim (1,2,1,1), \hspace{10mm}
%%%%%%%%%%%%%%%%%%%%
\chi_R = \begin{pmatrix}
 \chi_R^+ \\
 \chi_R^0 \\
\end{pmatrix} \sim (1,1,2,1), \nonumber \\[5pt]
%%%%%%%%%%%%%%%%%%%%%%%%%%%%
 %%%%%%%%%%%%%%%%%%%%%%%%%%%%%%%%
 \Phi &= \begin{pmatrix}
 \phi_1^0  & \phi_2^+  \\
 \phi_1^-  & \phi_2^0 \\
\end{pmatrix} \sim (1,2,2,0)\, , \hspace{15mm}
%%%%%%%%%%%%%%%%%%%%%%%%%5
\eta^+ \sim (1,1,1,2) .
\label{eq:Hspec}
\end{align}
%%%%%%%%%
The purpose of the $\chi_R$ Higgs field is to achieve $SU(2)_R \times U(1)_{B-L}$ symmetry breaking down to $U(1)_Y$.  The $\chi_L$ field is the parity partner of $\chi_R$, which takes part in electroweak symmetry breaking.  The $\Phi$ field is used to generate fermion masses.  Since the $\chi_R$ field cannot couple to the fermions, the $\nu_R$ fields would not acquire Majorana masses at tree-level.  The singlet scalar $\eta^+$ does have lepton number violating Yukawa couplings to $\nu_R$, which induce Majorana masses via two-loop diagrams (as well as sub-dominant one-loop diagrams), which we shall evaluate carefully in Sec. \ref{sec:scalar}.

All neutral components of the Higgs fields acquire nonzero VEVs, which are parameterized as follows:
\begin{align}
  \langle \Phi \rangle =  \frac{1}{\sqrt{2}}\begin{pmatrix}
 \kappa & 0 \\
 0 &\hspace{3mm} \ \kappa^\prime e^{i\alpha} \\
\end{pmatrix} \, , \hspace{7mm} 
%%%%%%%%%%%%%%%%%%%%%
\langle \chi_L \rangle = 
\frac{1}{\sqrt{2}}\begin{pmatrix}
 0 \\
  v_L e^{i\theta_L} \\
\end{pmatrix} \, , \hspace{7mm} 
%%%%%%%%%%%%%%%%%%%%%%%%%
\langle \chi_R \rangle = 
\frac{1}{\sqrt{2}} \begin{pmatrix}
 0 \\
 v_R \\
\end{pmatrix} \, .
\label{eq:vev}
\end{align}
%%%%%
Here the VEVs $\kappa$ and $v_R$, which can be complex in general, have been made real by $SU(2)_L$ and $SU(2)_R$ gauge transformations.  In order to accommodate the success of the standard $(V-A)$ theory of weak interactions, the VEVs should obey the hierarchy $v_R \gg \kappa, \kappa', v_L$.  Such a hierarchical structure would lead to the $W_R^\pm$ gauge boson being much heavier than the $W_L^\pm$ boson, which is a phenomenological requirement to satisfy low energy weak interactions constraints.  For example, $K^0-\overline{K^0}$ mixing constraint limits the mass of $W_R^\pm$ to be $M_{W_R} \geq 1.6$ TeV \cite{Beall:1981ze}. (The mass of the $W_R^\pm$ gauge boson is proportional to $v_R$, while that of the $W_L^\pm$ is proportional to $\sqrt{\kappa^2 + \kappa'^2 + v_L^2}$.)  Furthermore, direct searches for dijet resonances at the LHC has set a limit of 3.6 TeV on the mass of the $W_R^\pm$ boson \cite{Sirunyan:2019vgj}, which also suggests the VEV hierarchy.\footnote{This limit arises from high-mass resonance searches in the dijet channel, which is applicable to the model presented here.  A slightly more stringent limit arising from searches for same sign or opposite sign dilepton final states is not applicable to the present model, as the $\nu_R$ fields have MeV scale masses here and won't decay within the detector. See discussion in Sec. \ref{sec:collider}.}  

The most general Yukawa interaction of quark and leptons with the Higgs fields of the model is given by
\begin{align}
     -\mathcal{L}_Y &=  
     %%%%%%%%%%%%%%%%%%%%%%%%%
     \overline{\Psi}_{aL} \, (y_{ab}\  \Phi +\widetilde{y}_{ab}\ \widetilde{\Phi}) \, \Psi_{bR} + \overline{Q}_{aL} \, (Y_{ab}\ \Phi + \widetilde{Y}_{ab}\ \widetilde{\Phi}) \, Q_{bR} \nonumber\\[3pt]
     %%%%%%%%%%%%%%%%%%%%%%%%%%
   &+ f_{ab}^L\, (\Psi_{aL}^i C \Psi_{bL}^j) \, \epsilon_{ij} \, \eta^+ + f_{ab}^R \,  (\Psi_{aR}^i C \Psi_{bR}^j) \, \epsilon_{ij} \, \eta^+ + \text{H.c.} \, ,
   \label{eq:yuk}
\end{align}
where $\widetilde{\Phi} = \tau_2 \Phi^\star \tau_2$, $C$ is the charge conjugation matrix, and ($i,j$) and ($a,b$) stand respectively for $SU(2)$ and generation indices. The couplings $y$, $\widetilde{y}$, $Y$, $\widetilde{Y}$, $f^L$, and $f^R$ are 3 $\times$ 3 Yukawa coupling matrices, with $f_{ab}^{L,R} = - f_{ba}^{L,R}$ required by Lorentz symmetry. Under left-right Parity symmetry ($P$), the fermions and scalar fields transform as follows:
\begin{equation}
    \Phi \leftrightarrow \Phi^{\dagger}, \hspace{5mm} \widetilde{\Phi} \leftrightarrow \widetilde{\Phi}^{\dagger}, \hspace{5mm} \chi_L \leftrightarrow \chi_R , \hspace{5mm} \eta^+ \leftrightarrow \eta^+ ,  \hspace{5mm} Q_L \leftrightarrow Q_R, \hspace{5mm} \Psi_L \leftrightarrow \Psi_R \, 
    \label{eq:invariant}
\end{equation}
along with $W_L \leftrightarrow W_R$. For most of our discussions we shall assume $P$ to be exact, in which case the Yukawa coupling matrices obey the following relations:
\begin{equation}
    y = y^\dagger \, , \hspace{5mm} \widetilde{y}= \widetilde{y}^\dagger  \, , \hspace{5mm} Y = Y^\dagger \, , \hspace{5mm} \widetilde{Y} = \widetilde{Y}^\dagger \, , \hspace{5mm} f^L = f^R \equiv f \, , \hspace{5mm}
\end{equation}
%%%%
Once the Higgs fields acquire VEVs, fermion masses are generated with the mass matrices for up and down quarks ($M_u$ and $M_d$),  charged leptons ($M_\ell$), and Dirac neutrinos  ($M_{\nu^D}$) given by
\begin{eqnarray}
    M_u &=& \frac{1}{\sqrt{2}}\, (Y \, \kappa + \widetilde{Y} \, \kappa^{\prime} e^{-i \alpha} )\, , \hspace{15mm}
    M_d = \frac{1}{\sqrt{2}}\, (Y \, \kappa^\prime  e^{i \alpha} + \widetilde{Y} \, \kappa) \, , \label{eq:dM} \\[5pt]
    %%%%%%%%%%%%
    M_\ell &=& \frac{1}{\sqrt{2}}\, (y \, \kappa^\prime  e^{i \alpha} + \widetilde{y} \, \kappa) \, ,   \hspace{15mm}  
    M_{\nu^D} = \frac{1}{\sqrt{2}}\, (y \, \kappa + \widetilde{y} \, \kappa^{\prime}  e^{-i \alpha}) \, ,\label{eq:nuM}
\end{eqnarray}
%%%%%%%%%%%%%%%%%%%%%%%%%%%%%%%%%5
These relations can be inverted to express the Yukawa coupling matrices in terms of the mass matrices:
\begin{align}
    y &= \frac{\sqrt{2}}{\kappa^2 - \kappa'^2}\ (\kappa M_{\nu^D} - \kappa' e^{-i \alpha} M_\ell)  \, , \hspace{10mm} 
    \widetilde{y} = \frac{\sqrt{2}}{\kappa^2 - \kappa'^2}\ (\kappa M_{\ell} - \kappa' e^{i \alpha} M_{\nu^D}) \, , \label{eq:yuk1}\\
    Y &= \frac{\sqrt{2}}{\kappa^2 - \kappa'^2}\ (\kappa M_{u} - \kappa' e^{-i \alpha} M_d)  \, , \hspace{10mm}  
    \widetilde{Y} = \frac{\sqrt{2}}{\kappa^2 - \kappa'^2}\ (\kappa M_{d} - \kappa' e^{i \alpha} M_{u}) \, . 
    \label{eq:yukk}
    %%%%%%%
\end{align}
This assumes that $\kappa \neq \kappa'$, which has to be true for phenomenology, otherwise the masses of the up-type quarks would equal those of the down-type quarks.  These relations, Eq. (\ref{eq:yuk1}), provide important constraints on the loop-induced Majorana masses of the $\nu_R$ fields, especially when the $W_R^\pm$ mass is near the TeV scale.  In this case the {\it a priori} arbitrary Dirac neutrino mass matrix $M_{\nu^D}$ should have very small entries so that the light neutrino masses obtained from the seesaw formula are in the sub-eV range.  Thus, the $\nu_R$ masses will be solely proportional to the charged lepton masses, as shown in Eq. (\ref{op2}).  

Another observation about the Yukawa coupling relations of Eq. (\ref{eq:yukk}) is that the ratio $|\kappa'/\kappa|$, which can be taken to be $\leq 1$ without loss of generality, cannot be too close to 1, or else the top quark Yukawa coupling would become in the non-perturbative regime.  If we demand that the top Yukawa coupling not be larger than a reasonable perturbative value of $1.5$, we obtain an upper limit of $|\kappa'/\kappa| \leq (0.578,\,0.616,\, 0.645  )$, corresponding to the left-right symmetry breaking scale $v_R$ being $(1,\,10,\,100)$ TeV. These numbers are obtained by evolving the top quark Yukawa coupling, along with the standard model gauge couplings, from low energies to the scale $v_R$, which yields $Y_t = (0.865,\,0.793,\,0.736)$ at these scales.  These upper limits on $|\kappa'/\kappa|$ would be relevant in our discussion of $W_L^\pm-W_R^\pm$ mixing, especially in the context of supernova 1987a energy loss constraints, see Sec. \ref{Sec:sn}.

Since the $\nu_R$ fields acquire Majorana masses, the $6 \times 6$ neutrino mass matrix spanning $(\nu,\, \nu^c)$ fields can be written down as
\begin{equation}
    M_\nu = \left(
\begin{array}{cc}
 M_\nu^L  & M_{\nu^D}  \\
 M_{\nu^D}^T  &  M_\nu^R \\
\end{array}
\right) \, ,
\label{eq:SeesawMat}
\end{equation}
%%%%%%%%%%%%%%%%%%%%%5
where $M_{\nu^D}$ is given by Eq.~(\ref{eq:nuM}), and $ M_\nu^L$ and $M_\nu^R$ will arise through one-loop and two-loop radiative correction (cf. Sec. \ref{sec:numass}).
Assuming that  $M_{\nu^D} \ll M_\nu^R$, the $3 \times 3$ light neutrino mass matrix can be obtained as
\begin{equation}
    M_\nu^{\rm light} = M_\nu^L - M_{\nu^D} (M_\nu^R)^{-1} M_{\nu^D}^T \, ,
    \label{eq:lightnu}
\end{equation}
which explains the smallness of the neutrino mass.  The eigenvalues of the heavier states in Eq. (\ref{eq:lightnu}) are the same as the eigenvalues of $M_\nu^R$ in this approximation, which we shall evaluate in Sec. \ref{sec:numass}.  As for the light neutrino masses, if the second  (first) term in Eq. (\ref{eq:lightnu}) dominates over the first (second) term, it is the type-I (type-II) seesaw domination.  We shall investigate both options, but our results show that the model can support only the type-I seesaw scenario.

\section{Scalar Sector}\label{sec:scalar}

In this section we analyze the Higgs potential of the LR symmetric model without Higgs triplets. We shall assume Parity symmetry, as defined in Eq.~\eqref{eq:invariant}. The most general renormalizable Higgs potential involving $\Phi, \chi_L, \chi_R$, and $\eta^+$ fields is given by:
\vspace{1.5mm}
%%%%%%%%%%%%%%%%%%%%%%%%%
\begin{align}
  V \ =& -\mu _1^2\ tr(\Phi^{\dagger}\Phi) - \mu_2^2\ [tr(\tilde{\Phi}\Phi^{\dagger})+ tr(\tilde{\Phi}^{\dagger} \Phi)] - \mu_3^2\ [\chi_L^{\dagger} \chi_L + \chi_R^{\dagger} \chi_R] + \mu_\eta^2\ |\eta|^2  \nonumber \\[5pt]
    %%%%%
    &
 + \lambda_\eta\ |\eta|^4 +  \lambda_1\ tr(\Phi^{\dagger}\Phi)^2 + \lambda_2\ [ tr(\tilde{\Phi} \Phi^{\dagger})^2 
 + tr(\tilde{\Phi}^{\dagger} \Phi)^2 ]  + \lambda_3\ tr(\tilde{\Phi} \Phi^{\dagger}) \, tr(\tilde{\Phi}^{\dagger} \Phi)  \nonumber \\[5pt]
 %%%%%%
 & 
+ \lambda_4\ tr(\Phi^{\dagger}\Phi) \, [tr(\tilde{\Phi}\Phi^{\dagger})+ tr(\tilde{\Phi}^{\dagger}\Phi)]  + \rho_1\  [(\chi_L^{\dagger} \chi_L )^2 + (\chi_R^{\dagger} \chi_R )^2]
 + \rho_2\  \chi_L^{\dagger} \chi_L \chi_R^{\dagger}\chi_R       \nonumber\\[5pt]
 %%%%%
 &
 + \mu_4\  [\chi_L^{\dagger} \Phi \chi_R + \chi_R^{\dagger} \Phi^{\dagger} \chi_L] + \mu_5\  [\chi_L^{\dagger} \tilde{\Phi} \chi_R + \chi_R^{\dagger}\tilde{\Phi}^{\dagger}\chi_L ] + \alpha_1\ tr(\Phi^{\dagger} \Phi ) [\chi_L^{\dagger}\chi_L + \chi_R^{\dagger}\chi_R ] 
   \nonumber\\[5pt]
 %%%%%%
 &
 + \Big\{ \alpha_2 e^{i \delta}\ [\chi_L^{\dagger} \chi_L  tr(\tilde{\Phi} \Phi^{\dagger} ) + \chi_R^{\dagger} \chi_R  tr(\tilde{\Phi}^{\dagger} \Phi )] + {\rm H.c.} \Big\} + \alpha_3\ [\chi_L^{\dagger} \Phi \Phi^{\dagger}\chi_L + \chi_R^{\dagger} \Phi^{\dagger} \Phi  \chi_R  ]   \nonumber \\[5pt]
 %%%%%%%
 &  
  %%%
  + \Big\{ \alpha_4\ [\chi_L^T i \tau_2 \Phi \chi_R \eta^- + \chi_R^T i \tau_2 \Phi^\dagger \chi_L \eta^- ] + {\rm H.c.} \Big\} + \alpha_5\ |\eta|^2 tr(\Phi^\dagger \Phi) \nonumber \\[5pt]
  %%%%%%%%
  &
    + \alpha_6\ |\eta|^2 [tr(\tilde{\Phi} \Phi^\dagger) + tr(\tilde{\Phi}^\dagger \Phi)] + \alpha_7\ |\eta|^2 [\chi_L^\dagger \chi_L + \chi_R^\dagger \chi_R] \, .
  \label{eq:potential} 
\end{align}
%%%%%%%%%%%%%%%%%%%%%%%%%%%%%%%%5
%%%%%%%%%%%%%%%%%%%%%%%%%%%%%%%%%%%%%%%%5
Here all the couplings, save $\alpha_2$, have been made real by field redefinitions.   Certain additional invariants,  such as the one obtained from the $\alpha_4$ term by replacing $\Phi$ by $\tilde{\Phi}$, can be shown to be not independent. Inserting the VEVs of Eq.~(\ref{eq:vev}) in Eq.~(\ref{eq:potential}), we require the following conditions for the potential to be an extremum:
%%%%%%%%%%%%%%%%%%%%%%%%%%%%%%%%%%%%%%%%%%%
\begin{equation}
     \frac{\partial V}{\partial \kappa} = \frac{\partial V}{\partial \kappa'} = \frac{\partial V}{\partial v_L} = \frac{\partial V}{\partial v_R} =\frac{\partial V}{\partial \theta_L} =\frac{\partial V}{\partial \alpha} =  0
    \label{eq:mincond}
\end{equation}
%%%\cite{Deshpande:1990ip, Zhang:2007da}
These conditions lead to six relations among the VEVs and various Higgs potential parameters:
\begin{align}
    \hspace{12mm} 0\ =&\ \lambda_1 \kappa \kappa_+^2 - \kappa \mu_1^2 - 2 \kappa' \mu_2^2 \cos \alpha + \frac{1}{2} \alpha_1  \kappa (v_L^2 + v_R^2)  \nonumber\\
    %%%%%%%%%%
    & \hspace{1mm} + \kappa' \{ \lambda_4 \cos\alpha (3 \kappa^2 + \kappa'^2) + 2 \kappa \kappa' (\lambda_3 + 2 \lambda_2 \cos(2\alpha))\}  \nonumber\\
    %%%
    &\hspace{1mm} + \alpha_2 \kappa' \{ v_L^2 \cos(\delta-\alpha) + v_R^2 \cos(\delta+\alpha) \}  + \frac{\mu_5}{\sqrt{2}} v_L v_R \cos \theta_L \, ,  \\[3pt]
    %%%%%%%%%
    %%%%%%%%%%
    0\ =&\ \lambda_1 \kappa' \kappa_+^2 - \kappa' \mu_1^2 - 2 \kappa \mu_2^2 \cos \alpha + \frac{1}{2} (\alpha_1 + \alpha_3) \kappa' (v_L^2 + v_R^2)  \nonumber\\
    %%%%%%%%%%
    & \hspace{1mm} + \kappa \{ \lambda_4 \cos\alpha (3 \kappa'^2 + \kappa^2) + 2 \kappa \kappa' (\lambda_3 + 2 \lambda_2 \cos(2\alpha))\}  \nonumber\\
    %%%
    &\hspace{1mm} + \alpha_2 \kappa \{ v_L^2 \cos(\delta-\alpha) + v_R^2 \cos(\delta+\alpha) \} + \frac{\mu_4}{\sqrt{2}} v_L v_R \cos (\theta_L-\alpha) \, , \\[3pt]
        %%%%%%
        %%%%%%%%%%
    0\ =&\  \alpha_1 \kappa_+^2 v_L - 2 \mu_3^2 v_L + \sqrt{2} v_R \{ \kappa \mu_5 \cos \theta_L + \kappa' \mu_4 \cos(\theta_L - \alpha) \} \nonumber \\
   %%%%%%%%
    & \hspace{1mm} + v_L \{ 2 \rho_1 v_L^2 + \rho_2 v_R^2  + \alpha_3 \kappa'^2 + 4 \alpha_2 \kappa \kappa' \cos(\delta-\alpha) \} \, , \\[7pt]
  %%%%%%%%%%%%%%%%%%%
  %%%%%%%%%%%%%%
  %%%%%%%%%%%%%%
    0\ =&\  \alpha_1 \kappa_+^2 v_R - 2 \mu_3^2 v_R + \sqrt{2} v_L \{ \kappa \mu_5 \cos \theta_L + \kappa' \mu_4 \cos(\theta_L - \alpha) \} \nonumber \\
   %%%%%%%%
    & \hspace{1mm} + v_R \{ 2 \rho_1 v_R^2 + \rho_2 v_L^2  + \alpha_3 \kappa'^2 + 4 \alpha_2 \kappa \kappa' \cos(\delta+\alpha) \} \, , \\[7pt]
  %%%%%%%%%%%%%%%%%%%
  %%%%%%%%%%%%%%
  %%%%%%%%%%%%%%
  0\ =&\ v_L v_R \{ \kappa \mu_5 \sin\theta_L + \kappa' \mu_4 \sin(\theta_L-\alpha) \}  \, ,\\[7pt]
 %%%%%%%%%%%%%
  %%%%%%%%%%%%%%
  0\ =&\ 2 \kappa \kappa' \mu_2^2 \sin\alpha - 8 \kappa^2 \kappa'^2 \lambda_2 \cos\alpha \sin\alpha - \kappa \kappa' \kappa_+^2 \lambda_4 \sin\alpha \nonumber\\
  %%%%%%%%%%
  & \hspace{1mm} \alpha_2 \kappa \kappa' \{ v_L^2 \sin(\delta-\alpha) - v_R^2 \sin(\delta + \alpha) \} + \frac{1}{\sqrt{2}} \kappa' \mu_4 v_L v_R \sin(\theta_L - \alpha) \, . 
    \label{eq:Msq}
\end{align}
%%%%%%%%%%%%%%%%%%%%%%%%%%%%%%%%%%%%%
Here and in what follows  we shall take $|\kappa| \geq |\kappa'|$ without loss of generality and define
\begin{equation}
    \kappa_{\mp}^2 = \kappa^2 \mp \kappa^{\prime2} \, . \label{eq:kappa}
\end{equation}
For simplicity in  presenting the scalar mass spectrum, we shall assume that the Higgs potential parameters as well as the VEVs are all real. That is, we set $\alpha$ and $\theta_L$ of Eq.~\eqref{eq:vev} to zero, which is an allowed solution if the phase $\delta$ of Eq. (\ref{eq:potential}) is taken to be zero.  In this case the last two of Eq. (\ref{eq:Msq}) are automatically satisfied. From the remaining conditions of Eq. (\ref{eq:Msq}) we eliminate the mass parameters $\{\mu_1^2,\,\mu_2^2,\,\mu_3^2,\,\mu_5\}$ in favor of the VEVs $\{v_L,\,v_R,\,\kappa,\,\kappa'\}$, which are taken to be independent parameters and express the mass matrices in terms of these VEVs, the quartic couplings, one cubic scalar coupling parameter $\mu_4$, and $\mu_\eta^2$ which determines the mass of $\eta^+$.   

The mass matrix for the charged Higgs bosons $M_+^2$ is first constructed in a basis $\{ \phi_1^+, \phi_2^+, \chi_L^+,$ $\chi_R^+, \eta^+ \}$  by expanding the potential about the minimum given in  Eq.~(\ref{eq:vev}) to quadratic order.  This $5 \times 5$ matrix contains two massless modes, those associated with the massive gauge bosons $W_R^\pm$ and $W_L^\pm$, which we denote as $G_L^{ +}$ and  $G_R^{ +}$. We make a rotation by an orthogonal matrix $O^+$ that removes these two massless modes from the $5 \times 5$ matrix.  The intermediate states are denoted as $\{ G_L^{ +}, G_R^{ +}, h_1^{\prime +}, h_2^{\prime +}, h_3^{\prime +}\}$. The explicit rotation matrix $O^+$ to go to this intermediate basis is given in Eq. (\ref{eq:Op}) of Appendix \ref{sec:ChargedMass}. This rotation matrix depends only on the ratios of VEVs, which are our independent parameters. 
The $3 \times 3$ mass matrix for the remaining states $\{h_1^{\prime +}, h_2^{\prime +}, h_3^{\prime +}\}$ is presented in Eq. (\ref{eq:A6}) of Appendix \ref{sec:ChargedMass}. A subsequent rotation would bring this $3 \times 3$ matrix to a diagonal form, which is not explicitly carried out. We denote this rotation matrix as  $O'^+$.  The full transformation that takes the original charged scalar states to the mass eigenstates, which are denoted as $\{H_1^+, H_2^+, H_3^+\}$ is then  $V^+ = (O'^+ O^+)^T$.  

In an analogous fashion we remove the two Goldstone states $(G_1^{0},\, G_2^{0}$), corresponding to the $Z_L$ and $Z_R$ gauge bosons, from the $4 \times 4$ pseudoscalar mass matrix constructed in the initial basis $\{ \phi_1^{0i} , \phi_2^{0i}, \chi_L^{0i}, \chi_R^{0i} \}$. Here
the superscript $i$ refers to the imaginary components of the relevant fields.  This is achieved by a rotation matrix $O^i$, which is given in Eq. (\ref{eq:Oi}) of Appendix \ref{sec:neutralMass}.  The remaining $2 \times 2$ mass matrix is diagonalized by a second rotation matrix denoted as $O{'^i}$. The  mass eigenstates are denoted as $\{ G_1^{0} , G_2^{0}, A_1, A_2 \}$ and the $2 \times 2$ mass matrix for the massive pseudoscalar fields is given in Eq. (\ref{eq:B4}) of Appendix \ref{sec:neutralMass}. 

The $4 \times 4$ mass matrix for the real scalar bosons contains no zero modes. However, we rotate this matrix to an intermediate basis by a rotation matrix $O^r$ so that the SM-like Higgs boson is easily identifiable.  A second rotation by $O'^{r}$ would diagonalize this mass matrix. The physical states are denoted as
$\{ h^0 , H_1^{0}, H_2^{0}, H_3^{0} \}$.  The rotation matrix is given in Eq.~\eqref{eq:B4} and the mass matrix is given in \eqref{eq:B8} of Appendix. \ref{sec:neutralMass}. 

The full rotation that is performed in the various sectors can then be summarized as follows:
\begin{align}
    \begin{pmatrix}
    \phi_1^+ \\
    \phi_2^+ \\
    \chi_L^+ \\
    \chi_R^+ \\
     \eta^+ 
    \end{pmatrix} = V^+
    \begin{pmatrix}
    G_L^{ +} \\
    G_R^{ +} \\
    H_1^{+} \\
    H_2^{+} \\
    H_3^{+} \\
    \end{pmatrix} \, , \hspace{5mm}
    %%%%%%
    %%%%%%
    \begin{pmatrix}
    \phi_1^{0r} \\
    \phi_2^{0r} \\
    \chi_L^{0r} \\
    \chi_R^{0r} 
    \end{pmatrix} = V^r
    \begin{pmatrix}
    h^0 \\
    H_1^0 \\
    H_2^0 \\
    H_3^0
    \end{pmatrix} \, , \hspace{5mm}
    %%%%%
    %%%%%
    \begin{pmatrix}
    \phi_1^{0i} \\
    \phi_2^{0i} \\
    \chi_L^{0i} \\
    \chi_R^{0i} 
    \end{pmatrix} = V^i
    \begin{pmatrix}
    G_1^{0} \\
    G_2^{0} \\
    A_1 \\
    A_2
    \end{pmatrix} \, ,
    \label{eq:diagScal}
\end{align}
where
\begin{equation}
    V^+ = (O'^+ O^+)^T \, , \hspace{5mm} V^r = (O'^r O^r)^T \, , \hspace{5mm} V^i = (O'^i O^i)^T \, .
    \label{eq:Unit}
\end{equation}
%%%%%%%

We show in Table \ref{tab:Higgs} approximate expression for the physical Higgs states and their masses, in the approximation $v_R \gg v_L, \kappa, \kappa'$. 
We define the ratios
\begin{equation}
    \epsilon = \frac{\kappa'}{\kappa}\, , \hspace{10 mm} \epsilon' = \frac{v_L}{\kappa}.
    \label{eq:eep}
\end{equation}
and keep only liner terms in $\epsilon$ and $\epsilon'$ in the expressions given in Table \ref{tab:Higgs}.
In these limits, the Goldstone modes associated with charged scalars read as
\begin{align}
    G_1^+ &\simeq - \phi_1^+ + \epsilon \phi_2^+ + \epsilon' \chi_L^+ \nonumber \\
    G_2^+ &\simeq \chi_R^+~.
\end{align}
Similarly, the Goldstone modes associated with pseudoscalars read as
\begin{align}
    G_1^0 &\simeq - \phi_1^{0i} + \epsilon \phi_2^{0i} + \epsilon' \chi_L^{0i} \nonumber \\
    G_2^0 &\simeq \chi_R^{0i}~.
\end{align}
After reducing the charged, scalar and pseudoscalar mass matrices by removing the respective Goldstone modes, there still remains some mixing between heavy states. In the approximation made here, there is one mixing angle  denoted as $\omega$, which is defined as
\begin{equation}
    \tan 2\omega = \frac{2 \sqrt{2} \mu_4}{(\alpha_3+ (2\rho_1-\rho_2)) v_R} \, .
\end{equation}
The approximate mass eigenvalues of Table \ref{tab:Higgs} are functions of this angle.

\begin{table}[!t]
    \centering
    \begin{tabular}{|c|c|}
    \hline \hline
       Higgs state  & Mass  \\\hline \hline
         $H_1^+ \simeq ( \cos\omega\ \epsilon - \sin\omega\ \epsilon') \phi_1^+ + \cos\omega\  \phi_2^+ - \sin\omega\ \chi_L^+$  &  $m_{H_1^+}^2 \simeq \frac{v_R}{4}  \{ (\alpha_3 - \rho_{12}) v_R - \sqrt{A} \} $ \\[3pt]
         %%%%%
         $H_2^+ \simeq -(\sin\omega\ \epsilon + \cos\omega\ \epsilon') \phi_1^+ - \sin\omega\ \phi_2^+ - \cos\omega\ \chi_L^+$  & $m_{H_2^+}^2 \simeq \frac{v_R}{4}  \{ (\alpha_3 - \rho_{12}) v_R + \sqrt{A} \} $ \\[3pt]
         %%%%%
         $H_3^+ \simeq \eta^+$ & $m_{H_3^+}^2 \simeq \mu_\eta^2 + \frac{\alpha_7}{2} v_R^2$  \\[3pt]
         \hline
         %%%%
         $A_1 \simeq (\cos\omega\  \epsilon + \sin\omega\ \epsilon')   \phi_1^{0i} + \cos\omega\  \phi_2^{0i} + \sin\omega\ \chi_L^{0i}  $ & $m_{A_1}^2 \simeq m_{H_1^+}^2$\\[3pt]
        %%%%%%
         $A_2 \simeq (-\sin\omega\ \epsilon + \cos\omega\ \epsilon' ) \phi_1^{0i} - \sin\omega\ \phi_2^{0i} + \cos\omega\ \chi_L^{0i}  $ &  $m_{A_2}^2 \simeq m_{H_2^+}^2$ \\[3pt]
       \hline
       %%%%
        $h^{0} \simeq \phi_1^{0r} + \epsilon \phi_2^{0r} + \epsilon' \chi_L^{0r} - \frac{\alpha_1 \kappa}{2 \rho_1 v_R} \chi_R^{0r}$ & $m_{h^{0}}^2 \simeq 2 \kappa^2 (\lambda_1 + 4 \epsilon \lambda_4 - \frac{\alpha_1^2}{4 \rho_1})$ \\[3pt]
      %%%%%%
       $H_{1}^{0} \simeq (\cos\omega\ \epsilon + \sin\omega\ \epsilon') \phi_1^{0r} - \cos\omega\ \phi_2^{0r} - \sin\omega\ \chi_L^{0r}$ & $m_{H_1^0}^2 \simeq m_{H_1^+}^2$ \\[3pt]
     %%%%%%
       $H_{2}^{0} \simeq (\sin\omega\ \epsilon - \cos\omega\ \epsilon') \phi_1^{0r} - \sin\omega\ \phi_2^{0r} + \cos\omega\ \chi_L^{0r}$ & $m_{H_2^{0}}^2 \simeq m_{H_2^+}^2$ \\[3pt]
     %%%%%
      $H_{3}^{0} \simeq \chi_{R}^{0r} + 
      \frac{\alpha_1 \kappa}{2 \rho_1 v_R} (\phi_1^{0r} + \epsilon \phi_2^{0r} + \epsilon' \chi_L^{0r})$ & $m_{H_3^{0}}^2 \simeq 2 \rho_1 v_R^2 $ \\
         \hline \hline
    \end{tabular}
    \caption{Physical Higgs eigenstates and mass spectrum at the leading order with $v_R \gg v_L, \kappa, \kappa'$, keeping only the linear terms in $\epsilon = \frac{\kappa'}{\kappa}$, $\epsilon' = \frac{v_L}{\kappa}$, and $v_R >> v_L, \kappa, \kappa'$, $\rho_{12} = 2 \rho_1 - \rho_2$ and $A = 8 \mu_4^2 + ( \alpha_3 + \rho_{12})^2 v_R^2$. Here $H_1^{0}$ is the standard model-like Higgs.  }
    \label{tab:Higgs}
\end{table}

%%%%%%%%%%%%%%%%%%%%%%%%%%%%%%%%%%%5
%%%%%%%%%%%%%%%%%%%%%%%%%%%%%%%%%%%%%%%
\subsection{Scalar sector in the electroweak symmetric limit}\label{sec:ewcon}
In the electroweak conserving limit, the charged, real and pseudoscalar mass matrix can be obtained from the Higgs potential of Eq.~\eqref{eq:potential} by setting the electroweak breaking VEVs $\kappa, \kappa',$ and $v_L$ to zero. The mass matrices for the charged, real, and pseudoscalar have the same structure and is diagonalized by a single $3\times 3$ unitary matrix $V$ such that
\begin{equation}
    V^\dagger M^2 V = M^2_{{\rm diag}} 
\end{equation}
where $M^2_{{\rm diag}}$ contains the physical  masses of the Higgs field. The mass matrix $M^2$ in this limit is found to be
\begin{equation}
M^2 = \begin{pmatrix} 
          -\mu_1^2 + \frac{\alpha_2}{2} v_R^2\  &\hspace{3mm} 2 \mu_2^2 - \alpha_2 v_R^2\ & -\frac{\mu_5 v_R}{\sqrt{2}} \\
          %%%%
         2 \mu_2^2 - \alpha_2 v_R^2\   &\hspace{3mm} -\mu_1^2 + \frac{\alpha_1+ \alpha_3}{2} v_R^2\ & \frac{\mu_4 v_R}{\sqrt{2}} \\
          %%%%%
         -\frac{\mu_5 v_R}{\sqrt{2}} & \frac{\mu_4 v_R}{\sqrt{2}} & \hspace{3mm} \frac{1}{2} (\rho_2-2\rho_1)v_R^2
      \end{pmatrix}
      \label{eq:ewmass}
\end{equation}
All elements of the mass matrix in Eq. (\ref{eq:ewmass}) are of order $v_R^2$. However, we wish to make one of the Higgs doublets light so that it can trigger electroweak symmetry breaking at a lower scale $v_L$. This light doublet state is identified as the SM Higgs doublet.  Making this state light is achieved by demanding that the determinant of $M^2$ in Eq. (\ref{eq:ewmass}) is nearly zero. It is only this zero-mass doublet that acquires a vacuum expectation value.  Thus, the VEVs $(\kappa, \,\kappa', \, v_L)$ are related to the rotation matrix $V$ as
\begin{equation}
    \frac{V_{21}}{V_{11}} = \frac{\kappa'}{\kappa}\, , \hspace{10mm} \frac{V_{31}}{V_{11}} = \frac{v_L}{\kappa}~.
    \end{equation}
Using the unitary nature of $V$, these relations would lead to the relations
\begin{align}
    V_{11} = \frac{\kappa}{\kappa_L} \, , \hspace{10mm} V_{21} = \frac{\kappa'}{\kappa_L} \, , \hspace{10mm} V_{31} = \frac{v_L}{\kappa_L} \, ,
    \label{eq:rel}
\end{align}
where $\kappa_L$ is the electroweak VEV, defined as
\begin{equation}
    \kappa_L^2 = \kappa^2 + \kappa'^2 + v_L^2 \, .
    \label{eq:kL}
\end{equation}
%%%%%%%%
The masses of the scalar fields $\eta^+$ and $\chi_R^{0r}$ are given by
\begin{align}
    m_\eta^2 &= \mu_\eta^2 + \frac{\alpha_7}{2} v_R^2 \, , \\
    m_{\chi_R^{0r}}^2 &= 2 \rho_1 v_R^2 \, .
\end{align}
The remaining scalar fields $\chi_R^+$ and $\chi_R^{0i}$ are Goldstone modes.

%%%%%%%%%%%%%%%%%%%%%%%%%%%%%%%%%%%5
%%%%%%%%%%%%%%%%%%%%%%%%%%%%%%%%%%%%%%%
\section{Gauge Boson Sector} \label{sec:gauge}

In this section, we derive the physical gauge boson eigenstates and their masses arising from the symmetry breaking sector given in Eq.~(\ref{eq:vev}). Noting that $\chi_{L} \to U_{L} \chi_{L}$, $\chi_{R} \to U_{R} \chi_{R}$, and $\Phi \to U_L \Phi U_R^\dagger$ under $SU(2)_L \times SU(2)_R$ gauge transformations parameterized by unitary matrices $U_L$ and $U_R$, the  Lagrangian containing the covariant derivatives of the scalar fields can be written down as
\begin{equation}
    \mathcal{L}_{gauge} = (D_\mu \chi_L)^\dagger D_\mu \chi_L + (D_\mu \chi_R)^\dagger D_\mu \chi_R + tr[(D_\mu \Phi)^\dagger D_\mu \Phi] \, ,
    \label{eq:gaugeL}
\end{equation}
where
\begin{eqnarray}
     D_{\mu} \chi_{L} &=&\partial_{\mu} \chi_{L}-\frac{1}{2} i g_L \, \vec{\tau} \cdot \vec{W}_{\mu L} \, \chi_{L} - \frac{1}{2} i g_{B-L} \, \chi_L B_\mu \, ,\nonumber\\ 
     %%%%
     D_{\mu} \chi_{R}&=&\partial_{\mu} \chi_{R}-\frac{1}{2} i g_R \, \vec{\tau} \cdot \vec{W}_{\mu R} \, \chi_{R} - \frac{1}{2} i g_{B-L} \,  \chi_R B_\mu \, ,\nonumber\\ 
     %%%%%%
     D_{\mu} \Phi &=&\partial_{\mu} \Phi-\frac{1}{2} i g_L \, \vec{\tau} \cdot \vec{W}_{\mu L} \Phi + \frac{1}{2} i g_R \, \Phi \, \vec{\tau} \cdot \vec{W}_{\mu R} \, .
     \label{eq:coD}
\end{eqnarray}
%%%
Under left-right parity $W_L \leftrightarrow W_R$, which implies $g_L = g_R$. The gauge boson mass matrices  are then obtained by substituting Eq.~(\ref{eq:vev}) into Eq.~(\ref{eq:gaugeL}) and using Eq.~(\ref{eq:coD}).  In the charged gauge boson sector, in the basis $(W_L^+, \, W_R^+)$, the mass matrix reads as
\begin{equation}
M_{W_{LR}^+}^2 = 
\frac{1}{4} \left(\begin{array}{cc}
 {g_{L}^{2}\ \kappa_L^2} & {-2 \ g_{L} g_{R} \kappa \kappa^{\prime} e^{i \alpha} } \\[5pt] 
 %%%%%%%%%%%%%%%%%%%%%%%
{- 2 \ g_{L} g_{R} \kappa \kappa^{\prime} e^{-i \alpha}} & {g_{R}^{2}\ \kappa_R^2}
\end{array}\right) \, .
\label{eq:gaugecha}
\end{equation}
where $\kappa_L$ is given by Eq.~\eqref{eq:kL} and  we  have defined 
\begin{equation}
    \kappa_R^2 = \kappa^2 + \kappa^{\prime 2} + v_R^2~.
    \label{eq:KR}
\end{equation}
To diagonalize this matrix, one can first write it as
\begin{equation}
    M_{W_{LR}^+}^2 = P\ \hat{M}_{W_{LR}^+}^2\ P^\star \, ,
\end{equation}
where $P = {\rm diag}\ (1,\ e^{-i \alpha})$ and $\hat{M}_{W_{LR}^+}^2$ is a real symmetric matrix. The phase contained in $P$ can be absorbed into the $W_R^+$ field with a redefinition. The real symmetric matrix $\hat{M}_{W_{LR}^+}^2$ can be straightforwardly diagonalized. We let $W_{1,2}^{\pm}$ denote the mass eigenstates such that
\begin{align}
    W_1^+ &= \cos\zeta\ W_L^+ + \sin\zeta\  W_R^+ \, , \nonumber\\
    %%%
    W_2^+ &= -\sin\zeta\ W_L^+ + \cos\zeta\ W_R^+ \, .
\end{align}
The mixing angle $\zeta$ is then identified as
\begin{equation}
    \tan 2\zeta = \frac{4 g_L g_R \kappa \kappa'}{g_R^2 \kappa_R^2 - g_L^2 \kappa_L^2}
    \label{eq:Wmix}
\end{equation}
The mass eigenvalues are found in the limit of $v_R >> \kappa,\kappa^\prime, v_L$ as
\begin{equation}
    M_{W_1}^2 \simeq \frac{1}{4} g_L^2 \kappa_L^2, \hspace{20mm} M_{W_2}^2 \simeq \frac{1}{4} g_R^2 v_R^2 \, .
    \label{eq:MassW}
\end{equation}
%%%%
The mixing angle $\zeta$ is constrained to be $|\zeta|\leq 4 \times 10^{-3}$ from strangeness changing nonleptonic decays of hadrons \cite{Donoghue:1982mx}, as well as from $b \to s\gamma$ decay \cite{Babu:1993hx}, independent of the mass of $W_2$.

Similarly, in the neutral gauge boson sector, the states $(W_{\mu L}^3 , W_{\mu R}^3, B_\mu)$ field will mix to produce $A_\mu, Z_{\mu L},$ and $Z_{\mu R}$, in analogy with the SM. The photon field $A_\mu$ remains massless, while the two orthogonal field $Z_{\mu L}$ and $Z_{\mu R}$ mix. It is convenient to choose the following basis: 
\begin{align} 
A_{\mu} &=\frac{g_{L} g_{R} B_{\mu}+g_{B} g_{R} W_{\mu L}^{3}+g_{L} g_{B} W_{\mu R}^{3}}{\sqrt{g_{B}^{2}\left(g_{L}^{2}+g_{R}^{2}\right)+g_{L}^{2} g_{R}^{2}}} \, ,\nonumber\\
%%%%
Z_{\mu R} &=-\frac{g_{B} B_{\mu}+g_{R} W_{\mu R}^{3}}{\sqrt{g_{R}^{2}+g_{B}^{2}}} \, , \nonumber\\
%%%%
Z_{\mu L} &=\frac{g_{B} g_{R} B_{\mu}-g_{L} g_{R}\left(1+\frac{g_{R}^{2}}{g_{R}^{2}}\right) W_{\mu L}^{3}+g_{B}^{2} W_{\mu R}^{3}}{\sqrt{g_{B}^{2}+g_{R}^{2}} \sqrt{g_{B}^{2}+g_{L}^{2}+ \frac{g_B^2 g_L^2}{g_R^2}}}  \, ,
\end{align}
where $g_B \equiv g_{B-L}$. The photon field decouples from the rest in the mass matrix, while the $Z_L - Z_R$ fields mix with a mass matrix given by 
\begin{equation}
M_{Z_{LR}}^{2}=\frac{1}{4}
\begin{pmatrix}
\left(g_{Y}^{2}+g_{L}^{2}\right) \kappa_{L}^{2} &
%%%
\hspace{5mm} \left(g_R^2 \kappa_+^2 - g_Y^2 \kappa_{L}^{2} \right) \sqrt{\frac{g_{R}^{2}+g_{L}^{2}}{g_{R}^{2}-g_{Y}^{2}}}  \\[5pt]
%%%%%%%%%%%%%%%%%%%%%%%%%%
 \left(g_R^2 \kappa_+^2 - g_Y^2 \kappa_{L}^{2} \right) \sqrt{\frac{g_{R}^{2}+g_{L}^{2}}{g_{R}^{2}-g_{Y}^{2}}}  &
 %%%%
 \hspace{5mm} - \frac{2 g_R^2 g_Y^2 \kappa_+^2}{g_R^2 - g_Y^2} + \frac{g_{R}^{4} \kappa_{R}^{2}}{g_{R}^{2}-g_{Y}^{2}} +\frac{g_{Y}^{2}  \kappa_{L}^{2}}{g_{R}^{2}-g_{Y}^{2}}
\end{pmatrix} \, ,
\label{eq:gaugeN}
\end{equation}
We have used the relation between $SU(2)_R$, $U(1)_{B-L}$, and hypercharge coupling ($g_R, g_B, g_Y$) in Eq.~\eqref{eq:gaugeN} to eliminate $g_B$ in favor of $g_Y$ \cite{Babu:2018vrl}. 
\begin{equation}
    Y = T_{R}^3 + \frac{B-L}{2} \hspace{5mm} \Longrightarrow \hspace{5mm} \frac{1}{g_Y^2} = \frac{1}{g_R^2} + \frac{1}{g_B^2}~.
\end{equation}
%%%
We obtain easily the eigenvalues of the matrix of Eq.~(\ref{eq:gaugeN}) by writing the mass eigenstates as $Z_{1,2}$: %%
\begin{align}
    Z_1 &= \cos\xi\ Z_L + \sin\xi\ Z_R \nonumber \, , \\
    Z_2 &= -\sin\xi\ Z_L + \cos\xi\ Z_R \, .
\end{align}
with
\begin{equation}
    \tan 2\xi\ \simeq \frac{2\ (g_R^2 \kappa_+^2 - g_Y^2 \kappa_{L}^{2})\ \sqrt{(g_R^2- g_Y^2) (g_R^2+ g_L^2)}}{g_R^4 \kappa_R^2}
\end{equation}
%%%%%%%%%%%%%%%%%%%%%%%%%%%
Thus, in the approximation $v_R \gg \kappa, \kappa^\prime, v_L$, the masses of neutral gauge bosons read as
\begin{equation}
M_{Z_{1}}^{2} \simeq \frac{1}{4}\left(g_{Y}^{2}+g_{L}^{2}\right) \kappa_{L}^{2}, \quad \quad \quad M_{Z_{2}}^{2} \simeq \frac{1}{4} \frac{g_{R}^{4}}{(g_{R}^{2}-g_{Y}^{2})} v_{R}^{2} \, .
\end{equation}
%%%
Here $Z_1$ is identified as the $Z$ gauge boson. Note that the mass ratio $M_{Z_2}/M_{W_2} \simeq 1.19$ in this model. The mixing angle $\xi$ is constrained to be small, of order $10^{-3}$ from electroweak precision observables, but this limit is automatically satisfied once the lower limit on the mass of $Z_2$ of about 5 TeV from LHC searches is imposed.

%%%%%%%%%%%%%%%%%%%%%%%%%%%%%%%%%%
%%%%%%%%%%%%%%%%%%%%%%%%%%%%%%%%%
\section{Generation of Radiative Majorana Mass for \texorpdfstring{$\nu_R$}{nuR} }\label{sec:numass}
%%%%%%%%%%%%%

The LR symmetric model without Higgs triplets does not generate Majorana masses for the $\nu_R$ at the tree level.  However, their interactions with the $\eta^+$ field does lead to lepton number violation, and the $\nu_R$ fields will develop Majorana masses through loop corrections.  Such a mechanism for generating masses for the usual neutrinos radiatively is well studied  \cite{Zee:1980ai,Zee:1985id,Babu:1988ki,Cai:2017jrq,Babu:2019mfe}; here we apply such a scheme for inducing $\nu_R$ masses. As noted in the introduction, there are one-loop diagrams which induce $\nu_R$ masses proportional to electroweak symmetry breaking, which were studied in Ref. \cite{Babu:1988qv} and more recently in Ref. \cite{FileviezPerez:2016erl}.  We analyze these contributions in detail and show that they are sub-leading to the two-loop induced masses which do not require electroweak breaking effects.  

\subsection{One-loop radiative correction}\label{sec:1_2loop}

\begin{figure}[!t]
  \subfigure[]{
   \includegraphics[height=3.9cm, width=6.8cm]{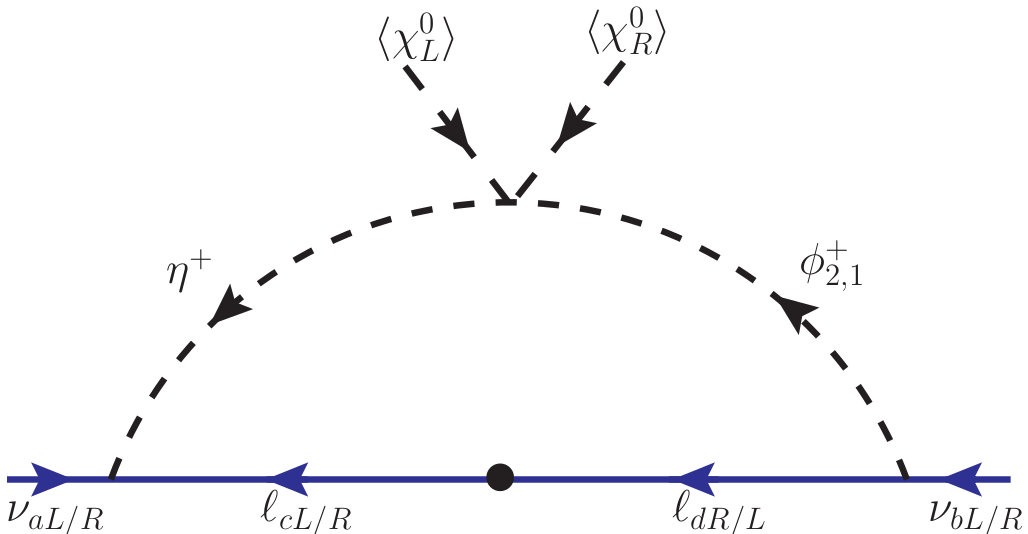}
   }\hspace{5mm}
  \subfigure[]{
    \includegraphics[height=4cm, width=7.0cm]{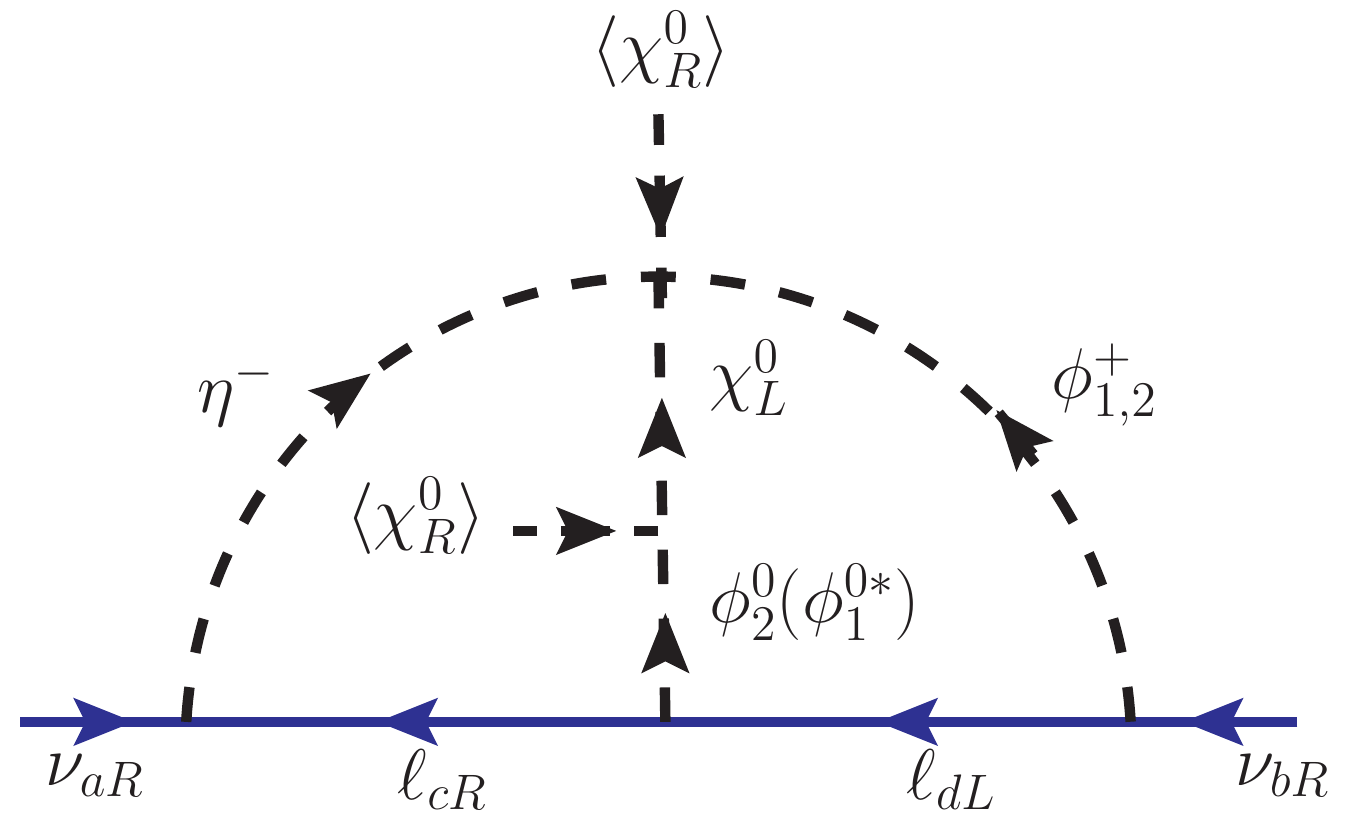}
    } 
  \caption{(a) Typical one-loop diagrams responsible to generate left/right-handed Majorana neutrino mass. (b) A typical two-loop diagram responsible to generate right-handed heavy Majorana neutrino mass.}
  \label{fig:loopdiag}
\end{figure}
%%%%%%%
%%%%%%
In this section we evaluate the one-loop contribution to right-handed Majorana neutrino masses. The relevant diagram is shown in Fig.~\ref{fig:loopdiag} (a). It is clear from this figure that the one-loop diagram requires two powers of electroweak symmetry breaking VEVs, one power arising from the charged lepton mass, and the other arising from $\eta^+\phi^+$ mixing.  Furthermore, one power of charged lepton Yukawa coupling of the $\phi^+$ scalar, which diminishes the induced mass.  While it is true that the $\phi^+$ Yukawa coupling has a contribution not proportional to the charged lepton Yukawa coupling, this contribution is proportional to the Dirac neutrino Yukawa coupling through the relations of Eq. (\ref{eq:yuk1}), which is even smaller for low scale $v_R$. These one-loop diagrams are suppressed by inverse powers of $v_R$, so raising $v_R$ will not make their contributions any large. These remarks are supported by our explicit computation, which we now summarize.

 The $5\times 5$ charged Higgs matrix is diagonalized by the unitary matrix $V^+$, as shown in Eq.~\eqref{eq:diagScal} and Eq.~\eqref{eq:Unit}. 
%%%%%%
The induced neutrino mass matrix arising from Fig.~\ref{fig:loopdiag} (a)  for both  $\nu_R$ and the light $\nu_L$ fields can be evaluated to be
\begin{align}
    (M_\nu^R)_{ab} &= \frac{1}{8 \pi^2}\ \big[f_{a\ell} M_\ell V_{5\beta}^+\ (y_{\ell b} V_{1 \beta}^{\star+} - \widetilde{y}_{\ell b} V_{2 \beta}^{\star+}) + (a \leftrightarrow b ) \big]\ \log\bigg(\frac{m_{H_1^+}^2}{m_{H_\beta^+}^2}\bigg) \, ,
    \label{eq:1loopvR}\\[5pt]
%%%%%%%%%%%%%%%%
    (M_\nu^L)_{ab} &= \frac{1}{8 \pi^2}\ \big[f_{a\ell} M_{\ell_k} V_{5\beta}^+\ (y_{k b}^\dagger V_{2 \beta}^{\star+} - \widetilde{y}_{k b}^\dagger V_{1 \beta}^{\star+}) + (a \leftrightarrow b ) \big]\ \log\bigg(\frac{m_{H_1^+}^2}{m_{H_\beta^+}^2}\bigg) \, . \label{eq:1loopvL} 
    %%%%%%%%%%%%%%%%%
\end{align}
where $H_{\beta}^+ = (H_1^+, H_2^+, H_3^+)$ with the index $\beta$ summed over the three different physical Higgs fields. Here we have assumed Parity symmetry so that $f^R = f^L = f$, which appears identically in the $\nu_R$ and the $\nu_L$ mass matrices.   Expanding Eq.~\eqref{eq:1loopvL} and Eq.~\eqref{eq:1loopvR}, the one-loop neutrino mass is obtained to be
\begin{align}
     M_\nu^R &= \frac{1}{8 \pi^2}\ \bigg[ (f M_\ell y^\dagger + y^\star M_\ell f^T) \bigg\{ O_{54}^{'+} (O_{31}^+ O_{34}^{'+} + O_{{41}}^+ O_{44}^{'+}) \log\bigg(\frac{m_{H_2^+}^2}{m_{{H_1^+}}^2}\bigg) \nonumber \\ 
     %%%
     & + O_{55}^{'+} (O_{31}^+ O_{35}^{'+} + O_{{41}}^+ O_{45}^{'+}) \log\bigg(\frac{m_{H_3^+}^2}{m_{{H_1^+}}^2}\bigg) \bigg\}  \nonumber \\
     %%%%%%%
     & - (f M_\ell \widetilde{y}^\dagger + \widetilde{y}^\star M_\ell f^T) \bigg\{ O_{54}^{'+} (O_{32}^+ O_{34}^{'+} + O_{{42}}^+ O_{44}^{'+}) \log\bigg(\frac{m_{H_2^+}^2}{m_{{H_1^+}}^2}\bigg)   \nonumber \\
     %%%%%%%%%%%
     & + O_{55}^{'+} (O_{32}^+ O_{35}^{'+} + O_{{42}}^+ O_{45}^{'+}) \log\bigg(\frac{m_{H_3^+}^2}{m_{{H_1^+}}^2}\bigg) \bigg\} \bigg] \, , \\[5pt]
     %%%%%%%%%%%%%%%%%%%%%%%%%%%%%%%%%%%%%55
     %%%%%%%%%%%%%%%%%%%%%%%%%%%%%%%%%%%%%%%%5
     M_\nu^L &= M_\nu^L\ (O_{32}^+ \leftrightarrow O_{31}^+\ , O_{42}^+ \leftrightarrow O_{41}^+ ) \, , 
\end{align}
%%%%%
where $O^+$ is the orthogonal matrix that takes the original charged scalar fields  to an intermediate basis, as shown in Eq.~\eqref{eq:Op} and $O'^+$ is the orthogonal matrix that takes the intermediate basis to the physical basis.  

We have computed the maximum allowed value of the $\nu_R$ mass arising from these one-loop diagrams by varying all parameters of the model within their allowed ranges.  Our results are plotted as a function of the $W_R^\pm$ mass in Fig. \ref{fig:1vs2}, along with the contributions arising from the two-loop diagrams.  Here we also show these one-loop induced masses when the assumption of Parity symmetry is relaxed, so that $f^L \neq f^R$.  It is clear from this figure that with the assumption of Parity, the maximum one-loop contribution to the $\nu_R$ mass is at most 10 eV, while without Parity this can be as large as an MeV or so, corresponding to TeV scale $W_R$.  Furthermore it is also clear from Fig. \ref{fig:1vs2} that the two-loop induced $\nu_R$ Majorana mass is always more important than the one-loop induced mass.  In our numerical evaluation of the mass, we have demanded that the Majorana masses of the $\nu_L$ fields arising from these diagrams do not exceed about 0.1 eV. This constraint restricts the maximum allowed one-loop $\nu_R$ mass significantly.

\subsection{Two-loop radiative corrections}

Now we analyze the two-loop induced $\nu_R$ Majorana masses in the model. The relevant diagram is shown in Fig. \ref{fig:1vs2} (b).  
The Yukawa  couplings and the Higgs potential coupling that are necessary to generate the neutrino mass at the two-loop level are given respectively in Eq.~(\ref{eq:yuk}) and in Eq.~(\ref{eq:potential}).  There are three more topologies similar to the one shown in Fig. \ref{fig:loopdiag} with the variation of the Higgs field inside the loop which are shown in  Fig. \ref{fig:Topology} of Appendix \ref{sec:Topology}. Since the external neutrino in the diagram are Majorana particles, each diagram has another set with internal particles replaced by their charge conjugates. The sum of these pairs of diagrams makes the neutrino mass matrix symmetric. The two-loop diagrams do not require electroweak symmetry symmetry breaking, so we work in this limit. In the electroweak conserving limit, the $\eta^+$ field does not mix with the doublet fields, and the charged and the neutral scalar mass matrices of the three $SU(2)_L$ doublets are identical, as shown given in Eq. (\ref{eq:ewmass}). Thus matrix is diagonalized by a single $3\times3$ unitary matrix, $V$, with the scalar $\eta^+$ remaining a mass eigenstate, and with the $\chi_R^+$ and $\chi_R^{0i}$ being Goldstone modes:
%%%%%
\begin{align}
\begin{pmatrix}
    \phi_1^+ \\
    \phi_2^+ \\
    \chi_L^+
    \end{pmatrix} = V
    \begin{pmatrix}
    G^+ \\
    H_1^{+} \\
    H_2^{+} 
    \end{pmatrix} \, , \hspace{5mm}
    %%%%%%
    %%%%%%
    \begin{pmatrix}
    \phi_1^{0r} \\
    \phi_2^{0r} \\
    \chi_L^{0r} 
    \end{pmatrix} = V
    \begin{pmatrix}
    h^0 \\
    H_1^0 \\
    H_2^0
    \end{pmatrix} \, , \hspace{5mm}
    %%%%%
    %%%%%
    \begin{pmatrix}
    \phi_1^{0i} \\
    \phi_2^{0i} \\
    \chi_L^{0i} 
    \end{pmatrix} = V
    \begin{pmatrix}
    G^{0} \\
    A_1 \\
    A_2
    \end{pmatrix} \, ,
    \label{eq:diagScalp}
\end{align}
%%%%%%%%%%%%%%

We carry out the evaluation of neutrino mass in Feynman gauge.  Here we must keep the Goldstone contributions, which however, have the same structure as the contributions from physical scalars. It is sufficient to identify the Goldstone boson masses as those of the $W$ and the $Z$ boson.  We also set the mass of the $h^0$ field to be 125 GeV.  In this gauge, we should also include possible contributions from the gauge bosons.  However, these contributions always require electroweak symmetry breaking, which will result in much smaller contributions.  

The most general Majorana mass matrix for the $\nu_R$ fields arising from Fig.~\ref{fig:loopdiag} (see Fig.~\ref{fig:Topology} for the complete set of diagrams) can be written as
\begin{align}
    (M_\nu^R)_{ab} = \sqrt{2}\ \alpha_4 v_R\ (A_{1ab} + A_{2ab} + A_{3ab} ) \, ,
    \label{eq:2loopMR}
\end{align}
where
\begin{align}
    A_{1ab} &= \Big\{ f_{ac}\ \big[ y_{cd}^\star\ V_{2\gamma}^\star \{ -V_{3\gamma} V_{1\beta} - V_{3\gamma} V_{2 \beta} + V_{2\gamma} V_{3\beta}\} - \widetilde{y}_{cd}^\star V_{1\gamma} V_{3\beta} V_{1\gamma}^\star \big] \nonumber\\
    %%%%%%%%
    &\hspace{5mm} \big( y_{db} V_{1\beta}^\star - \widetilde{y}_{db} V_{2\beta}^\star \big) +  (a \leftrightarrow b) \Big\}\ I_{cd}^{\eta \gamma \beta}\, ,\nonumber\\[5pt]
    %%%%%%%%%%%%%%%%
    %%%%%%%%%%%%%%%%
    A_{2ab} &= \Big\{ f_{ac}\ \big( y_{cd}^\star V_{2\beta}^\star - \widetilde{y}_{cd}^\star V_{1\beta}^\star \big)\ \big[ \widetilde{y}_{db}\ V_{2\gamma}^\star \{ -V_{3\gamma} V_{1\beta} - V_{3\gamma} V_{2 \beta} + V_{2\gamma} V_{3\beta}\} \nonumber\\
    %%%%%
    &\hspace{5mm} - y_{db} V_{1\gamma} V_{3\beta} V_{1\gamma}^\star \big]\ + (a \leftrightarrow b) \Big\}\  I_{cd}^{\eta \beta \gamma} \, , \nonumber\\[5pt]
    %%%%%%%%%%%%%%%%%%%%%5
    %%%%%%%%%%%%%%%%%%%%%%%
    A_{3ab} &= \Big\{ \big( y_{ca} V_{1\beta}^\star - \widetilde{y}_{ca} V_{2\beta}^\star \big)\ f_{cd}\ \big[ \widetilde{y}_{db}\ V_{2\gamma}^\star \{ -V_{3\gamma} V_{1\beta} - V_{3\gamma} V_{2 \beta} + V_{2\gamma} V_{3\beta}\} \nonumber\\
    %%%%%
    &\hspace{5mm} - y_{db} V_{1\gamma} V_{3\beta} V_{1\gamma}^\star \big]\ + (a \leftrightarrow b) \Big\}\ I_{cd}^{\beta \eta \gamma} \, .
    %%%%%%%%%%%%%%%%%%%%%5
    \label{eq:flavorst}
\end{align}
Here we have defined the two-loop integrals as
\begin{align}
    I_{cd}^{\eta \gamma \beta} &= \int \int \frac{d^4 p}{(2 \pi)^4} \frac{d^4 q}{(2 \pi)^4} \frac{q.p}{(q^2 - m_{\eta}^2)(q^2 - m_d^2)(p^2 - m_{H_\beta^+}^2)(p^2 - m_c^2)((p-q)^2 -  m_{H_\gamma^0}^2)} \, , \nonumber\\
    %%%%%%%%%
    I_{cd}^{\eta \beta \gamma} &= I_{cd}^{\eta \gamma \beta} (\gamma \leftrightarrow \beta)  \, , \nonumber\\
    I_{cd}^{\beta \eta \gamma} &=I_{cd}^{\eta \beta \gamma} (\eta \leftrightarrow \beta) \, .
    %%%%%%%%%%%
    \label{eq:Ir123}
\end{align}
%%%%%%%%%%%%%%%
Here $m_c$ and $m_d$ are the charged lepton masses. ($m_{\eta}$, $m_{H_\beta^+}$) and $m_{H_\gamma^0}$ are the masses of the charged and neutral scalars, respectively. The calculation of $I_{cd}^{\eta \gamma \beta}$ \cite{tHooft:1972tcz, vanderBij:1983bw, Broadhurst:1987ei, Ghinculov:1994sd,  Choudhury:1994vr, Adams:2013kgc, Kreimer:1991jv,Kreimer:1992zv, Frink:1997sg, Usyukina:1994eg} is carried out in Appendix \ref{sec:Icd}\footnote{In Appendix \ref{sec:Icd}, integral $I_{45}^{132} \equiv I_{cd}^{\alpha \gamma \beta}$ is evaluated as it is the most general case including electroweak symmetry breaking, whence $\eta^+$ mixes with the other charged Higgs fields.}.
%%%%%%%%%%%
In Eq. (\ref{eq:Ir123}) the indices $\beta$ and $\gamma$ are summed over different Higgs states, except $\eta^+$, which remains unmixed with the other states in this limit. $\alpha_4$ is the quartic coupling given in Eq.~(\ref{eq:potential}). $f$ and $y$ are Yukawa couplings given by Eq.~(\ref{eq:yuk}).  Here $A_1, A_2, A_3$ contain the flavor structures associated with each of the diagram in  Fig.~\ref{fig:Topology} (a), (b), and (c). Note that the flavor structure $(yfy^T)$ or $(\widetilde{y}f\widetilde{y}^T)$ has no contribution to $A_{3ab}$ in Eq.~\eqref{eq:flavorst} as the Yukawa  coupling matrix $f$ is antisymmetric in flavor and therefore these terms vanish when the conjugate diagrams are included.  

%%%%
\subsubsection{Case with \texorpdfstring{$M_{\nu^D} \ll M_\ell$}{mddd}}\label{sec:mdml}

To illustrate the calculation of two-loop induced  $\nu_R$ Majorana mass, we take $M_{\nu^D} \ll M_\ell$ which is realized in the case of low scale $W_R^\pm$ (cf. Sec.~\ref{sec:lowLR}). This makes both Yukawa couplings $y$ and $\widetilde{y}$ proportional to the charged lepton masses, simplifying the flavor structure of Eq.~\eqref{eq:flavorst}. Although each diagram is divergent in the physical basis of scalars, we show explicitly that the divergent piece of the integral vanishes due to unitarity conditions.
%%%%%%%%%%%%%%%%%%%%%%%%
%%%%%%%%%%%%%%%%%%%%%%%%%%%%%
The induced $\nu_R$ Majorana neutrino mass matrix takes the form in this limit given by
\begin{equation}
    (M_\nu^R)_{ab} =  \frac{2\sqrt{2}\ \alpha_4 v_R }{\kappa^2 (1-\epsilon^2)^2}\ (f M_\ell^2 + M_\ell^2 f^T)
    %%%%
     \, \Big( C_{\beta \gamma}\ I^{\eta \gamma \beta} + C'_{\beta \gamma}\ I^{\eta \beta \gamma} \Big)\, ,
    \label{eq:MajR}
 \end{equation}
where $I^{\eta \beta \gamma}$ stands for  $I_{cd}^{\eta \beta \gamma}$, as we neglect the charged lepton masses in comparison to the scalar masses in evaluating the integrals. Here we have defined
\begin{align}
    C_{\beta \gamma} = \epsilon^2\ \lambda_{\beta \gamma}^{1'} -\epsilon\ \lambda_{\beta \gamma}^{2'} - \epsilon\ \lambda_{\beta \gamma}^{1} +  \lambda_{\beta \gamma}^{2} \, , \nonumber \\
    %%%%%%%%%
     C'_{\beta \gamma} = -\epsilon^2\ \lambda_{\beta \gamma}^{2} +\epsilon\ \lambda_{\beta \gamma}^{2'} + \epsilon\ \lambda_{\beta \gamma}^{1} -  \lambda_{\beta \gamma}^{1'} \, .
     \label{eq:CCp}
\end{align}
and
\begin{align}
    \lambda_{\beta \gamma}^{1} &= |V_{1\gamma}|^2 V_{3\beta} V_{1\beta}^\star  \, , \nonumber\\
    %%%%%%%
    \lambda_{\beta \gamma}^{2} &= |V_{1\gamma}|^2 V_{3\beta} V_{2\beta}^\star  \, , \nonumber\\
    %%%%%%%
    \lambda_{\beta \gamma}^{1'} &=  -V_{2\gamma}^\star |V_{1\beta}|^2 V_{3\gamma} - V_{2\gamma}^\star V_{1\beta}^\star V_{3\gamma} V_{2 \beta} + |V_{2\gamma}|^2 V_{1\beta}^\star V_{3\beta} \, , \nonumber \\
    %%%%%%%
    \lambda_{\beta \gamma}^{2'} &=  -V_{2\gamma}^\star |V_{2\beta}|^2 V_{3\gamma} - V_{2\gamma}^\star V_{1\beta} V_{3\gamma} V_{2 \beta}^\star + |V_{2\gamma}|^2 V_{2\beta}^\star V_{3\beta} \, .
    \label{eq:coeffInt}
\end{align}
%%%%%%%%%%%%%%%%%%%%
Here the parameter $\epsilon$ is defined in Eq.~\eqref{eq:eep}. 

Each diagram with the variation of index $\beta$ and $\gamma$ is divergent. However, the divergent piece is independent of $\beta$ and $\gamma$; therefore, the sum of the diagrams will be convergent because of the following unitarity condition
\begin{equation}
   \sum_{\beta \gamma } \lambda_{\beta \gamma}^{1} = \sum_{\beta \gamma } \lambda_{\beta \gamma}^{2} = \sum_{\beta \gamma } \lambda_{\beta \gamma}^{1'} = \sum_{\beta \gamma } \lambda_{\beta \gamma}^{2'} = 0\,  .
\end{equation}
%%%%%
%%%%%%%%%%%
\begin{figure}[!t]
\centering
\captionsetup{justification=centering}
    \includegraphics[scale=0.36]{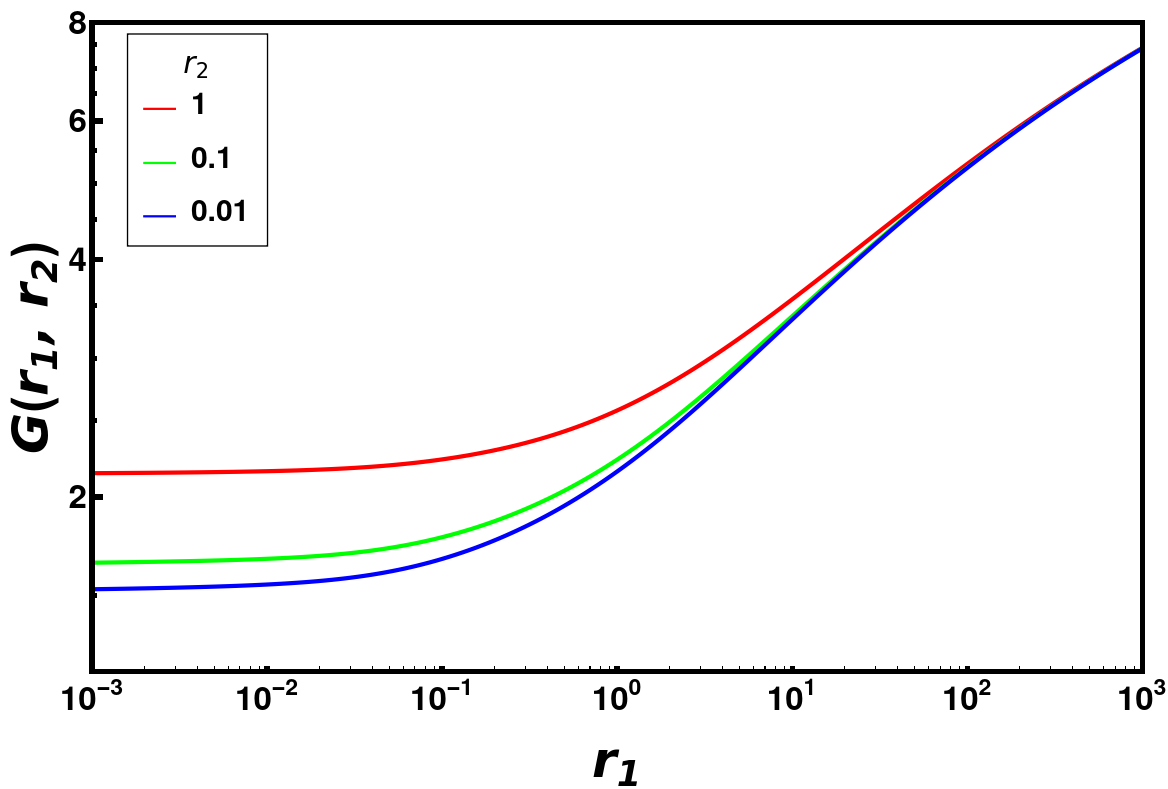}
\caption{Integral function in Eq.~(\ref{eq:Gplot}) as a function of mass ratios, $r_1$ and $r_2$. }
    \label{fig:integral_plot}
\end{figure}
%%%%%%%%%%%%%%%%%%%%%%%%%%%5
Summing over possible values of  $\beta$ and $\gamma$, the dimensionless quantity $I_{cd}^{\eta \gamma \beta}$ can be expressed as the ratio of scalar masses. In evaluating the integral $I_{cd}^{\eta \gamma \beta}$, we neglect terms that are proportional to the masses of charged lepton since they are much smaller than the masses of scalar fields. Thus, in the limit of $m_c = m_d = 0$, the $\nu_R$ mass matrix can be expressed in terms of two mass ratios of scalars as
\begin{align}
     (M_{\nu_R})_{ab} &=   \frac{2\sqrt{2}\ \alpha_4 v_R }{\kappa^2 (1-\epsilon^2)^2\ (16 \pi^2)^2}\ (f M_\ell^2 + M_\ell^2 f^T) \nonumber \\
     %%%%%%%%
    & \hspace{4 mm} \Big\{ C_{ \beta \gamma}\  G\bigg( \frac{m_{\eta}^2}{m_{H_\gamma^0}^2}, \frac{m_{H_\beta^+}^2}{m_{H_\gamma^0}^2} \bigg) + C'_{ \beta \gamma}\  G\bigg( \frac{m_{\eta}^2}{m_{H_\beta^+}^2}, \frac{m_{H_\gamma^0}^2}{m_{H_\beta^+}^2} \bigg) \Big\}\, . 
     \label{eq:Rnumass}
\end{align}
We have evaluated he function $G\big( r_1, r_2 \big)$  analytically to be. 
\begin{eqnarray}
    G(r_1, r_2) &=& -\frac{1}{2}\log\big(r_1 r_2 \big) + \frac{1}{4} \log^2\bigg(\frac{r_1}{r_2} \bigg) + \frac{1}{2 r_1 r_2} \Bigg[- \frac{\pi^{2}}{6}-\left(r_{1}-1\right)\Big(\mathrm{Li}_{2}(1-r_1) \nonumber\\
    %%%%%%%%%%%%%
   &+&  \mathrm{Li}_{2}(1-1/r_1) r_{1}\Big) +\left(r_{1}+r_{2}-1\right)\left(F\left[r_{2}, r_{1}\right]+F\left[\frac{r_{2}}{r_{1}}, \frac{1}{r_{1}}\right] r_{1}+F\left[\frac{r_{1}}{r_{2}}, \frac{1}{r_{2}}\right] r_{2}\right) \nonumber\\
   %%%%%%%%%%%%%%%%
   &-& \left(r_{2}-1\right)\Big(\mathrm{Li}_{2}(1-r_2)+\mathrm{Li}_{2}(1-1/r_2) r_{2}\Big) \Bigg] \, .
   \label{eq:Gplot}
\end{eqnarray}
%%%%%%%%%%%%%%%%%%%%%%%%%%%%
In getting Eq.~(\ref{eq:Gplot}), we define $\mu^2 = m_3^2 \equiv m_{H_\gamma^0}^2$ in Eq.~(\ref{eq:D20}). This function is plotted in Fig.~\ref{fig:integral_plot} as functions of the mass ratios.  
Red, green, and blue lines in Fig.~\ref{fig:integral_plot} show values of the function $G(r_1,r_2)$ as a function of $r_1$ for specific choices of $r_2=1$, $r_2=0.1$, and $r_2=0.01$.  There are simple asymptotic limits of the function as given below:

\begin{equation}
G(r_1, r_2)\ \rightarrow\  \left\{\begin{array}{ll}
-\frac{3}{2} -\frac{r}{2} +\frac{7}{4} r \log r &\hspace{5mm} \text { for } r_1=r_2=r,\ r<<1 \\[5pt]
-\frac{1}{2} -\frac{\pi^2}{6} +\frac{1}{4} r_1 \log r_1  &\hspace{5mm} \text { for } r_2=1,\  r_1 << 1 \\[5pt]
\end{array}\right.
\end{equation}
%%%%%%
We have also discussed the details of asymptotic forms in terms of masses eigenstate for two cases: $m_{H_\gamma^0} \gg m_{\eta} = m_{H_\beta^+}$ ($r_1 = r_2 = r, 0 < r < 1$) and $m_{H_\gamma^0} = m_{H_\beta^+} \gg m_{\eta}$, $(r_2=1, r_1 < 1)$ in Appendix~\ref{sec:AIcd}.

%%%%%%%%%%%%%%%%%%%%%%%%%%%%%%%%%%%%%%%%%%%%%%%
%%%%%%%%%%%%%%%%%%%%%%%%%%%%%%%%%%%%%5
%%%%%%%%%%%%%%%%%%%%%%%%%%%%%%%%%%%%%%

%%%%%%%%%%%%%%%%%%%%%%%%%%%%%%%%%%%%%%%%%%%%%%%
%%%%%%%%%%%%%%%%%%%%%%%%%%%%%%%%%%%%%5
%%%%%%%%%%%%%%%%%%%%%%%%%%%%%%%%%%%%%%
\subsection{Comparing One-loop vs two-loop neutrino mass}\label{sec:1vs2}

Here we proceed to compare the one-loop induced $\nu_R$ mass with the two-loop contribution. We found that the  two-loop contribution is always dominant over the one-loop contribution if parity is an exact symmetry.
%%%%%%%%%%%%%%%%%%%%%%%
\begin{figure}
    \centering
    \includegraphics[scale=0.4]{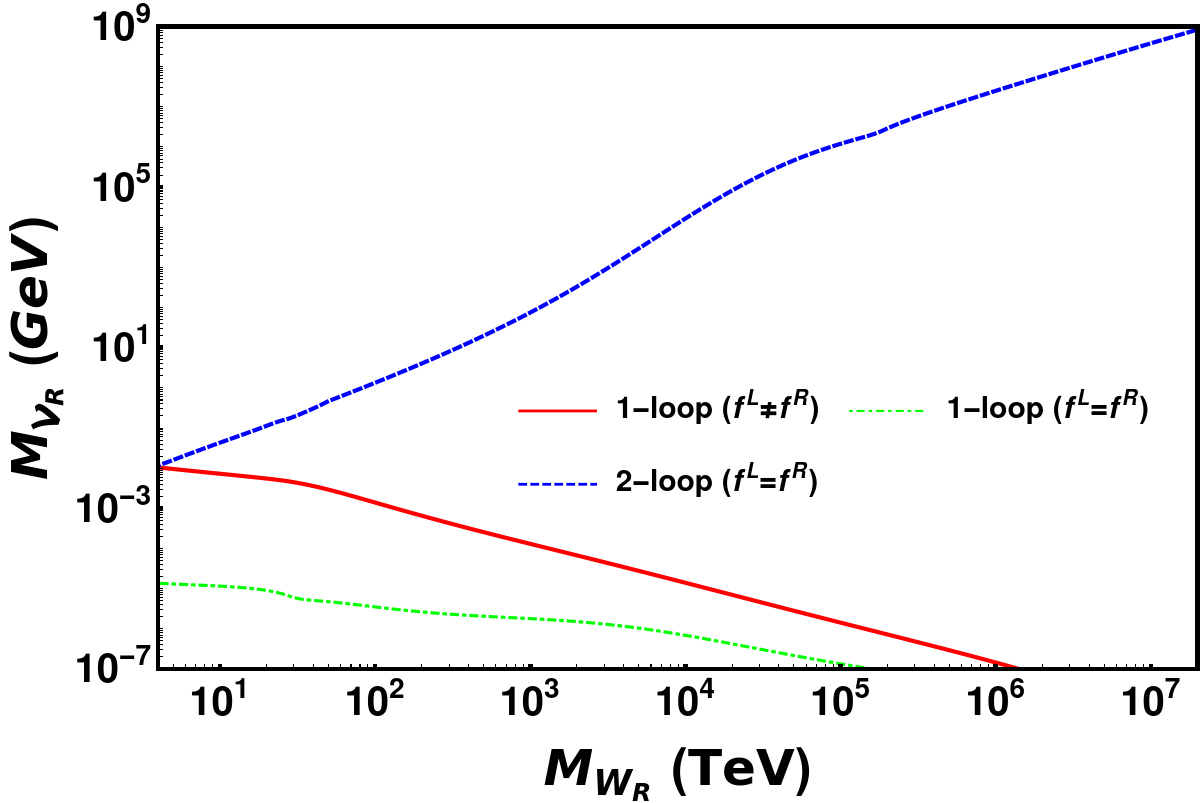}
    \caption{Maximum values of the one-loop induced and two-loop induced $M_{\nu_R}$ as a function of $M_{W_R}$. The dashed blue line corresponds to the two-loop contribution, while the dashed green and the solid red lines respectively represent one-loop contributions to the $\nu_R$ mass with and without Parity symmetry. }
    \label{fig:1vs2}
\end{figure}
Fig.~\ref{fig:1vs2} shows the maximum one-loop and two-loop contribution to RH Majorana neutrino mass as a function of $M_{W_R}$. The solid blue dashed line corresponds to the contribution generated from the full two-loop function of RH Majorana mass given in  Eq.~\eqref{eq:2loopMR} when the parity is assumed to be exact, i.e., there is no breaking of parity in the Yukawa coupling $f$. The solid red line represents one-loop contribution from Eq.~\eqref{eq:1loopvR} to the $\nu_R$ Majorana mass without assuming parity, i.e., when $f^L \neq f^R$ in Eq.~\eqref{eq:yuk}. In contrast, the dashed green line represents the maximum $\nu_R$ mass taking parity symmetry to be exact. The two-loop contribution is proportional to the choice of quartic coupling $\alpha_4$. Here  we have chosen the maximum allowed value of $\alpha_4 = 4 \pi$ from perturbative unitarity limit in generating Fig.~\ref{fig:1vs2}. Note that choosing $\alpha_4$ to a smaller value, such as $\alpha_4 = 3$, the crossover between contribution from the two-loop and the one-loop contributions happens at 8 TeV $M_{W_R}$ mass. In contrast, with exact parity symmetry, the two-loop contribution is always more significant than the one-loop contribution, which is clear from the green dashed line in Fig.~\ref{fig:1vs2}. 

We now summarize the procedure adopted to obtain the maximum $\nu_R$ mass from the two-loop diagrams. We first diagonalize the scalar mass matrix given in Eq.~(\ref{eq:ewmass}). As mentioned in Sec.~\ref{sec:ewcon}, we first identify the eigenstate corresponding to the lightest eigenvalue and fix that mass at $100$ GeV. After this identification, the rotation matrix that takes the original basis to an intermediate basis can be parametrized by three parameters, $\kappa, \kappa'$, and $v_L$, see Eq. (\ref{eq:rel}). One can then rotate the remaining $2\times2$ matrix to get the masses of the heavier Higgs fields. To obtain the maximum $\nu_R$ mass we  diagonalizing the mass matrix given in Eq.~\eqref{eq:2loopMR}. Furthermore, the Yukawa couplings $y$ and $\widetilde{y}$ appearing in Eq.~\eqref{eq:flavorst} are expressed in terms of charged lepton mass and Dirac neutrino mass as given in Eq.~\eqref{eq:yukk}. We take account of the  running of quark and lepton masses as a function of $SU(2)_R$ breaking scale $v_R$. We take the Dirac neutrino masses to be arbitrary, but demand that this be less than the $\nu_R$ Majorana mass. A detailed numerical scan is then done with these constraints to obtain the maximum $\nu_R$ mass. For instance, with the quartic coupling $\alpha_4$ taking value as large as $4\pi$, the maximum allowed right-handed neutrino mass in this model is about 16 MeV for $M_{W_R}$ of 5 TeV. A similar approach is taken in evaluating the maximum $\nu_R$ mass arising from the one-loop diagrams as given by Eq.~\eqref{eq:1loopvR}. Since the one-loop neutrino mass generation requires electroweak symmetry breaking, we numerically diagonalize the full $5\times 5$ charged scalar matrix given in Eq.~\eqref{eq:diagScal} and perform a scan over the parameters.

\begin{table}[]
    \centering
    \begin{tabular}{|c|c|c|c|c|c|c|c| }
    \hline
        $M_{W_R}$ (TeV) & 5 & 10 & 15 & 30 & 50 & 100 & $10^4$    \\ \hline
        $M_{\nu_R}$ (GeV) & 0.042 & 0.010 & 0.020 & 0.05 & 0.11 & 0.36 & $4.2\times 10^3$ \\
        \hline
    \end{tabular}
    \caption{Maximum contribution to the Majorana mass of  $\nu_R$  $M_{\nu_R}$ as a function of the $W_R^\pm$ gauge boson mass $M_{W_R}$. Here we set of $\alpha_4 = 3.0$ and $f_{\mu\tau} \simeq f_{e\tau} = 1.0$.}
    \label{tab:MnuRMWR}
\end{table}

In Table \ref{tab:MnuRMWR} we have listed the maximum possible $\nu_R$ masses within the model as a function of the $W_R$ mass with the quartic coupling $\alpha_4 = 3.0$ and $f_{\mu_\tau} \simeq f_{\mu\tau} = 1.0$ fixed. We see that for low scale $W_R^\pm$, the model predicts $\nu_R$ mass in the $(1-100)$ MeV range or below.  This can lead to interesting phenomenological consequences, which are discussed in the next section.
It is also worth mentioning that  with the small mixing among the charged scalars, the two-loop contribution to the $\nu_R$ mass begins to dominate over the one-loop contribution with the following relation:
\begin{equation}
    v_R \gtrsim \sqrt{16 \pi^2 \kappa M_\ell} \, .
\end{equation}
Taking $\kappa = 246$ GeV, and $M_\ell = m_\tau$, the two-loop contribution exceeds the one-loop contribution above $v_R \gtrsim 26$ GeV. 
%%%%%%%%%%%%%%%%%%%%%%%%%%%%%%%%%%%%%%%%%%%
%%%%%%%%%%%%%%%%%%%%%%%%%%%%%%%%%%%%%%%%%%%
%%%%%%%%%%%%%%%%%%%%%%%%%%%%%%%%%%%%%%%%%%%
\section{Realizing Low Scale Left-Right Symmetry}\label{sec:lowLR}
%%%%%%%

We have seen that the mass of the $\nu_R$ field is in the tens of MeV range if the mass of the $W_R^\pm$ is in the multi-TeV range in the model.  In this section we pursue the possibility that $W_R^\pm$ is within reach of collider experiments in the near future.  Keeping this in mind we seek a fit to the neutrino oscillation data with $\nu_R$ masses in the $(1-100)$ MeV range.  One should ensure that this scenario is not in conflict with experimental constraints as well as constraints from cosmology and astrophysics.  We illustrate here that all constraints can be satisfied in such a low-scale $W_R^\pm$ scheme.

Since the radiative correction to the LH neutrino mass is always suppressed in comparison to the usual seesaw (cf. Sec.~\ref{sec:typeII}), one can take $M_\nu^L = 0$ in Eq.~\eqref{eq:SeesawMat}. One can then diagonalize RH neutrino mass matrix (cf. Eq.~\eqref{eq:diagMR}) and write the unitary matrix that transforms the weak eigenstates $\nu_L$ and $\nu_R$ to the mass eigenstates $\nu_j\ (j=1,2,3)$ and $N_\alpha\ (\alpha=4,5,6)$ as
\begin{equation}
   U =   \left(
\begin{array}{cc}
 U_{\nu\nu} &  U_{\nu N} \\
  U_{N\nu} &  U_{NN} \\
\end{array}
\right)
\end{equation}
Note that $U_{\nu \nu}, U_{\nu N}, U_{N \nu}$, and $U_{NN}$ are not separately unitary. The unitary matrix $U$ diagonalizes the $6\times 6$ neutrino mass matrix as
%%%%%%%%%
\begin{equation}
    U^\dagger M_\nu \, U^\star = \left(
\begin{array}{cc}
  m_{\nu_j} &  0 \\
  0 &  M_{N_\alpha} \\
\end{array}
\right)
\end{equation}
where 
\begin{equation}
    m_{\nu_j} = {\rm Diag}\ (m_1,m_2,m_3) \hspace{20mm} M_{N_\alpha} = {\rm Diag}\ (M_1, M_2, M_3)~.
\end{equation}
%%%%%%%%%%%%%%%%%%%%%%%%%%5
Here $U_{\nu \nu}^\star$ is the usual PMNS matrix characterizing the mixing among light neutrinos. The three neutrino oscillation observables $\theta_{12}$, $\theta_{13}$, and $\theta_{23}$ are obtained from the following relations: 
\begin{equation}
    s_{12}^2 = \frac{|U_{e2}|^2}{1-|U_{e3}|^2}, \hspace{5mm} s_{13}^2 = |U_{e3}|^2, \hspace{5mm}  s_{23}^2 = \frac{|U_{\mu3}|^2}{1-|U_{e3}|^2} \, .
\end{equation}
where $s_{ij} = \sin{\theta_{ij}}$, with $\theta_{ij}$ being the mixing angles among different flavor eigenstates $i$ and $j$. The magnitude of Dirac CP violation $\delta$ is determined by the Jarlskog invariant $J_{cp}$ \cite{Jarlskog:1985ht}:
%\vspace{-2mm}
\begin{eqnarray}
    J_{cp} &=& \operatorname{Im}(U_{\mu3} U_{e3}^\star U_{e2} U_{\mu 2}^\star) \nonumber\\
    &=& \frac{1}{8} \cos\theta_{13} \, \sin2\theta_{12} \, \sin2\theta_{23} \, \sin2\theta_{13} \, \sin\delta \, .
    \label{eq:Jarlskog}
\end{eqnarray}

Next, we analyze various low energy constraints on sterile neutrino with a mass of 1 MeV to 100 MeV range that mixes with the light neutrinos. There are various constraints one needs to consider, such as lepton universality, invisible $Z$-boson decay, neutrinoless double beta decay, magnetic and electric dipole moments, neutrino oscillation constraints, and cosmological constraints. We will discuss astrophysical constraints in the next section arising from the energy loss in $\nu_R$ in supernovae. Here we focus on the most stringent constraints for the mixing of sterile neutrino as a function of its mass in the range 1 to 100 MeV.

%%%%%%%%%%%%%%%%%%%%%%%%%%%%%%%5
%%%%%%%%%%%%%%%%%%%%%%%%%%%%
%%%%%%%%%%%%%%%%%%%%%%%%
\subsection{Direct experimental constraints}
%%%
\begin{table}[!t]
    \centering
    \begin{tabular}{|c|c|c|c|c|c|c|}
    \hline
       Mass  & 1 MeV & 5 MeV  & 10 MeV  & 30 MeV  & 50 MeV  & 100 MeV  \\ \hline
        $|U_{eN}|^2$  & 2.6$\times10^{-4}$ & 1.1$\times 10^{-5}$& 3.5$\times 10^{-6}$  & $4.4\times 10^{-7}$ & 1.2$\times 10^{-7}$ & 7.1$\times 10^{-9}$ \\ 
        %%%%
        & {\tt BD2} & {\tt BOREXINO} & {\tt BOREXINO} & {\tt PIENU} & {\tt $\pi_{e2}$ PIENU} & {\tt PIENU} \\ \hline
        %%%%
        %%%%
        $|U_{\mu N}|^2$ & 1.1$ \times 10^{-2}$ & 2.75$ \times 10^{-4}$ & 2.06$ \times 10^{-4}$ & 8.6$ \times 10^{-6}$ & 2.35$ \times 10^{-4}$ & 3.76$ \times 10^{-6}$ \\
        %%%%%%%
         & {\tt $\pi_{\mu2}$ PSI} & {\tt $\pi_{\mu2}$ PSI} & {\tt $\pi_{\mu2}$ PSI} & {\tt $\pi_{\mu2}$ PIENU} & {\tt $K_{\mu2}$ KEK} & {\tt $K_{\mu2}$ KEK} \\ \hline
         %%%%
        %%%%
        $|U_{\tau N}|^2$ & $-$ & $-$ & 0.49 & 0.021 & 4.9$ \times 10^{-3}$ & 5.1$ \times 10^{-4}$ \\
        %%%%%%%
         &  & & {\tt CHARM} & {\tt CHARM} & {\tt CHARM} & {\tt CHARM} \\ \hline
    \end{tabular}
    \caption{Constraint on active and sterile mixing $U_{eN}$, $U_{\mu N}$ and $U_{\tau N}$ for  different masses of the sterile neutrino. }
    \label{tab:lowconst}
\end{table}
%%%%
One can obtain constraints on active--sterile neutrino mixing \cite{deGouvea:2015euy, Bolton:2019pcu} by looking for visible final state particles in beta-decay, pion decay, kaon decay, muon decay, etc for sterile neutrino mass in the $(1-100)$ MeV range. There are several dedicated searches for the existence of a sterile neutrino. We quote various direct experimental constraints in Table.~\ref{tab:lowconst} in this mass range for the sterile neutrino. The {\tt TRIUMF PIENU} \cite{PIENU:2011aa} experiment performed a kinematic search for sterile neutrino produced in pion decay and set the limit on the mixing $U_{\nu N}$ $(\nu = e, \mu)$ for the mass range of few MeV to tens of MeV. For example, the collaboration set limits at the level of $|U_{eN}| < 10^{-8}$ in the mass range of sterile neutrino 60 MeV to 129 MeV \cite{Aguilar-Arevalo:2017vlf, Bryman:2019ssi, Bryman:2019bjg}
Reactor neutrino experiments put bounds to sterile neutrino with masses from 1 MeV to 10 MeV as it can decay to electron pair and a neutrino ($N_\alpha \to e^+ e^- \nu$). Experiments like {\tt Rovno} \cite{Derbin:1993wy} and {\tt Bugey} \cite{Hagner:1995bn} reactors have set the limits on mixing $U_{eN}$. We note the most stringent limit of a sterile neutrino mixing  with $\nu_e$ from the {\tt BOREXINO} experiment \cite{Bellini:2013uui} in Table.~\ref{tab:lowconst}, which looked for neutrinos produced in Sun with masses up to 14 MeV. 

The limit on $\nu_\mu$-sterile neutrino mixing for the mass range of 1 MeV to 100 MeV is provided by various experiments such as {\tt PSI} \cite{Minehart:1981fv}, {\tt PIENU} \cite{Aguilar-Arevalo:2019owf}, {\tt KEK} \cite{Hayano:1982wu, Yamazaki:1984sj} and measurement of muon decay spectrum \cite{Shrock:1981wq}. Similarly, the upper limits on the mixing of $\nu_\tau$-sterile neutrino in the relevant mass region is provided by {\tt NOMAD} \cite{Astier:2001ck} and {\tt CHARM} \cite{Orloff:2002de}. The most stringent limits are summarized in Table.~\ref{tab:lowconst}.

%%%%%%%%%%%%%%%%%%%%%%%%%%%%%%%%5
\subsection{Neutrinoless double beta decay}
%%%%%%%%%%%%%%%%%%%%%%%%%%%%%%%%%%%
Neutrinoless double beta decay provides important limits on the active-sterile mixing as a function of sterile neutrino mass. The inverse half-life $T_{1/2}^{0\nu}$ for $0\nu\beta\beta$ can be expressed as \cite{Kovalenko:2009td, Faessler:2014kka, Bolton:2020ncv}
\begin{align}
    \frac{1}{T_{1/2}^{0\nu}} = G^{0\nu} g_A^4 |M^{0\nu}|^2 \left| \frac{m_{1}}{m_e} + \frac{\langle {\bf p}^2 \rangle}{m_e} \sum_{\alpha = 4}^{6} \frac{U_{eN_\alpha}^2 M_{N_\alpha}}{\langle {\bf p}^2 \rangle + M_{N_{\alpha}}^2} \right|^2 \, .
    %%%
    \label{eq:betadecay}
\end{align}
%%%%%%%
Here $G^{0 \nu}$ and $g_A$ are the phase factor and the axial vector coupling relevant for the decay.  $M^{0\nu}$ is the light neutrino exchange nuclear matrix element, whereas $\langle {\bf p} \rangle$ is the average momentum transfer of the process. It should be noted that if the heavy sterile state and the light states are in a complete seesaw formalism, then the effective neutrino mass is zero.
\begin{equation}
    (M_\nu)_{11} = m_1 + \sum_{i} U_{ei}^2 M_{N_\alpha} = 0~.
\end{equation}  In the low $W_R^\pm$ scenario of the LRSM, the model predicts $\nu_R$ masses of a few MeV; thus, momentum transfer can be much heavier than sterile neutrino mass, suppressing the $0\nu\beta\beta$ decay rate. In this case we find that the next order contribution does give important constraint, but this is much weaker than the one resulting from  $(M_\nu)_{11}$ were it not zero.  We show two different cases as presented in {\tt Fit1} and {\tt Fit2} with the variation of sterile neutrino mass and show that they are consistent with effective Majorana neutrino mass constraint from neutrinoless double beta decay. 

%%%%%%%%%%%%%%%%%%%%%%5
\subsection{Cosmological constraints}
%%%%%%%%%%%%%%%%%%%%%%%5

A sterile neutrino in the mass range of $(1-100)$ MeV can potentially upset the successful predictions of big bang cosmology.  If these neutrinos are long-lived, they will contribute to the effective number of neutrino species, which is constrained by Planck data \cite{Aghanim:2018eyx}.  A long-lived sterile neutrino can also over-close the universe, in contradiction with observations.  Here we show that the model with low $W_R^\pm$ indeed satisfies all the cosmological constraints.

The MeV mass sterile neutrino (denoted here as $N$) can decay into three neutrinos, $\nu_i e^+ e^-$ or into $\nu_i \gamma$.  The three body decays arise through the mixing of $N$ with the active neutrino $\nu_i$.  The rates for these decays are given by

\begin{align}
    \Gamma (N_\alpha \to e^+ e^- \nu ) &= \sum_j \Gamma_j (N_\alpha \to e^+ e^- \nu_j )    \nonumber \\
    & =
     2\ \sum_{j} |U_{j\alpha}|^2 \frac{ G_F^2 M_{N_\alpha}^5}{192 \pi^3}\ \Big[ \Big\{ \delta_{je} + (- \frac{1}{4}+ \frac{1}{2} \sin^2\theta_w) \Big\}^2 + \frac{1}{4} \sin^4\theta_w \Big] \, ,\\[5pt]
    %%%%%%%%%
    \Gamma (N_\alpha \to 3 \nu) &= \sum_{ij} \Gamma_j^{ii}\ (N_\alpha \to  \bar{\nu}_{i} \nu_{i} \nu_j ) = 2 \sum_{j} |U_{j\alpha}|^2\  \frac{1}{4}\ \frac{G_F^2 M_{N_\alpha}^5}{192 \pi^3}\ (1+2+1) \, .
\end{align}
Here a factor 2 appears to account for the Majoran nature of $N$.  The $e^+e^-\nu$ decay is mediated by $Z$ boson as well as by $W$ boson.  It turns out that the new contributions to these decays arising from $W_R^\pm$ and $Z_R$ are not significant, if these gauge bosons have masses at or above 5 TeV.  

The radiative decay $N \to \nu \gamma$ receives contribution from the exchange of $\eta^+$ which is enhanced compared to the $W^\pm$ and mixed $W_L-W_R$ contribution.  The decay rate is given by \cite{Pal:1981rm,Lavoura:2003xp}

\begin{align}
   \Gamma (N_\alpha \to \nu \gamma) &= \sum_j \Gamma_j (N_\alpha \to \nu_j \gamma) \nonumber \\
   &= 2 \times \left( \frac{\alpha M_{N_\alpha}^3 m_\tau^2}{128 \pi^4} \right) \left[ \bigg| \frac{f_{e\tau}^2 + f_{\mu \tau}^2 }{m_\eta^2} \Big\{ 1+ \log \Big(\frac{m_\tau^2}{m_\eta^2} \Big) \Big\} \bigg|^2 + \bigg| \frac{g^2\ \rho }{2 M_{W_R}^2} \bigg|^2 \right] 
\end{align}
%%%%%%%%%%%%%%%%%%%%%%%%%%%%%%%%%%%%%%%%%%%
where $\rho = \zeta M_{W_R}^2/M_{W_L}^2$ and we have kept only the magnetic moment contribution proportional to the $\tau$ lepton mass \cite{Babu:1987be,Fukugita:1987ti}, which turn out to be the most dominant. We shall use these formulas to estimate the lifetime of the sterile neutrino and show that the radiative decay mediated by the $\eta^+$ scalar can lead to a lifetime of order 1 second, which would make it consistent with big bang cosmology.

As it turns out, one of the right-handed neutrinos will have a mass much smaller than the other two in the LRSM due to the flavor structure of the induced mass matrix.  This lightest $\nu_R$ cannot decay fast enough to satisfy the lifetime limit of 1 second.  If this $\nu_R$ has negligible mixing with the active neutrinos, its contribution to the effective number of neutrinos would be about 0.1, which is not inconsistent with Planck observations \cite{Aghanim:2018eyx}.  This reduction in $N_\nu^{\rm eff}$  arises since the right-handed neutrino decoupled from the plasma above QCD phase transition, when the number of degrees of freedom was around 67.  $N_{\nu}^{\rm eff} = [g^*( 400 \,{\rm MeV)}/g^*(1~{\rm MeV})]^{4/3} \simeq 0.1$, where 400 MeV is the typical decoupling temperature of $\nu_R$ if the $W_R$ mass is of order 5 TeV. If such a decoupled light sterile neutrino has a mass of order eV of less, it would not over-close the universe, thus making it consistent with cosmology.

\subsection{Neutrino oscillation fit consistent with low \texorpdfstring{$W_R$}{WRR} mass}\label{sec:lowWR}
%%%%%%%%%%%%%%%%%%%%%%%%%%%%%%%%%%%%%%%%%%%%%%%%%%%%%%%%%%%%%%%
%%%%%%%%%%%%%%%%%%%%%%%%%%%%%%%%%%%%%%%%%%%%%%%%%%%%%%%%%%%%%%%
Here we provide a fit to the neutrino oscillation data in the context of low scale $W_R^\pm$ that satisfies all of the experimental and cosmological constraints on MeV scale sterile neutrinos.

The Yukawa coupling matrix $f$ that couples left- and right-handed lepton doublets with charged scalar singlet $\eta^+$ can be made real by the phase redefinitions.  The Yukawa coupling matrices $y$ and $\widetilde{y}$ in Eq.~\eqref{eq:yuk} are hermitian due to Parity symmetry. Thus the theory would appear to have enough parameters to easily satisfy the neutrino oscillation data. However, in the low $v_R$ ($M_{W_R}$) regime, one can safely ignore the Dirac neutrino mass $M_{\nu^D}$ contributions in evaluating right-handed neutrino mass. This reduces the input parameters making the Yukawa couplings $y$ and $\widetilde{y}$ in Eq.~\eqref{eq:yukk} proportional to the charged lepton mass matrix (cf. Sec.~\ref{sec:mdml}). Moreover, the $\nu_R$ Majorana mass matrix of Eq.~\eqref{eq:MajR} has the following structure, taking advantage of the hierarchy $\frac{m_e^2}{m_\tau^2} \ll  \frac{m_\mu^2}{m_\tau^2} << 1$:
\begin{equation}
    M_{\nu_R} = \mathcal{J} 
    \begin{pmatrix}
    0 & \frac{m_\mu^2\ x_1}{m_\tau^2}  & 1 \\
    \frac{m_\mu^2\ x_1}{m_\tau^2} & 0 & x_2 - \frac{m_\mu^2\ x_2}{m_\tau^2} \\
   1 &  x_2 - \frac{m_\mu^2\ x_2}{m_\tau^2} & 0
    \end{pmatrix}
    \label{eq:fitMR}
\end{equation}
with
\begin{equation}
    \mathcal{J} =\frac{2\sqrt{2}\ \alpha_4 v_R m_\tau^2 f_{e\tau}}{\kappa^2 (1-\epsilon^2)^2\ (16 \pi^2)^2}\
     %%%%%%%%
   \Big\{ C_{ \beta \gamma}\  G\bigg( \frac{m_{\eta}^2}{m_{H_\gamma^0}^2}, \frac{m_{H_\beta^+}^2}{m_{H_\gamma^0}^2} \bigg) + C'_{ \beta \gamma}\  G\bigg( \frac{m_{\eta}^2}{m_{H_\beta^+}^2}, \frac{m_{H_\gamma^0}^2}{m_{H_\beta^+}^2} \bigg) \Big\}\, . 
\end{equation}
Here $x_1 = f_{e\mu}/f_{e \tau}$, $x_2 = f_{ \mu \tau}/f_{e \tau}$, and $\epsilon = \frac{\kappa'}{\kappa}$. ($C_{\beta \gamma}$, $C'_{\beta \gamma}$) are given in Eq.~\eqref{eq:CCp} and the function $G$ is given in Eq.~\eqref{eq:Gplot}. Note that we have taken Parity to be exact so that the two-loop diagram is always dominant compared to one-loop diagram (cf. Fig.~\ref{fig:1vs2}). Moreover, we take $x_1 = 0$ by choosing $f_{e\mu} = 0$. As discussed in the context of cosmology, the lightest $\nu_R$ can be decoupled from the $6 \times 6$ mass matrix so that it does not cause problems with $N_\nu^{\rm eff}$. 
This structure has an interesting feature as one of the eigenvalues of the $\nu_R$ mass matrix is precisely zero, while the two others become degenerate. Furthermore, one of the light neutrino mass eigenvalue also becomes zero in this case. 
%%%%%%%%%%%%%%%%%%%%%55
%%%%%%%%%%%%%%%%%%%%%%55
\begin{table}[t!]
%\vspace{-4mm}
\centering
\begin{tabular}{|c|c|c|c|}
\hline \hline
 {\textbf{Oscillation}}   &   {\textbf{$3\sigma$ allowed range}} &  \multicolumn{2}{c|}{\textbf{Model Fits}}\\
 \cline{2-4}
 {\bf parameters} & \bf{{\tt NuFit5.0}}~\cite{Esteban:2020cvm} & {\tt Fit1 } & {\tt Fit2}   \\ \hline \hline
 $\Delta m_{21}^2 (10^{-5}~{\rm eV}^2$)  &   6.82 - 8.04  & 7.40 & 7.45    \\ \hline
 $\Delta m_{31}^2 (10^{-3}~{\rm eV}^2) $ &   2.435 - 2.598   & 2.49 & 2.48  \\ \hline
  $\sin^2{\theta_{12}}$   &   0.269 - 0.343 & 0.325  & 0.316  \\ \hline
% $\sin^2{\theta_{23}}$  (IH) &   0.419 - 0.617  & - &    0.589   \\
 $\sin^2{\theta_{23}}$   &   0.415 - 0.616  & 0.537 &  0.561  \\ \hline 
%  $\sin^2{\theta_{13}} $  (IH) &   0.02052 - 0.02428  & - &  0.0228 \\
  $\sin^2{\theta_{13}}  $  &   0.02032 - 0.02410  & 0.0221 &   0.0220    \\ \hline
%  $\delta_{CP}/^\circ$ (IH) & 193 - 352 & - & 318  \\ 
  $\delta_{CP}/^\circ$  & 120 - 369 & 274 &  275 
  %%%%%%%
  \\\hline \hline
\end{tabular}
\caption{$3\sigma$ allowed ranges of the neutrino oscillation parameters from a recent global-fit~\cite{Esteban:2020cvm}, along with the model predictions, as described in Sec. \ref{sec:lowWR}. }
\label{nuTab3}
%%%%%
%%%%%%%%%%%%%%%%%%%%%%
\vspace{5mm}
 \centering
\begin{tabular}{|c|c|c|c|c|c|c|c|c|c|}
\hline \hline
  & $m_{12}$ (MeV) & $m_{23}$ (MeV) & $m_{33}$ (MeV) & $\theta/~^\circ$  &$\varphi_{12}/~^\circ$ & $\varphi_{13}/~^\circ$ & $\varphi_{23}/~^\circ$   \\ \hline 
  %%%%%
  {\tt Fit1}  & -$2.54 \times 10^{-4}$ & -$4.22 \times 10^{-4}$ & $6.12 \times 10^{-4}$ & $42.1$ & $296$ & $182$ & $328$   \\ \hline
  %%%%%%
{\tt Fit2} & -4.30$\times10^{-3}$ & 7.36$\times10^{-3}$ & -9.11$\times10^{-3}$ & $319$ & $249$ & $192$ & $275$   \\ \hline \hline
%%%%%%%%%%%
\end{tabular}
\caption{Values of parameters that gives Fit1 and Fit2, as prescribed in Table \ref{nuTab1} to fit the neutrino oscillation data.  }
\label{nuTab4}
%\vspace{-4mm}
\end{table}
%%%%%%%%%%%%%%%%%%%%%%%%%%%%%%%
\begin{table}[]
    \centering
    \begin{tabular}{|c|c|c|c|c|c|c|c|c|}
        \hline \hline
     &  $m_\eta$ (TeV) & $m_{\nu_R}$ (MeV) & $M_{W_R}$ (TeV) & $\alpha_4$ & $\tau$ (s) & $m_{\beta\beta}$ (eV) \\ \hline 
  %%%%%
    {\tt Fit1}  & 4.0  & 4.2  & 4.0  & 3.0 & 0.97  & 0.009  \\ \hline
  %%%%%%
    {\tt Fit2}    & 4.0 & 10 & 6.0 & 4.0 & 0.072 & 0.017 \\ \hline \hline
    \end{tabular}
    \caption{Masses of $\eta$, $\nu_R$, and $W_R$ that are consistent with {\tt Fit1} and {\tt Fit2}. Here $\tau$ stands for the lifetime of the sterile neutrino and $m_{\beta \beta}$ corresponds to the effective Majorana neutrino mass.}
    \label{tab:param}
\end{table}
%%%%%%%%%%%%%%%%%%%%%%%%%%%%%%%%%%%%%%%%%%%%%%%%%%%%%%5
%%%%%%%%%%%%%%%%%%%%%%%%%%%%%%%%%%%%%%%%%%%%%%%%%%%%%%5

We make the following unitary transformation to diagonalize the $\nu_R$ Majorana mass matrix of Eq.~\eqref{eq:fitMR}:
\begin{equation}
    O M_{\nu_R} O^T = M_{\nu_R}^{{\rm diag}}
    \label{eq:diagMR}
\end{equation}
%%%%%%
\begin{equation}
    O = \begin{pmatrix}
        -\cos\theta & \sin\theta & 0 \\
        -\frac{\sin\theta}{\sqrt{2}} & -\frac{\sin\theta}{\sqrt{2}}  & \frac{1}{\sqrt{2}} \\
        -\frac{\sin\theta}{\sqrt{2}} & -\frac{\sin\theta}{\sqrt{2}} & \frac{1}{\sqrt{2}}
     \end{pmatrix} \, , \hspace{10 mm}
     %%%%%%%%
M_{\nu_R}^{{\rm diag}} =  \begin{pmatrix}
            0 &  0 & 0 \\
            0 & -m_{R} & 0 \\
            0 & 0 & m_{R}
           \end{pmatrix} \, ,
\end{equation} 
where 
\begin{align}
    m_{R} &= \mathcal{J}\ \sqrt{1 + x_2^2 \Big(1- \frac{m_\mu^2}{m_\tau^2} \Big)^2} \, , \nonumber \\
    \tan\theta &= \frac{m_\tau^2}{x_2 \Big(m_\tau^2- m_\mu^2 \Big)} \, .
\end{align}
This orthogonal rotation takes $\nu_R$ fields into a new basis such that the arbitrary hermitian Dirac neutrino mass matrix is modified as follows:
\begin{equation}
   \hat{M}_{\nu_D} = 
    \begin{pmatrix}
    0 & \hat{m}_{12} & \hat{m}_{13}  \\
    0 & \hat{m}_{22} &  \hat{m}_{23} \\
    0 & \hat{m}_{32} & \hat{m}_{33}
    \end{pmatrix} \, ,
\end{equation}
where
\begin{align}
  \hat{m}_{12} &= \frac{1}{2} \sec\theta \Big\{ -\sqrt{2}\ m_{12} e^{i \varphi_{12}} - (e^{i (2\varphi_{13}- \varphi_{23})} m_{23} + e^{i2 \varphi_{23}} m_{33}) \tan\theta \Big\} \, , \nonumber \\
  %%%%%%%%%
  \hat{m}_{13} &= \frac{1}{2} \Big\{ \sqrt{2}\ m_{12} e^{i \varphi_{12}} - (e^{i (2\varphi_{13}- \varphi_{23})} m_{23} + e^{i2 \varphi_{23}} m_{33}) \tan\theta \Big\} \, , \nonumber \\
  %%%%%%%%%
  \hat{m}_{13} &= e^{-i \varphi_{12}} m_{12} \csc\theta \, , \nonumber \\
  %%%%%%%%%%
  \hat{m}_{23} &= \sqrt{2}\ e^{i \varphi_{23}} m_{23} - e^{-i \varphi_{12}} m_{12} \cot\theta \, , \nonumber \\
  %%%%%%%%%%
  \hat{m}_{32} &= \frac{1}{\sqrt{2}}\ \sec\theta (-e^{-i \varphi_{23}} m_{23} + m_{33}) \, , \nonumber  \\
  %%%%%%%
  \hat{m}_{33} &= \frac{1}{\sqrt{2}} (e^{-i \varphi_{23}} m_{23} + m_{33})
\end{align}
%%%%%%%%%%%%%%
where $\varphi_{ij}$ corresponds to the phase of $m_{ij}$. We make the following substitution to get the first column zero, so that one $\nu_R$ decouples from the seesaw setup: 
\begin{align}
    m_{11} &= \frac{1}{2}\ \tan\theta \Big\{ -\sqrt{2}\ m_{12} e^{i \varphi_{12}} - (e^{i (2\varphi_{13}- \varphi_{23})} m_{23} + e^{i2 \varphi_{23}} m_{33}) \tan\theta \Big\} \, , \nonumber \\
    %%%%%%%%%%%%%
    m_{22} &= e^{i \varphi_{23}} m_{23} - \sqrt{2}\ e^{-i \varphi_{12}} m_{12} \cot\theta \, , \nonumber \\
    %%%%%%%%%%%
    m_{13} &= \frac{1}{\sqrt{2}} e^{i (\varphi_{13} - \varphi_{23})} \tan\theta\ (m_{33} e^{i \varphi_{23}} - m_{23}) \, .
\end{align}
The analysis reduces the neutrino mass matrix into the $5\times 5$ matrix, with one zero eigenvalue corresponding to an active neutrino, and two nearly degenerate $\nu_R$ states. The fit of neutrino oscillation data to this $5\times 5$ mass matrix is shown in Table~\ref{nuTab3}, with the corresponding input parameter given in Table~\ref{nuTab4}. The active and sterile mixing matrix for {\tt Fit1} and {\tt Fit2} is found to be:
\begin{equation}
    U_{\nu N} = 
   10^{-3} \times \begin{pmatrix}
   0.89 + 0.53 i &\hspace{5mm} - 0.69 + 0.11 i \\
   0.59 - 1.23 i &\hspace{5mm} 3.16 + 0.59 i \\
   -0.21 + 0.68 i &\hspace{5mm} 2.73 -0.011 i 
    \end{pmatrix}  \hspace{1cm} {\tt Fit1}
\end{equation}
%%%%%
\begin{equation}
    U_{\nu N} = 
   10^{-3} \times \begin{pmatrix}
    -0.41 + 0.15 i &\hspace{5mm} -0.069 + 0.29 i \\
   0.48 - 0.15 i &\hspace{5mm} -1.12 - 0.69 i \\
   -0.31 + 0.19 i &\hspace{5mm} -1.11 - 0.078 i 
    \end{pmatrix}  \hspace{1cm} {\tt Fit2}
\end{equation}

It is easy to verify that with these fits, all the direct active-neutrino oscillation constraints listed in Table \ref{tab:lowconst} are satisfied.  Furthermore, with these choices of parameters, the lifetime of the heavier $\nu_R$ fields are found to be less than 1 second, showing consistency with cosmology.
%%%%%%%%%%%%%%%%%%%%%%%%%%%%%%%%%%%%%%%%%%%%%%%%%%%%%%5
%%%%%%%%%%%%%%%%%%%%%%%%%%%%%%%%%%%%%%%%%%%%%%%%%%%%%%5%%%%%%%%%%%%%%%%%%%%%%%%%%%%%%%%%%%%%%%%%%%%%%%%%%%%%%5
%%%%%%%%%%%%%%%%%%%%%%%%%%%%%%%%%%%%%%%%%%%%%%%%%%%%%%5
\section{Supernova Energy Loss Constraints}
\label{Sec:sn}

Supernova dynamics may be significantly altered in presence of right-handed charged current interactions, provided that the right-handed neutrinos are lighter than about 10 MeV, which is the case in our model with TeV scale $W_R^\pm$. Barbieri and Mohapatra have derived a lower limit of 23 TeV on the $W_R^\pm$ mass by demanding that the $\nu_R$ luminosity not exceed $10^{53}$ erg/sec for supernova 1987a \cite{Barbieri:1988av}.  We have reexamined this limit carefully and found that this may be significantly weaker, with the lower limit on $W_R^\pm$ as low as 4.6 TeV.  We have made three improvements over the estimate of Ref. \cite{Barbieri:1988av}.  First, we computed the exact cross section for the production of $\nu_R$ inside supernova via the reaction $e^- + p \to \nu_R + n$. Our cross section turns out to be a factor of 3.3 smaller compared to the naive cross section valid for low energy neutrino scattering used in the estimate in Ref. \cite{Barbieri:1988av}. Secondly, we have included an important interference effect between the $W_R^\pm$ contribution and the $W_L^\pm-W_R^\pm$ mixed contribution in the production cross section that further reduces the cross section compared to Ref. \cite{Barbieri:1988av} for one sign of the mixing parameter.  And third, we have used the average electron energy to be $\sim 150$ MeV, as opposed to $300$ MeV used in \cite{Barbieri:1988av}, which appears to be reasonable, given that the core temperature of supernovae is $(30-70)$ MeV. We now detail the improvements we have made.

Light right-handed neutrinos with masses less than about 10 MeV may be produced inside supernovae through the process $e + p \rightarrow \nu_R + n$ mediated by the $W_R^\pm$ gauge boson.  Unless the $W_R$ is lighter than about 600 GeV, the $\nu_R$ produced this way would not thermalize and will escape,  carrying energy with them, which could be in conflict with the energy loss mechanism inferred from sn1987a.  The effective interactions involving the leptons and quarks in this model is given by
\begin{equation}
{\cal L} = \frac{4 G_F \cos\theta_C}{\sqrt{2}}\left[ -\sin\zeta\, \overline{d_L} \gamma^\mu u_L + \cos\zeta\, \frac{M_{W_L}^2}{M_{W_R}^2}\, \overline{d_R} \gamma^\mu u_R  \right] (\overline{\nu}_R\gamma_\mu e_R)
\end{equation}
where $G_F$ is Fermi coupling, $\theta_C$ is the cabibbo angle, and $\zeta$ is the
the $W_L-W_R$ mixing angle defined in Eq. (\ref{eq:Wmix}).  This Lagrangian needs to be converted to hadronic Lagrangian involving the proton and the neutron.  Since strong interactions are parity conserving, we infer that the left-handed and the right-handed quark currents will yield the same hadronic matrix elements.  From quasi-elastic neutrino-nucleon cross section calculations \cite{LlewellynSmith:1971uhs} we obtain the matrix elements  for both terms.  Compared to the Fermi coupling this operator will have a suppression factor given by
\begin{equation}
B = -\sin \zeta + \cos \zeta\, \frac{M_{W_L}^2}{M_{W_R}^2}~.
\end{equation}
Note that the two terms here would interfere, which was not accounted for in Ref. \cite{Barbieri:1988av}.  Note also that the two terms are comparable in magnitude with an unknown relative sign, see Eq. (\ref{eq:Wmix}). For one sign of the mixing parameter this interference can reduce the $\nu_R$ production cross section.

%%%%%%%%%%%%%%%%%%
\begin{figure}
    \centering
    \includegraphics[scale=0.3]{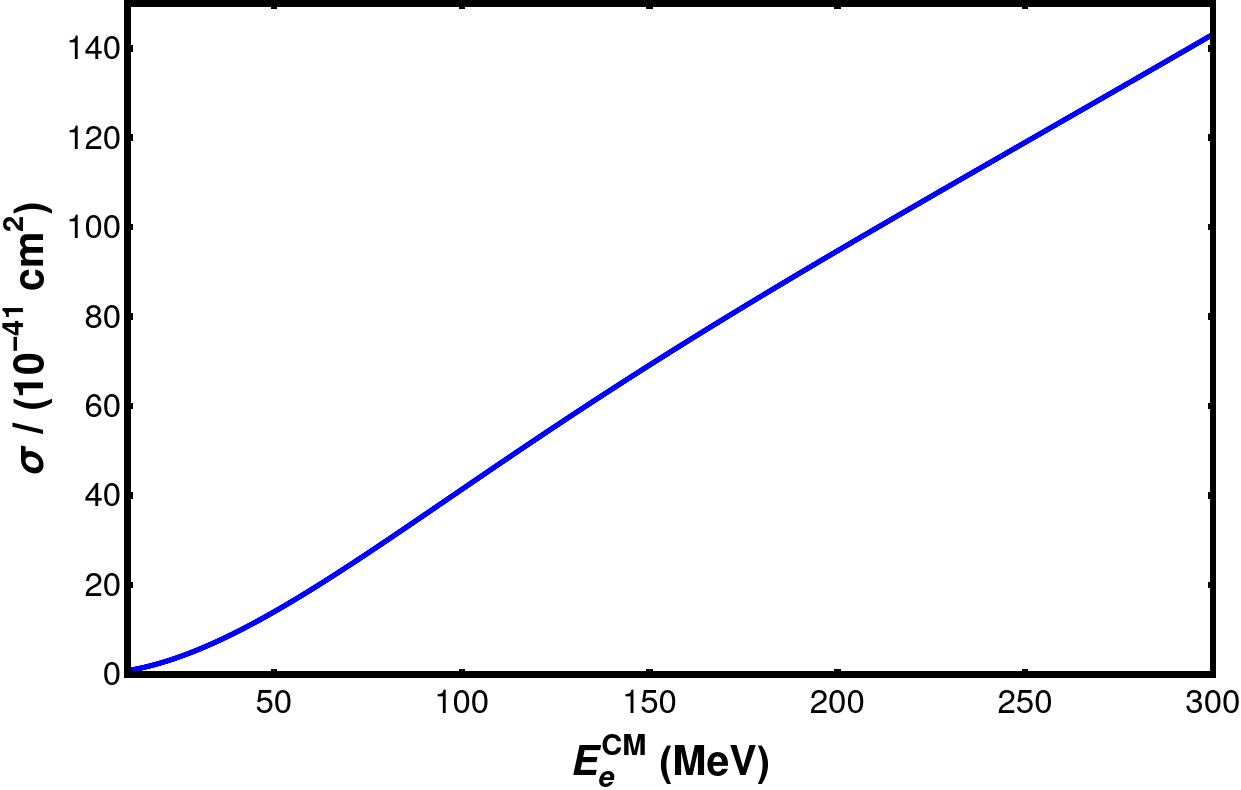}
    \caption{Total cross section for $\nu_R$ production via $e + p \rightarrow \nu_R + n$ at supernova energies as a function of the CM energy of the electron.}
    \label{fig:sigEe}
\end{figure}
%%%%%%%%%%%%%
We have worked out the cross section for the scattering process 
\begin{equation}
    e^-\ (p_p) + p\ (p_p) \to \nu\ (p_\nu) + n\ (p_n)
\end{equation}
explicitly.  
The differential cross section is given by
\begin{equation}
    \frac{d\sigma}{dt} = \frac{1}{64 \pi} \frac{G_F^2 \cos^2\theta_C |B|^2}{(s-m_p^2-m_e^2)^2 - 4 m_p^2 m_e^2}\ |\mathcal{M}^2|
\end{equation}
Here  $\mathcal{M}$ is the invariant amplitude expressed in terms of leptonic and hadronic currents as
\begin{equation}
    \mathcal{M} = \bar{u}_{\nu} \gamma^{\alpha} \left(1+\gamma_{5}\right) u_{e} \cdot \bar{u}_{n}\left(f_{1} \gamma_{\alpha} + g_{1} \gamma_{\alpha} \gamma_{5} + i f_{2} \sigma_{\alpha \beta} \frac{ q^{\beta}}{2 M} + g_{2} \frac{q_{\alpha}}{M} \gamma_{5}\right) u_{p}
\end{equation}
The following definitions of the Mandelstam variables are used here:
\begin{align}
    s &= (p_e + p_p)^2 = (p_\nu + p_n)^2 \, , \nonumber \\
    %%%%%%
    t &= (p_e - p_\nu)^2 = (p_n - p_p)^2 \, , \nonumber \\
    %%%%%%
    u &= (p_e - p_n)^2 = (p_\nu - p_p)^2 \, .
\end{align}
with $s+u+t = m_e^2 + m_p^2 + m_\nu^2 + m_n^2$ and $q^\mu = (p_n - p_p)^\mu = (p_e - p_\nu)^\mu$. After some tedious but straightforward algebra, we find the spin-averaged and summed amplitude-square to be
%%%%%%%%%
\begin{equation} 
    2\ |\mathcal{M}^2| = A(t) + (s-u) B(t) + (s-u)^2 C(t)
\end{equation}
where
%%%%%%%%
\begin{align}
   A(t) &= 4 |f_1^{2}| \Big\{ -\left(m_{e}^{2}-m_{\nu}^{2}\right)^{2} - 4 M^{2}\left(m_{e}^{2}+m_{\nu}^{2} - t\right) + t^{2} \Big\}+ 4 |g_1^2| \Big\{ -(m_e^2 - m_{\nu_R}^2)^2 \nonumber \\
   %%%%%%
   & + 4 M^2 (m_e^2+ m_{\nu_R}^2 - t) + t^2 \Big\} +  |f_2^2| \Big\{ - (m_e^2 + m_{\nu_R}^2 -t) \frac{t^2}{M^2} - 4 (m_e^2 - m_{\nu_R}^2)^2 + t^2 \Big\} \nonumber \\
   %%%%%%
   &+ 4 \frac{|g_2^2| t}{M^2} \Big\{- (m_e^2- m_{\nu_R}^2)^2 + (m_e^2 + m_{\nu_R}^2) t \Big\} + 8 {\rm Re}(f_1 f_2^\star) \Big\{2 t^2 - (m_e^2 - m_{\nu_R}^2)^2 \nonumber \\
   %%%%%%%
   & - (m_e^2 + m_{\nu_R}^2) t \Big\} + 16\ {\rm Re(g_1 g_2^\star)} \Big\{- (m_e^2 - m_{\nu_R}^2)^2 + (m_e^2 + m_{\nu_R}^2) t \Big\} \nonumber \\
   %%%%%%
   %%%%%%
   &- 4 \Delta^2 \bigg[ \left(|f_1^2| + \frac{|f_2^2| t}{4M^2}\right) (4 M^2 - m_e^2 - m_{\nu_R}^2 +t ) + |g_1^2| (4 M^2 + m_e^2 + m_{\nu_R}^2 - t)  \nonumber \\
   &+ \frac{|g_2^2|}{M^2} \Big\{ (m_e^2 + m_{\nu_R}^2) t - (m_e^2-m_{\nu_R}^2)^2 \Big\} + 4 (m_e^2 + m_{\nu_R}^2) {\rm Re}(g_1 g_2^\star) \nonumber \\
   &+ 2 (2t - m_e^2 - m_{\nu_R}^2) {\rm Re}(f_1 f_2^\star) \bigg]
   %%%%%%%%%
   + 64 M \Delta (m_e^2 - m_{\nu_R}^2) {\rm Re}(f_2^\star g_1 + f_1 g_1^\star) \\[5pt]
   %%%%%%%%%%%%%%%%%55
   %%%%%%%%%%%%%%%%%%
   B(t) &= -16 t {\rm Re}(f_2^\star g_1 + f_1 g_1^\star) + 4 \Delta \frac{(m_e^2-m_{\nu_R}^2)}{M} (|f_2^2| + {\rm Re}(f_1 f_2^\star + g_1 g_2^\star)) \\[5pt]
   %%%%%%%%%%%%%%%%%%%%%
   %%%%%%%%%%%%%%%%%%%%%%%
   C(t) &= 4 \left(|f_1^2| + |g_1^2|\right) - \frac{|f_2^2| t}{M^2}
   \label{eq:amp}
\end{align}
where we have used the definitions 
\begin{equation}
    \Delta = m_n - m_p \, , \hspace{20 mm} M = \frac{m_p + m_n}{2} \, . 
\end{equation}
Furthermore the form factors $(f_i, g_i)$ are real functions and are given by 
\begin{align}
    f_1 &= \frac{1- (1+ \xi)t / 4 M^2}{(1-t/4M^2) (1-t/M_V^2)^2} \, , \hspace{10mm}  f_2 = \frac{\xi}{(1-t/4M^2) (1-t/M_V^2)^2} \nonumber \\
    %%%%%
    g_1 &= \frac{g_1(0)}{(1-t/M_A^2)^2} \, , \hspace{31mm} g_2 = \frac{2 M^2 g_1}{m_\pi^2 -t}
\end{align}
where $g_1(0) = -1.270 \pm 0.003$, $M_V^2 = 0.71$ GeV$^2$, $M_A^2 \simeq 1 \, {\rm GeV}^2$, and $\xi = \kappa_p - \kappa_n = 3.706$ is the difference between the proton and neutron anomalous magnetic moments in units of the nuclear magneton. 

These results are in agreement with the results of Ref. \cite{LlewellynSmith:1971uhs} as well as Ref. \cite{Strumia:2003zx} derived for quasi-elastic neutrino and anti-neutrino scattering on nucleons if $m_{\nu_R}$ in Eq. (\ref{eq:amp}) is set to zero, and if the signs of the terms with coefficients $f_1^* g_1$ and $f_2 g_1^*$ are flipped.  This sign flip arises due to the $V+A$ nature of the leptonic current in the present case.  We have numerically evaluated the total cross section as a function of the electron center of mass energy, which is plotted in Fig. \ref{fig:sigEe}. From this figure  one can read off the cross section to be $\sigma = 69.2 \times 10^{-41} {\rm cm}^2$ for electron energy of 150 MeV.  This value is a factor of 3.3 smaller compared to the estimate used in Ref. \cite{Barbieri:1988av}.  This difference can be attributed primarily to the low energy approximation used as well as due to the absence  $g2$ and $f_2$ form factors and the momentum dependence of the form factors in the naive estimate.  
Since the core temperature of supernova is in the range of (30-70) MeV, we find it reasonable to choose the average electron energy to be about 150 MeV, in contrast to the energy of 300 GeV used in Ref. \cite{Barbieri:1988av}.  Following the same rough model of supernova dynamics, we have derived the mass limit on $W_R$, which can be as low as about 4.6 TeV.  For this estimate we also used the fact that $|\sin\zeta| < 0.95 \times M_{W_L}^2/M_{W_R}^2$, which arises from the requirement that the top-quark Yukawa coupling not exceed about 1.5 (see discussions following Eq. (\ref{eq:yukk}).) Although the estimate is very rough, we conclude that the model with low $W_R$ mass may be compatible with supernova supernova constraints.

There is one other source of energy loss in supernovae in the presence of a $\nu_R$ field with a mass less than 10 MeV.  This arises through the transition magnetic moment interactions which could produce $\nu_R$ via $\nu_L e \to \nu_R e$ and $\nu_L p \to \nu_R p$.  Once produced this way, the $\nu_R$ will escape, thus providing a source for supernova energy loss. Since in the model with low $W_R$, the $\nu_R$ decays into a $\nu_L + \gamma$, the transition magnetic moment is sizeable \cite{Babu:1987be,Fukugita:1987ti}.  For the decay lifetime to be $< 1$ sec, we find that the transition moment is about $\mu_{\nu_R\nu_L} \sim  1 \times 10^{-11} \mu_B$. Ref. \cite{Barbieri:1988nh} has estimated an upper limit of $(0.2 - 0.8) \times 10^{-11} \mu_B$ from the energy loss argument, which may be just about consistent with the needed value within the model.

%%%%%%%%%%%%%%%%%%%%%%%%%%%%%%%%%%%%%%%%%%%
%%%%%%%%%%%%%%%%%%%%%%%%%%%%%%%%%%%%%%%%%%%
\section{High Scale Left-Right Symmetry}\label{sec:nufit}
%%%%%%%
We now wish to show that the model is consistent if the $W_R^\pm$ mass is very high, well above the LHC reach, by fitting the model with neutrino oscillation data. Here one does not need to decouple one of the $\nu_R$ fields from the rest as was done in the low scale $W_R$ scheme. We explore both the type-I and type-II seesaw scenarios, and show that the model only supports the former case. We adopt the usual seesaw assumption  by taking $M_{\nu^D} \ll M_{\nu_R}$.

Here one can have Dirac neutrino mass $M_{\nu^D}$ arbitrary and large, unlike the low scale $W_R^\pm$ scheme discussed in Sec.~\ref{sec:lowLR}. Thus, it has enough parameters to fit the light neutrino oscillation observables. A simplifying assumption is to take $\kappa^\prime = 0$, so that the charged lepton masses and Dirac neutrino masses in Eq.~\eqref{eq:nuM} become
\begin{align}
    M_\ell = \frac{1}{\sqrt{2}}\ \widetilde{y} \kappa \, , ~~~
    %%%%%%
    M_{\nu^D} = \frac{1}{\sqrt{2}}\ y \kappa \, . 
    \label{eq:typeID}
\end{align}
One can choose to work in a basis where $M_\ell$ is diagonal, in which case $\widetilde{y}$ is also diagonal. Thus, $M_{\nu_R}$ has the same structure as in Eq.~\eqref{eq:fitMR} with a modified overall factor $\mathcal{J'}$ given as
 \begin{equation}
     \mathcal{J'} = \frac{2\sqrt{2}\ \alpha_4 v_R f_{e\tau} m_\tau^2}{\kappa^2} \big( \lambda_{ \beta \gamma}^2\ I_{cd}^{\eta \gamma \beta} - \lambda_{ \beta \gamma}^{1'}\ I_{cd}^{\eta \beta \gamma} \big)
 \end{equation}
where $\lambda_{ \beta \gamma}^2$ and $\lambda_{ \beta \gamma}^{1'}$ are given in Eq.~\eqref{eq:coeffInt}. There is ample freedom in the choice of the Dirac neutrino mass texture; we choose a specific form given by
\begin{equation}
    %%%%%%%
    M_{\nu^D} = m_{33} \begin{pmatrix}
     z_{11} & z_{12} e^{i \varphi} & z_{13}  \\
   z_{12} e^{-i \varphi} & z_{22} & 0 \\
   z_{13} & 0 & 1
    \end{pmatrix} \, .
    \label{eq:Diractex}
\end{equation}
From this structure we obtain a symmetric light neutrino mass matrix through the type-I seesaw formula as
\begin{equation}
   (M_\nu^{\rm light})_{ij} = m_0\ a_{ij} \, , \hspace{10 mm} 
   %%%%%%%%~.
\label{eq:MLighta}
\end{equation}
Here $a_{ij} = a_{ji}\ (i,j = 1,2,3)$ are  obtained by inserting Eq.~\eqref{eq:fitMR} and Eq.~\eqref{eq:Diractex} into type-I seesaw equation given in Eq.~\eqref{eq:lightnu}. Here $m_0$ fixes the overall scale of the light neutrino masses. For this structure of light neutrino mass matrix, the model provides excellent fits for both the normal hierarchy and the inverted hierarchy as shown in Table \ref{nuTab1} as Fit3 (NH) and Fit4 (IH) along with the $3\sigma$ allowed ranges taken from a recent NuFit5.0 global analysis \cite{Esteban:2020cvm}. The choice of parameters that gives these fits are tabulated in Table.~\ref{nuTab2}. These fits are in perfect agreement with the observed experimental values.

The overall scale $m_0$ determines the scale for the $SU(2)_R$ breaking, which depends on the choice of the Dirac mass entry $m_{33}$. With the benchmark parameters of Fit1 (NH) and Fit2 (IH) represented in Table ~\eqref{nuTab2}, we can simply write the right-handed neutrino masses as
\begin{align}
    M_{\nu_R} &=  1.25\times 10^{11}\ {\rm Diag}\{ 1.318, 1.315, 0.0035 \} \Big(\frac{m_{33}^2\ }{{\rm GeV}}\Big)\  \, , \hspace{10mm} {\tt Fit1\ (NH)} \\
    %%%%
     M_{\nu_R} &=  6.9 \times 10^{8}\ {\rm Diag}\{ 1.690, 1.630, 0.0380 \} \Big(\frac{m_{33}^2\ }{{\rm GeV}}\Big)\ \, . \hspace{13mm} {\tt Fit2\ (IH)}
\end{align}

%%%%%%%%%%%%%%%%%%%%%%%%%%%%%%%%%%%%%%%%%%%%%%%%%%%%%%%%%%%%%%%
%%%%%%%%%%%%%%%%%%%%%%%%%%%%%%%%%%%%%%%%%%%%%%%%%%%%%%%%%%%%%%%
    \color{black}
\begin{table}[t!]
%\vspace{-4mm}
\centering
\begin{tabular}{|c|c|c|c|}
\hline \hline
 {\textbf{Oscillation}}   &   {\textbf{$3\sigma$ allowed range}} &  \multicolumn{2}{c|}{\textbf{Model Fit}}\\
 \cline{2-4}
 {\bf parameters} & \bf{{\tt NuFit5.0}}~\cite{Esteban:2020cvm} & {\tt Fit3 (NH)} & {\tt Fit4 (IH)}   \\ \hline \hline
 $\Delta m_{21}^2 (10^{-5}~{\rm eV}^2$)  &   6.82 - 8.04  & 7.40 & 7.42    \\ \hline
 $\Delta m_{23}^2 (10^{-3}~{\rm eV}^2) $(IH) &   2.414 - 2.581  & - & 2.48   \\ 
 $\Delta m_{31}^2 (10^{-3}~{\rm eV}^2) $(NH) &   2.435 - 2.598   & 2.517 & -  \\ \hline
  $\sin^2{\theta_{12}}$   &   0.269 - 0.343 & 0.314  & 0.311  \\ \hline
 $\sin^2{\theta_{23}}$  (IH) &   0.419 - 0.617  & - &    0.589   \\
 $\sin^2{\theta_{23}}$ (NH)  &   0.415 - 0.616  & 0.570 & -   \\ \hline 
  $\sin^2{\theta_{13}} $  (IH) &   0.02052 - 0.02428  & - &  0.0228 \\
  $\sin^2{\theta_{13}}  $(NH)  &   0.02032 - 0.02410  & 0.0217 & -      \\ \hline
  $\delta_{CP}/^\circ$ (IH) & 193 - 352 & - & 318  \\ 
  $\delta_{CP}/^\circ$ (NH) & 120 - 369 & 317 & -  
  %%%%%%%
  \\\hline \hline
\end{tabular}
\caption{$3\sigma$ allowed ranges of the neutrino oscillation parameters from a recent global-fit~\cite{Esteban:2020cvm}, along with the model predictions for both normal (NH) and inverted (IH) hierarchy scenarios. }
\label{nuTab1}
%%%%%
%%%%%%%%%%%%%%%%%%%%%%
\vspace{5mm}
 \centering
\begin{tabular}{|c|c|c|c|c|c|c|c|c|c|c|}
\hline \hline
  & $x_1$ & $x_2$ & $z_{11}$ & $z_{12}$ & $z_{13}$ & $z_{22}$ & $\varphi$ & $m_0$ (eV)   \\ \hline 
  %%%%%
  {\tt Fit3 (NH)}  & $1$ & $-0.860$ & $0.649$ & $-0.562$ & $-0.139$  & $0.617$ & $2.42^\circ$ & $0.008$   \\ \hline
  %%%%%%
{\tt Fit4 (IH)}  & $11.17$ & $-1.316$ & $-0.0205$ & $-0.0416$ & 0 & $0.0448$ & $5.10^\circ$ & 1.45  \\ \hline \hline
%%%%%%%%%%%
\end{tabular}
\caption{Best fit values of parameters that yield Fit1 and Fit2, as prescribed in Table \ref{nuTab1} to fit the neutrino oscillation data.  }
\label{nuTab2}
%\vspace{-4mm}
\end{table}
%%%%%%%%%%%%%%%%%%%%%
%%%%%%%%%%%%%%%%%%%%%%%%%%%%%%%%%%%%%%%%%%%%%%%%%%%%%%%%%%%%%%%
%%%%%%%%%%%%%%%%%%%%%%%%%%%%%%%%%%%%%%%%
%%%%%%%%%%%%%%%%%%%%%%%%%%%%%%%%%%%%%%%%
%\vspace{-5mm}
\begin{figure}[!t]
$$
    \includegraphics[scale=0.28] {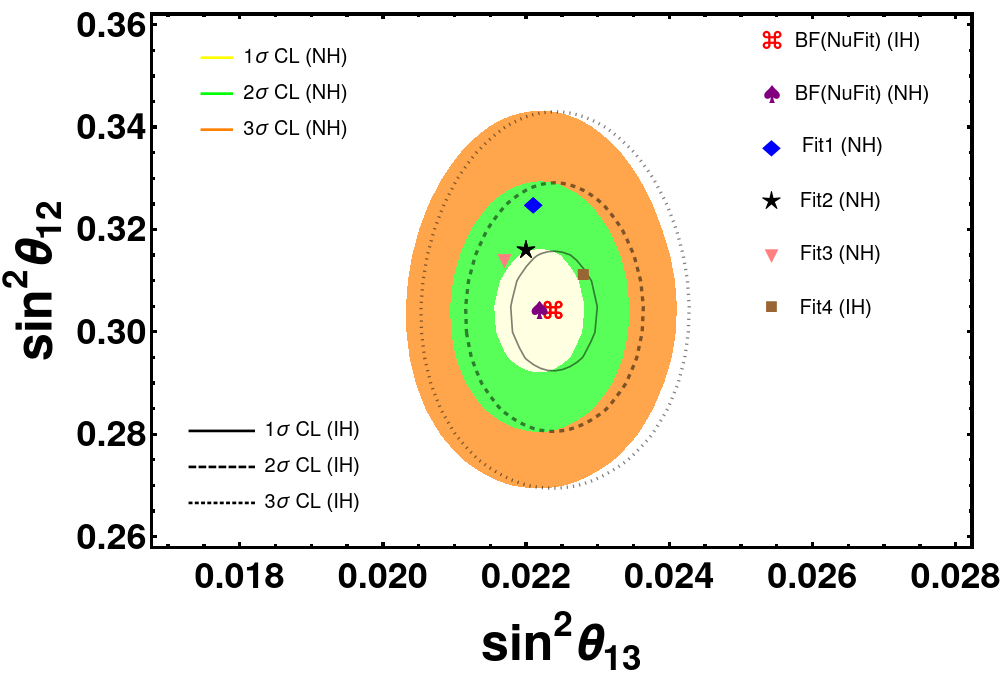}
    \includegraphics[scale=0.28] {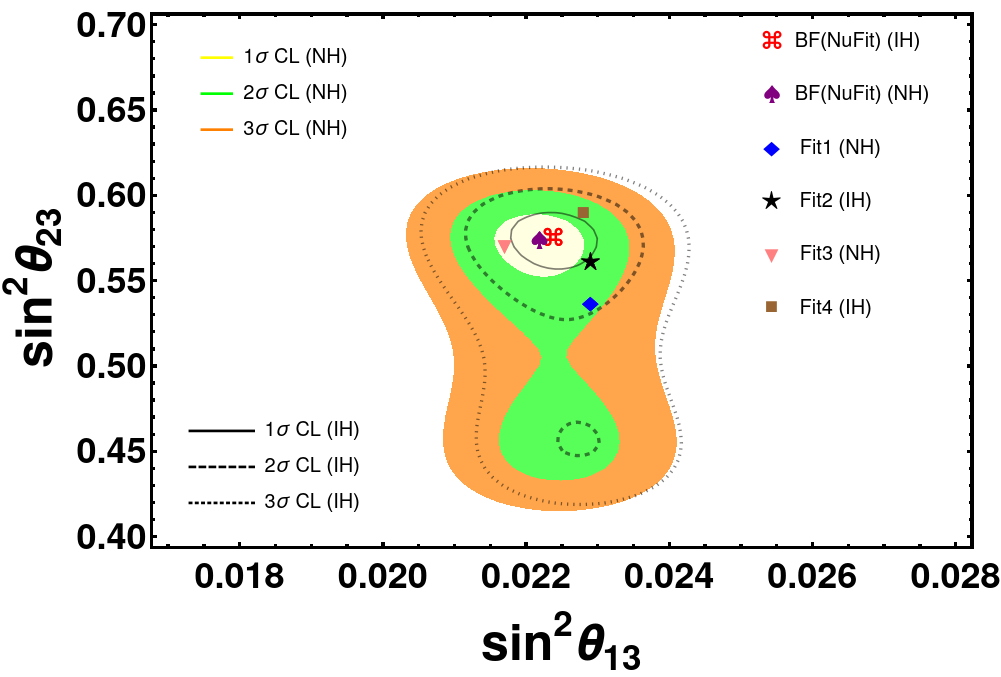}
$$  
$$
    \includegraphics[scale=0.28] {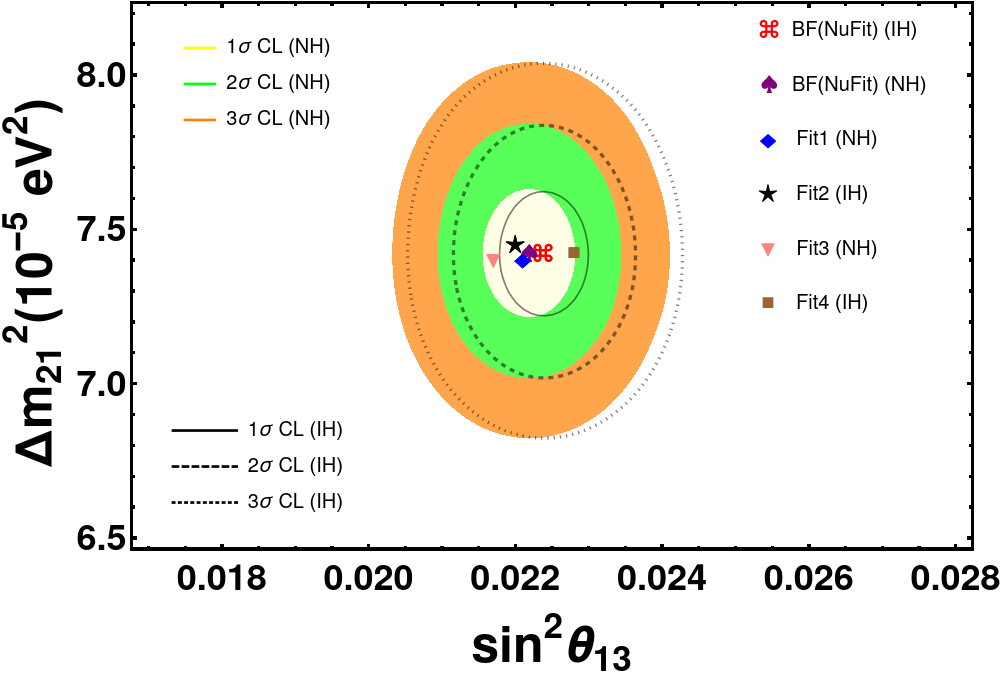}
    \includegraphics[scale=0.28] {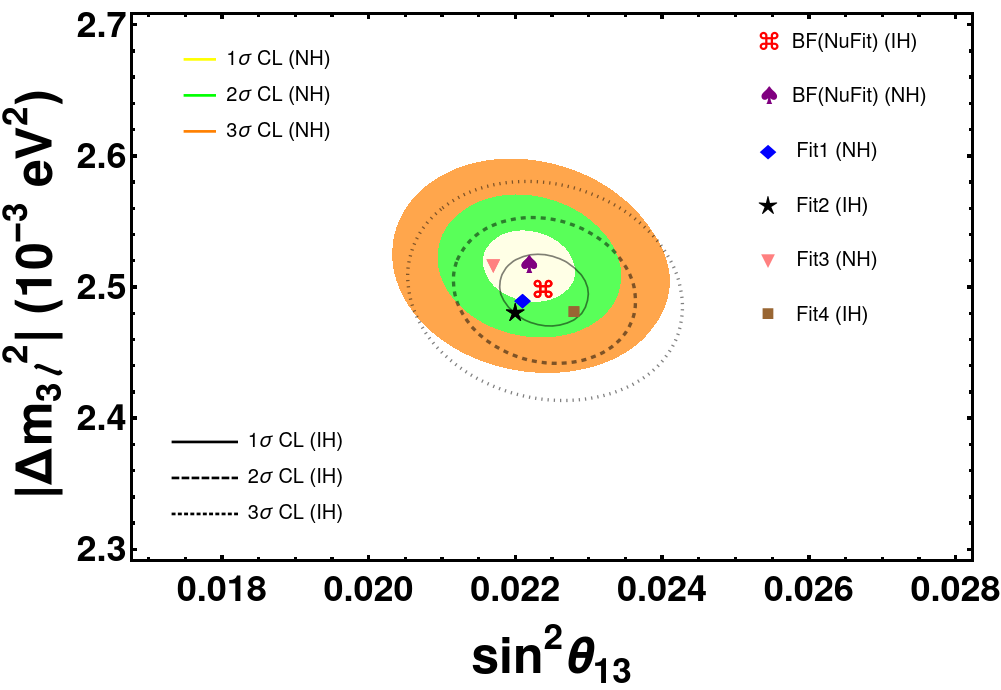}
$$
    \caption{Global oscillation analysis obtained from NuFit5.0 \cite{Esteban:2020cvm} for both normal hierarchy (NH) and inverted hierarchy (IH) compared with our model benchmark points (Fit1 and Fit2). Yellow, Green, and orange colored contours represent $1\sigma$, $2\sigma$, and $3\sigma$ CL allowed regions for NH, whereas solid, dashed, and dotted lines respectively represent $1\sigma$, $2\sigma$, and $3\sigma$ CL allowed regions for IH. Red, purple, and (blue, black, pink, brown) markers are best-fit from NuFit for IH and NH, and benchmark points Fit1, Fit2, Fit3, and Fit4. }
    \label{fig:chisq}
\end{figure}
%%%%%%%%%%%%%%%%%%%%%%%%%%%%%%%%%%%%%%%%%%%%
%%%%%%%%%%%%%%%%%%%%%%%%%%%%%%%%%%%%%%%%%%%%%%
In addition to the best-fit results in the tabulated format, we also display them in Fig. \ref{fig:chisq} in the two-dimensional projections of $1\sigma$, $2\sigma$, and $3\sigma$ confidence regions of the global fit results \cite{Esteban:2020cvm} (with the inclusion of the Super-K atmospheric $\Delta \chi^2$-data). The global-fit best-fit points, along with the model predictions for each benchmark point are shown for comparison. The theoretical predictions are in good agreement within the observed experimental data. 

%%%%%%%%%%%%%%%%%%%%%%%%%%%%%%%%%%%%%%%%%%%%%%%%55
%%%%%%%%%%%%%%%%%%%%%%%%%%%%%%%%%%%%%%%%%%%%%%%%%5

%%%%%%%%%%%%%%%%%%%%%%%%%%%%%%%%%%%%%%%%%%%%%%%%%%%%%%5
%%%%%%%%%%%%%%%%%%%%%%%%%%%%%%%%%%%%%%%%%%%%%%%%%%%%%%5
\subsection{Inconsistency with the type-II seesaw scenario}\label{sec:typeII}
%%%%%%%%%%%%%%%%%%%%%%%%%%%%%%%%%%%%%%%%%%%%%%%%%%%%%%5
%%%%%%%%%%%%%%%%%%%%%%%%%%%%%%%%%%%%%%%%%%%%%%%%%%%%%%5
%%%%%%%%
In the limit of small mixing between scalars, i.e. in the limit when the flavor eigenstate and mass eigenstate coincide, we take Eq.~(\ref{eq:SeesawMat}) and obtain the condition when $M_\nu^{I}$ dominates $M_\nu^{II}$. Note that for the type-II contribution, $M_\nu^L$ is given by one-loop diagrams which is shown in Fig.~\ref{fig:loopdiag} (a) and evaluated in Eq.~\eqref{eq:1loopvL}. Now Eq.~(\ref{eq:SeesawMat}) can be written as
\begin{equation}
    M = \begin{pmatrix}
      \varepsilon \frac{\kappa + v_L}{v_R}  \mathcal{F'}  &\hspace{5mm} M_{\nu^D}  \\[5pt]
 M_{\nu^D}^T  & \varepsilon^2 v_R \alpha_4 \mathcal{F}\\
    \end{pmatrix}
\end{equation}
where $\mathcal{F'}$ and $\mathcal{F}$ is the flavor structure given in Eq.~\eqref{eq:1loopvL} and Eq.~(\ref{eq:MajR}). $\varepsilon$ is the one-loop factor, which is equal to $1/(16 \pi^2)$. After some straightforward algebra, we obtain the following relation between type-I and type-II:
\begin{equation}
    M_{\nu}^{II} \lesssim \frac{\varepsilon^3 M_\ell}{\kappa}\ M_{\nu}^{I}
    \label{eq:condII}
\end{equation}
%%%%%%%%%%%%%%%%%%%%%%%%%%555
It is clear from the above equation that the type-I contribution will always dominates type-II contribution. Similar analysis can be performed with different scenarios such as $y<\widetilde{y}, \,\kappa < \kappa^\prime$ and vice-versa. However, it turns out there is no solution in any of these cases where the type-II contribution dominates. It is worth mentioning that  fine-tuning $y\kappa = - \widetilde{y} \kappa^\prime$ can lead type-II dominance; however, the diagonal elements of the light neutrino mass matrix will be all zero in this case. With diagonal elements zero in the neutrino mass matrix, in a basis where the charged lepton masses are diagonal, one cannot obtain the correct neutrino oscillation pattern \cite{Zee:1985id, Zee:1980ai, Wolfenstein:1980sy}. 
%%%%%%%%%%%%%%%%%%%%%%%%%%%%%%%%%%%%%%%5
%%%%%%%%%%%%%%%%%%%%%%%%%%%%%%%%%%%%%%%%%%%%%%%%%%%%%
%%%%%%%%%%%%%%%%%%%%%%%%%%%%%%%%%%%%%%%%%%%%%%%%%%%%%%%%%%%%%%%%%%%%%%%%

%%%%%%%%%%%%%%%%%%%%%%%%%%%%%%%%%%%%%%%5
%%%%%%%%%%%%%%%%%%%%%%%%%%%%%%%%%%%%%%%%%%%%%%%%%%%%%
%%%%%%%%%%%%%%%%%%%%%%%%%%%%%%%%%%%%%%%%%%%%%%%%%%%%%%%%%%%%%%%%%%%%%%%%
\section{Collider Implications} \label{sec:collider}
%%%%%%%%%%%%%%%%%%%%%%%%%%%

The $W_R^\pm$ gauge bosons as well as other new particles in the model can be produced at the Large Hadron Collider experiments, if they are sufficiently light.  In the standard left-right symmetric model, the $W_R$ boson can be resonantly produced when kinematically allowed, which then decays into a charged lepton plus right-handed neutrino, or a pair of jets.  CMS has obtained a lower limit of 3.6 TeV on the $W_R$ mass from resonant searches in the dijet channel \cite{Sirunyan:2019vgj}. This limit is applicable to the present model.  There are also somewhat stronger limits derived from searches for the decays of $W_R$ into leptonic final state of the same sign or opposite sign \cite{Aaboud:2018spl,Sirunyan:2018pom}. These occur in the standard left-right symmetric model as the heavy $\nu_R$ decays into leptons plus jets \cite{Keung:1983uu}.  In our model, however, when the $W_R$ mass is in the TeV range, the mass of $\nu_R$ is of order 10 MeV, which would mean that it won't decay within the detector. Thus these leptonic constraints on the $W_R^\pm$ mass are not applicable to our scenario.

There are however, other ways of testing the model at the LHC.  We focus on the discovery potential of the right-handed neutrino as well as the $\eta^\pm$ scalar present in the model.  $\eta^+ \eta^-$ can be pair-produced via the Drell-Yan process mediated by the $Z$ and photon, which has a significant cross section at LHC energies for relatively low mass $\eta^+$.  Assuming that the $W_R^\pm$ is quite heavy, the mass of the $\nu_R$ can be in the few to hundred GeV range in the model.  The $\eta^+$ would then decay into $\ell_R^+ \nu_R$ as well as into $\ell^+_L \nu_L$ final states, with roughly equal branching ratios.  The $\nu_R$ would then decay through a virtual $\eta^\pm$ exchange into $\ell_R + \ell_L + \nu_L$.  This would lead to interesting multi-lepton signals that has already been searched for by CMS. Here we carry out an analysis of this process and derive bounds from the LHC and estimate the reach of high luminosity LHC.

%%%%%%%%%%%%%%%%%%%%%%%%%%%%5
\begin{figure}
 $$
    \includegraphics[scale=0.4]{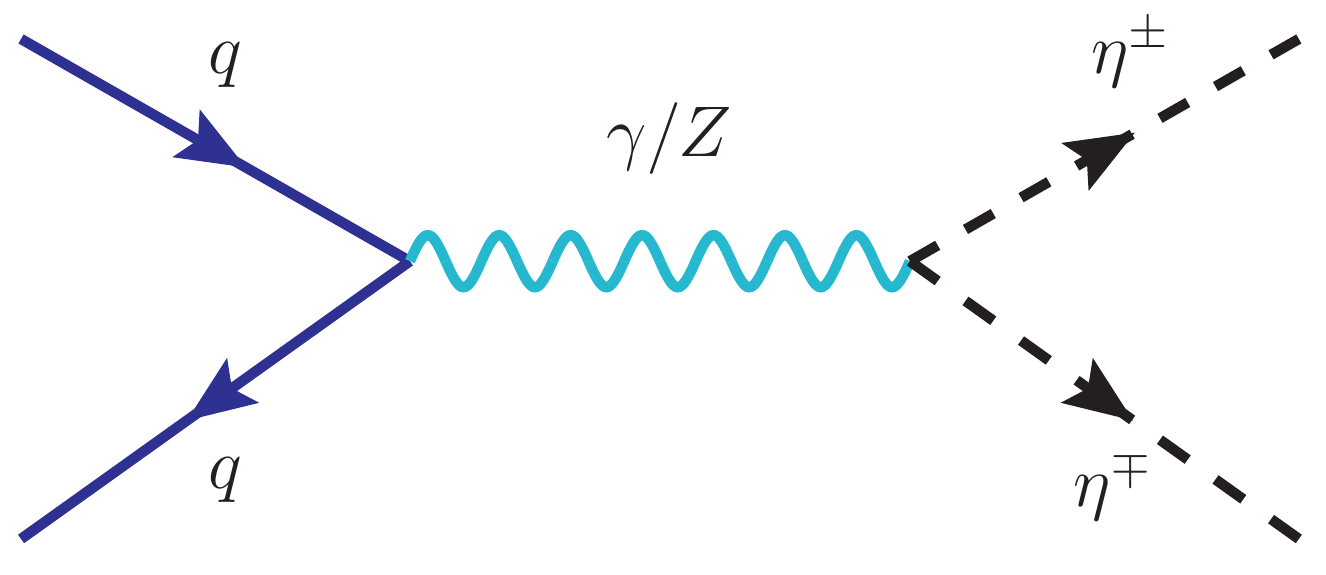} \hspace{20mm}
    \includegraphics[scale=0.4]{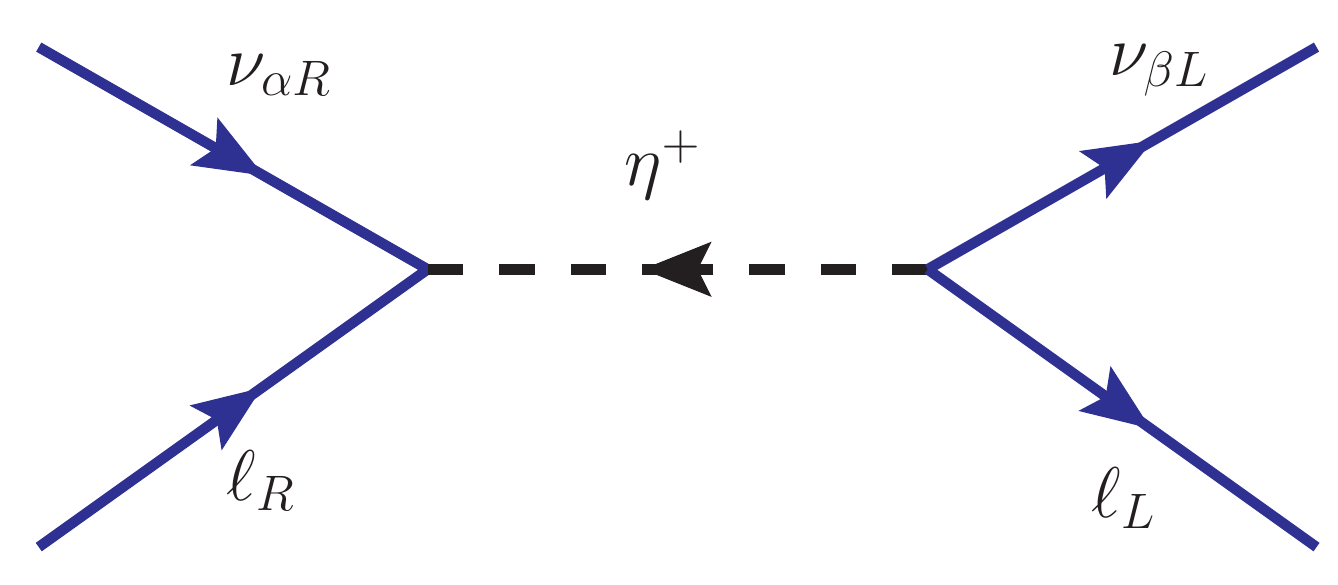}
    $$
    \caption{Feynman diagram for pair production of singly-charged scalar $\eta^{\pm}$ (left) and decay of RH neutrino $\nu_{ R} \to \ell^+ \ell^- \nu$  (right) at LHC (note: $\alpha, \beta, \ell = e, \mu \, (\beta, \alpha \neq \ell)$) .}
    \label{fig:feynLHC}
\end{figure}
%%%%%%%%%%%%%%%%%%%%%%%%%%5%
%%%%%%
\begin{figure}
\centering
    \includegraphics[scale=0.4]{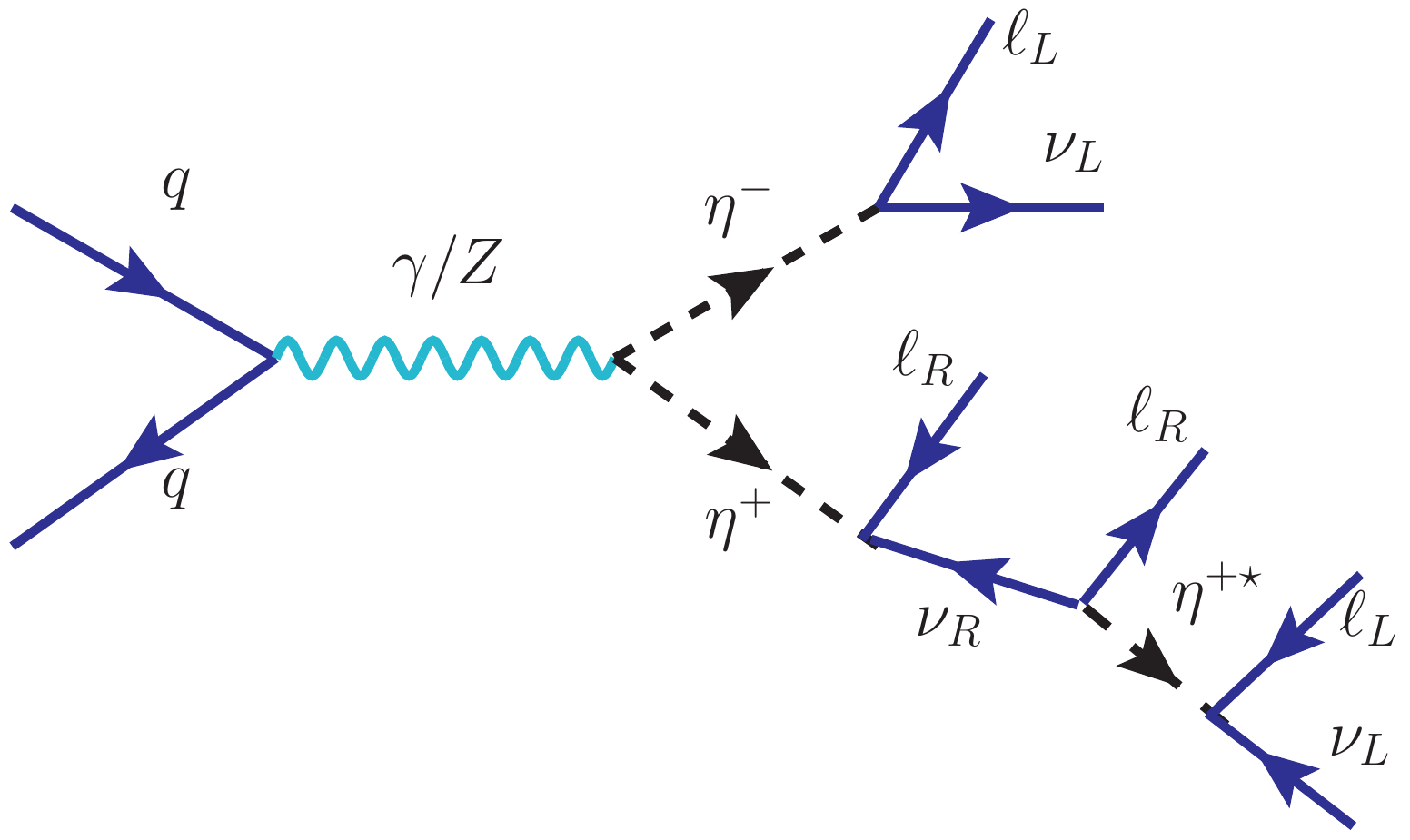}
    \caption{Feynman diagram for the production of 4-lepton + $\slashed{E}_T$ through Drell-Yan process with pair production of $\eta^{\pm}$. $\nu_R$ decays to lepton pair and neutrino via virtual $\eta$.
    }
    \label{fig:feyn4lep}
\end{figure}
%%%%%%

It is worth mentioning that $M_{\nu_R} \ll M_{W_R}$ in the model due to the two-loop suppression in $\nu_R$ mass.  Thus it is natural to expect $\nu_R$ to be light, even when the $W_R$ gauge boson is heavy. The decay of $\nu_R$ into SM leptons is suppressed due to their small mixing, typically of the order $m_{\nu^D}/M_{\nu_R}$.  Furthermore, the decay of $\nu_R$ to $\nu_L \phi_1^0 (\phi_2^0)$ is also suppressed, as the coupling with the SM Higgs is small, while the other neutral scalar should be heavy, of order 10 TeV or higher, in order to satisfy flavor changing neutral current constraints \cite{Guadagnoli:2010sd}.  

We consider the case where in addition to the $\nu_R$, the scalar $\eta^+$ is also light, which opens up the possibility of production of $\nu_R$ via $\eta^+$, as shown in Eq.~(\ref{eq:yuk}). At the LHC  $\eta^+$ is produced through the $s$-channel Drell-Yan process $pp \to \gamma / Z \to \eta^{+} \eta^{-}$, as shown in Fig. \ref{fig:feynLHC} (left) followed by $\eta^{\pm}$ decaying into leptons \footnote{The decay of $\eta^{\pm}$ to quarks is generally suppressed since the mixing angle of $\eta^+$ with other charged scalars is small.}. We take the cleanest channels $e, \mu $ in the final state from the decay of $\eta^{\pm}$. This can be achieved by setting the Yukawa coupling that couples $\eta^{\pm}$ to $\nu_\alpha \tau$ to be small, as shown by Table \ref{nuTab2}, obtained from neutrino oscillation fit. We also set the coupling $f_{e \mu}$ at $0.01$ to suppress the muon decay constraint and the single production of $\eta^{\pm}$. Then,  $\eta^+$ decays into  $\Bar{\nu}_e \mu^+, \Bar{\nu}_\mu e^+, \Bar{\nu}_{eR} \, \mu_R^+$, and $\Bar{\nu}_{\mu R} e_R^+$ with a branching ratio (BR) of 1/4 for each process. The right-handed neutrino $\nu_{eR}$ and $\nu_{\mu R}$  can decay into a pair of leptons and neutrinos via virtual $\eta$, as shown in Fig. \ref{fig:feynLHC} (right) with branching ratio 1/2 for each process (i.e. $\nu_{eR} \to \mu^- \mu^+ \nu_e, \, e^- \mu^+ \nu_\mu$ and $\nu_{\mu R} \to \mu^- e^+ \nu_e, \, e^+ e^- \nu_\mu$). This means that the Drell-Yan production of $\eta^+ \eta^-$ has $2l+\slashed{E}_T$, $4l+\slashed{E}_T$, and $6l+\slashed{E}_T$ as possible final states. 

We note that although dilepton + MET search ($p p \to \eta^+ \eta^- \to \ell^+ \ell^- \slashed{E}_T$) is appealing as it has a higher cross-section, the background is much harder to suppress. Thus, we study the process with 4-leptons and 6-leptons final states, i.e, $p p \to 4l + \slashed{E}_T$ (Ref: Fig. \ref{fig:feyn4lep}) and $p p \to 6l + \slashed{E}_T$. 4-leptons and 6-leptons final states will have a suppressed background in contrast to 2-lepton final states. We implement our model file in the {\tt FeynRules} package \cite{Christensen:2008py} and compute all the cross-sections at the parton-level using {\tt MadGraph5} event generator \cite{Alwall:2014hca}. As a consistency check, we have also generated a model file in {\tt SARAH} \cite{Staub:2013tta} and obtained results in agreement. 

%%%%%%%%%%%%%%%%%%%%%%%55
\begin{figure}
\centering
    \includegraphics[scale =0.4]{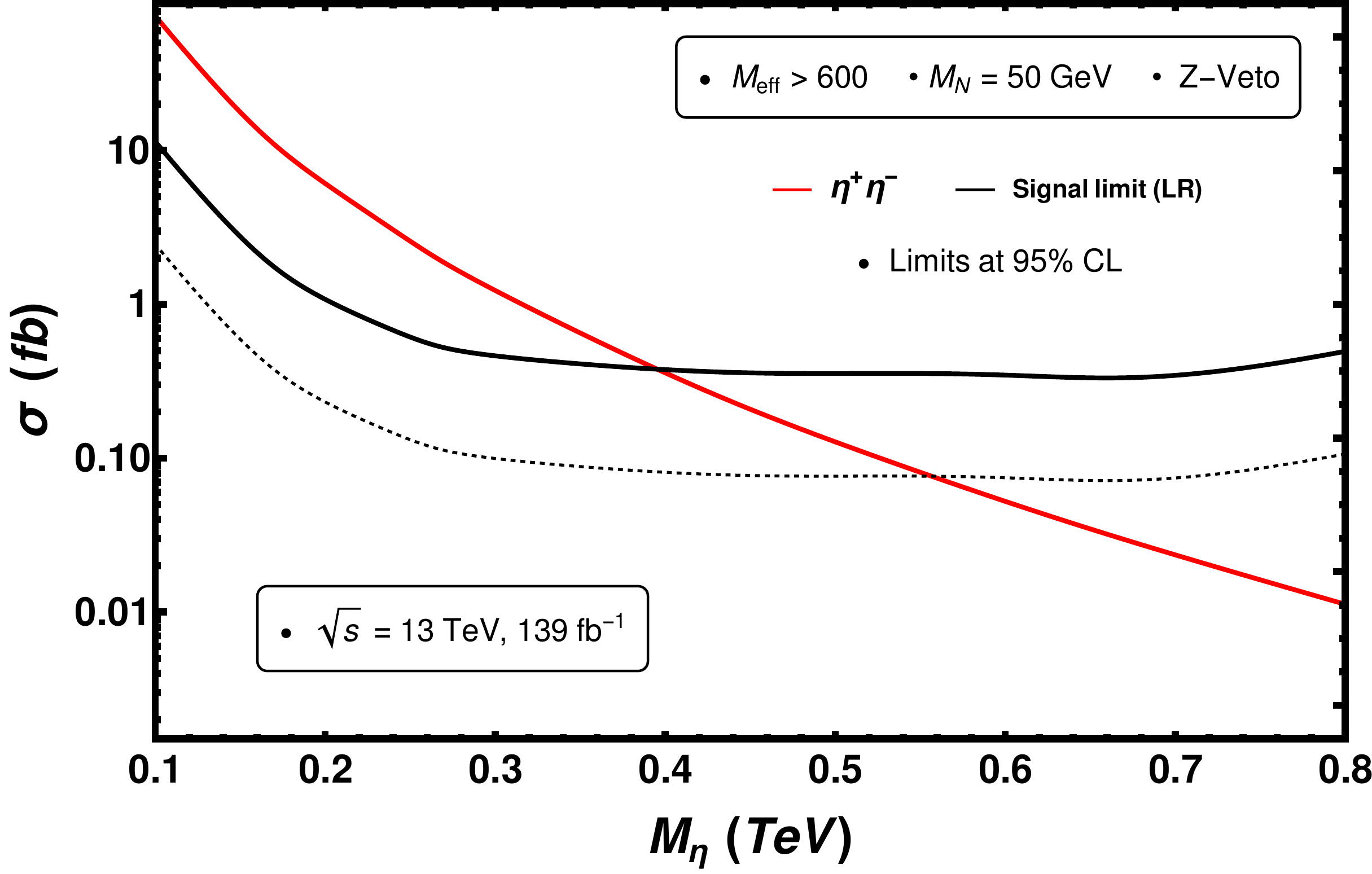} 
    \caption{The observed $95\%$ C.L. limit of the production cross section, $\sigma (p p \to \eta^+ \eta^-)$, as a function of mass of $\eta^+$ scalar obtained from four leptons searches at $\sqrt{s} = 13$ TeV pp collisions by the  ATLAS experiment \cite{ATLAS:2020fdg}. Dotted line corresponds to the future sensitivity limit at the high luminosity LHC with 3 ab$^{-1}$ data.}
    \label{fig:pp_eta}
\end{figure}
%%%%%%%%%%%%%%%%%%%%%%%%%%%%

Following the search done by ATLAS \cite{ATLAS:2020fdg} with four or more leptons, we reproduce the background for signal region (SR0$^{{\rm loose}}_{{\rm bveto}}$) with $139$ fb$^{-1}$ integrated luminosity at $\sqrt{s} = 13$ TeV proton-proton collisions. This same effective cut is implemented on the signal region for our model.  We implement $Z$-veto cut that rejects events where any SFOS
lepton pair has an invariant mass close to $Z$ boson mass, i.e., in the mass range of
$81.2-101.2$ GeV. To suppress the radiative $Z$ boson decays into four leptons, $Z$ veto also takes into account combinations of any SFOS LL pair with an additional lepton or with second SFOS LL pair. Also, to separate background and left-right (LR) model signals, $m_{eff} > 600$ GeV cut is used. Here $m_{eff}$ is defined as 
\begin{equation}
    m_{eff} = \sum_{leptons} p_T + \sum_{jets} p_T ( > 40 GeV) + E_T^{miss}~.
\end{equation}
where $E_T^{miss}$ is the missing transverse energy, and the $p_T > 40$ GeV requirement suppresses contribution from pileup and the underlying events \cite{Aad:2015ina}.
The observed signal limit, %of fiducial cross-section  $0.3324$ fb
as shown in Ref. \cite{ATLAS:2020fdg} is used to evaluate constraints on our model. Fig. \ref{fig:pp_eta} shows the limit on the production cross-section of $\eta^+ \eta^-$ as a function of the mass of the scalar $\eta^+$ assuming the mass of right-handed neutrino mass at 50 GeV. We see that the current limit on the mass of $\eta^{\pm}$ is 390 GeV from LHC, as shown in Fig. \ref{fig:pp_eta}. We also show in Fig.~\ref{fig:pp_eta} the mass reach for 3 ab$^{-1}$ integrated luminosity by rescaling and assuming the same efficiency. We find that the reach for $\eta$ mass is 555 GeV. 

For 6-leptons final state, both $\eta^+$ and $\eta^-$, as shown in Fig. \ref{fig:feynLHC} and Fig. \ref{fig:feyn4lep} decays into right-handed neutrinos $\nu_R$; both $\nu_R$ then decay to a pair of leptons and a light neutrino. There are no current 6-leptons $+$ MET searches available in the literature. It would be of interest to do such a study as we expect half the number of events in this channel in comparison to 4-leptons + MET searches. For instance, for $\eta^+$ mass of 390 GeV, 6-lepton final state would have a cross-section of 0.544 fb with the basic selection cuts form the ATLAS searches \cite{Aaboud:2018zeb, ATLAS:2020fdg}. It is beyond the scope of this work to do the full analysis as it requires detailed reducible background simulation, which leads to events with fake leptons, as shown in  Ref. \cite{ATLAS:2020fdg}.
%%%%%%%%%%%%%%%%%%%%%%%%%%%%%%%%%%%%%%%%%%%%%%%%%%%%%
%%%%%%%%%%%%%%%%%%%%%%%%%%%%%%%%%%%%%%%%%%%%%%%%%%%%%%%%%%%%%%%%%%%%%%%%
\section{Summary and Discussions}\label{sec:summary}

In this paper, we have presented a simple and minimal left-right symmetric model which does not use the conventional Higgs triplets.  Gauge symmetry breaking is achieved by Higgs doublets and a Higgs bi-doublet.  Majorana masses for the right-handed neutrinos are induced through two-loop diagrams involving a singly charged scalar field $\eta^+$.  This model naturally exhibits a hierarchy in the masses of $\nu_R$ and $W_R$.  If the $W_R$ gauge boson has a mass in the $(5-20)$ TeV range, the $\nu_R$ fields will have masses of a few tens of MeV.  We have shown that such a scenario is consistent with low energy constraints, as well as constraints arising from cosmology and astrophysics. 

The model presented admits type-I seesaw mechanism for the entire range of $W_R$ mass ranging from a few TeV to the GUT scale of order $10^{16}$ GeV. Prior analysis of left-right symmetric models with this Higgs spectrum focused on the one-loop induced $\nu_R$ Majorana masses, which turn out to be sub-dominant. For the entire parameter space of the model we have shown that the dominant contributions to $\nu_R$ masses would arise from two-loop diagrams, which do not rely on electroweak symmetry breaking, unlike the one-loop diagrams.  We have found excellent fits to neutrino oscillation parameters for low $W_R$ scenario as well for high $W_R$ scenario.  

We have explored the multi-lepton signals at colliders arising from the production and decays of the $\eta^+$ scalar, assuming that it is kinematically accessible to the LHC.  While the current limit on the $\eta^+$ mass is found to be 350 GeV, we estimate that at the high luminosity run of the LHC this limit can be improved to 555 GeV.

%%%%%%%%%%%%%%%%%%%%%%%%%%%%%%%%%%%%%%%%%%%%%%%%%%%%%
%%%%%%%%%%%%%%%%%%%%%%%%%%%%%%%%%%%%%%%%%%%%%%%%%%%%%%%%%%%%%%%%%%%%%%%%

\section*{Acknowledgement}
We thank Ahmed Ismail for discussions. 
This work is supported in part by the US Department of Energy Grant No. DE-SC 0016013.

 % reset counter 
%\clearpage
\appendix
\section*{Appendices} 
\addcontentsline{toc}{section}{Appendices}
\renewcommand{\thesubsection}{\Alph{subsection}}

\subsection{Evaluation of the charged scalar mass matrix}\label{sec:ChargedMass}

\renewcommand{\theequation}{A.\arabic{equation}}
  % redefine the command that creates the equation no.
  \setcounter{equation}{0} 
%%%%%%%%%%%%%%%%%%
In evaluating the masses  of charged scalar we first identify the Goldstone bosons as:
\begin{eqnarray}
     G_L^+ &=& \frac{v_L \chi_L^+ + \kappa^\prime \phi_2^+ - \kappa \phi_1^+}{\sqrt{v_L^2 + \kappa^{\prime 2} + \kappa^2}} \, ,  \label{eq:goldstoneL} \\
     %%%%%%
     G_R^+ &=& \frac{\kappa_L^2 v_R \chi_R^+ + 2 \kappa \kappa^\prime v_L \chi_L^+ + \kappa (\kappa_-^2 + v_L^2) \phi_2^+  - \kappa^\prime (\kappa_-^2 + v_L^2) \phi_1^+}{\sqrt{(\kappa_+^2 +  v_L^2)(\kappa_-^4 + v_L^2 v_R^2 + k_+^2 (v_L^2 + v_R^2))}} \, .
    \label{eq:goldstoneR}
\end{eqnarray}
$G_L^+$ and $G_R^+$ have been chosen to be orthogonal to each other. We choose the following orthogonal matrix $O$ which transforms the original basis $\{\phi_1^+ , \phi_2^+, \chi_L^+, \chi_R^+, \eta^+ \}$ to a new basis $\{ G_L^+, G_R^+, h_1^{\prime +}, h_2^{\prime +},  h_3^{\prime +} \}$.  
\begin{equation}
O^+ = \left(
    \begin{array}{ccccc}
        -\frac{\kappa}{\kappa_L} & \frac{\kappa^\prime}{\kappa_L}  & \frac{v_L}{\kappa_L} & 0 & 0 \\
        %%%%%%%%%%%%%%%%%%55
        \frac{- \kappa^\prime (\kappa_-^2 - v_L^2)}{\sqrt{N_2}} &  \frac{ \kappa (\kappa_-^2 + v_L^2)}{\sqrt{N_2}} &  \frac{2 \kappa \kappa^\prime v_L}{\sqrt{N_2}} &  \frac{\kappa_L^2 v_R}{\sqrt{N_2}} & 0 \\
         %%%%%%%%%%%%%%%%%%%%%
        \frac{\kappa^\prime v_R}{\sqrt{\kappa_-^4 + \kappa_+^2 v_R^2}}  & \frac{\kappa v_R}{\sqrt{\kappa_-^4 + \kappa_+^2 v_R^2}} & 0 & \frac{\kappa_-^2}{\sqrt{\kappa_-^4 + \kappa_+^2 v_R^2}} & 0 \\
        %%%%%%%%%%%%%%%%%%%%5
         \frac{\kappa v_L (\kappa_-^2 + v_R^2)}{\sqrt{N_1}}  & \frac{ \kappa^\prime v_L (-\kappa_-^2 + v_R^2)}{\sqrt{N_1}}  & \frac{-k_-^4 - k_+^2 v_R^2 }{\sqrt{N_1}} & \frac{2 \kappa \kappa^\prime v_L v_R }{\sqrt{N_1}}  & 0 \\
         %%%%%%%%%%%%%%%%%
         0 & 0 & 0 & 0 & 1 
    \end{array} 
    \right) \, ,
    \label{eq:Op}
\end{equation}
where $\kappa_L$ is given in Eq.~\eqref{eq:kL} and where we have defined normalization factors as
\begin{align}
    N_1 &= (\kappa_-^4 + \kappa_+^2 v_R^2)(\kappa_-^4 + v_L^2 v_R^2 + k_+^2 (v_L^2 + v_R^2)) \, , \nonumber \\
    %%%%%%%%
    N_2 &= (\kappa_+^2 +  v_L^2)(\kappa_-^4 + v_L^2 v_R^2 + k_+^2 (v_L^2 + v_R^2)) \, .
\end{align}
To get the mass spectrum for charged Higgs, mass matrix $M_+^2$ in the basis $\{ \phi_1^+, \phi_2^+, \chi_L^+, \chi_R^+, \eta^+ \}$ is first constructed from the bilinear terms by expanding the potential given in Eq.~(\ref{eq:potential}) around the VEVs shown in Eq.~(\ref{eq:vev}). One can write the mass matrix in the new basis by performing the transformation
\begin{equation}
    O^+ M_+^2 O^{+T} = \widetilde{M}_+^2 \, .
    \label{eq:chargD1}
\end{equation}
The elements of the symmetric mass matrix  $\widetilde{M}_{ij}^+ = \widetilde{M}_{ji}^+$ are given by
%%%%
\begin{align}
     \hspace{8mm} \widetilde{M}_{11}^+ &= \frac{1}{2(\kappa_-^6 + \kappa_-^2 \kappa_+^2 v_R^2)} \Big\{ \alpha_3 \Big( 4 \kappa^2 \kappa'^2 v_L^2 v_R^2 + (k_-^4 + \kappa_+^2 v_R^2)^2 \Big)   \nonumber\\
     %%%%%%%%%%
     &\hspace{2mm}+ 4\sqrt{2}\kappa' \mu_4 v_L v_R (\kappa_-^4 + \kappa_+^2 v_R^2) - \rho_{12} v_L^2 \Big(\kappa_-^6 + 2 \kappa_-^4 v_R^2  \Big) \nonumber \\
     %%%%%%%%%%%%%
     &\hspace{2mm} - \rho_{12} v_L^2 v_R^4 (3 \kappa'^2 + \kappa^{2}) \Big\}  \, ,\nonumber\\[7pt]
     %%%%%%%%%%%%%%%%%%
     %%%%%%%%%%%%%%%%%%%
    \widetilde{M}_{12}^+ &= \frac{\sqrt{N_1}}{2 \kappa (\kappa_-^4 + \kappa_+^2 v_R^2)^{3/2}} \Big\{ \kappa' v_L v_R (-2 \alpha_3 \kappa^2 + \rho_{12} (-\kappa_-^2 + v_R^2))  \nonumber \\
    &\hspace{1mm} - \sqrt{2} \mu_4 (\kappa_-^4 + \kappa_+^2 v_R^2) \Big\} 
    \, ,\nonumber\\[7pt]
    %%%%%%%%%%%%%%%
    %%%%%%%%%%%%%%%
     \widetilde{M}_{22}^+ &= \frac{1}{2(\kappa_-^4 + \kappa_+^2 v_R^2)^2} \Big\{ \Big(\alpha_3 \kappa_-^2 - \rho_{12} v_R^2 \Big) N_1 \Big\} \, ,\nonumber \\[7pt]
     %%%%%%%%%%%%%%%
     %%%%%%%%%%%%%%
     \widetilde{M}_{13}^+ &= \frac{1}{2 \sqrt{\kappa_-^4 + \kappa_+^2 v_R^2}} \Big\{ - \alpha_4 (\kappa + \kappa^\prime) v_L \Big( (\kappa-\kappa^{\prime})^2 + v_R^2 \Big) \Big\}  \, ,\nonumber \\[7pt]
     %%%%%%%%%%%%%%%%%%%
     %%%%%%%%%%%%%%%%%%
     \widetilde{M}_{23}^+ &= \frac{1}{2 (\kappa_-^4 + \kappa_+^2 v_R^2)} \alpha_4 v_R (\kappa- \kappa^\prime )\sqrt{N_1}  \, ,\nonumber \\[7pt]
     %%%%%%%%%%%%%%%%%%%
     %%%%%%%%%%%%%%%%%%
     \widetilde{M}_{33}^+ &= 2 \alpha_6 \kappa \kappa^{\prime} + \frac{\alpha_5}{2} \kappa_+^2 + \mu_\eta^2 + \frac{\alpha_7}{2} (v_L^2 + v_R^2)  \, ,
     \label{eq:A6}
\end{align}
where $\rho_{12}= 2\rho_1 - \rho_2$. One can further rotate $\widetilde{M}_+^2$ to the physical basis $\{ H_1^{+}, H_2^{ +}, H_3^{+}\}$ to get the masses of charged scalars such that
\begin{equation}
    O'^+ \widetilde{M}_{+}^2 O'^{+T} = \hat{M}_{+}^2 \, 
    \label{eq:chargD2}
\end{equation}
with $\hat{M}_{+}^2$ being a diagonal matrix.

%%%%%%%%%%%%%%
%%%%%%%%%%%%%%%%%%
\subsection{Evaluation of the neutral scalar mass matrices}\label{sec:neutralMass}
\renewcommand{\theequation}{B.\arabic{equation}}
  % redefine the command that creates the equation no.
  \setcounter{equation}{0} 
%%%%%%%%%%%%%%%%%%
The basis $\{ \phi_1^0 , \phi_2^0, \chi_L^0, \chi_R^0 \}$  consisting of complex fields can be broken up into real and imaginary components which are rotated into new basis $\{ h_1^{0r} , h_2^{0r}, h_3^{0r}, h_4^{0r} \}$ and  $\{ G_1^0, G_2^0, h_1^{0i} , h_2^{0i}\}$. For simplicity, we turn off all the phases, which ensures that there is no mixing between scalars and pseudo-scalars. We first construct the scalar and pseudoscalar mass matrices from the Higgs potential of Eq.~(\ref{eq:potential}) in the original basis states.

In the pseudoscalar sector we make a rotation from the original basis to an intermediate basis denoted as $\{ G_1^0, G_2^0, h_1^{0i} , h_2^{0i}\}$, where the Goldstone bosons are identified as
%%%%%%%%%
\begin{align}
 G_1^0 &= \frac{ -\kappa \phi_{1}^{0i} + \kappa' \phi_{2}^{0i} + v_L \chi_L^{0i} }{\sqrt{v_L^2 + \kappa^{\prime 2} + \kappa^2}}  \\
 %%%%%%
 G_2^0 &= \frac{ \kappa v_L^2 \phi_{1}^{0i} - \kappa' v_L^2 \phi_{2}^{0i}  + \kappa_+^2 v_L \chi_L^{0i} + \kappa_L^2 v_R   \chi_R^{0i} }{\sqrt{(v_L^2 + \kappa^{\prime 2} + \kappa^2)  (\kappa_+^2 v_L^2 + \kappa_L^2 v_R^2}) } 
\end{align}
We orthogonal rotation to go to this intermediate basis is denoted as $O^i$, which is chosen to be
%%%%%%%%%%%
\begin{equation}
    O^i M_i^2 O^{i T} = \widetilde{M}_i^2 \, ,
\end{equation}
where $M_i^2$ is the mass matrix in the $\{ \phi_1^{0i} , \phi_2^{0i}, \chi_L^{0i}, \chi_R^{0i} \}$ basis, $\widetilde{M}_i^2$ is in new basis, and
\begin{equation}
O^i = \left(
    \begin{array}{cccc}
        -\frac{\kappa}{\kappa_L} & \frac{\kappa^\prime}{\kappa_L}  & \frac{v_L}{\kappa_L} & 0  \\
        %%%%%%%%%%%%%%%%%%%%%%%%
        \frac{\kappa v_L^2}{\kappa_L \sqrt{N_3}}  & \frac{- \kappa^\prime v_L^2 }{\kappa_L \sqrt{N_3}}  & \frac{\kappa_+^2 v_L }{\kappa_L \sqrt{N_3}} & \frac{\kappa_L v_R }{\sqrt{N_3}} \\
        %%%%%%%%%%%%%%%%%%5
        \frac{\kappa^\prime }{\kappa_+}  & \frac{\kappa }{\kappa_+} & 0 & 0  \\
        %%%%%%%%%%%%%%%%%%%%
         \frac{\kappa v_L v_R}{\kappa_+ \sqrt{N_3}}  & \frac{- \kappa^\prime v_L v_R}{\kappa_+ \sqrt{N_3}}  & \frac{\kappa_+ v_R }{\sqrt{N_3}} & \frac{-\kappa_+ v_L }{\sqrt{N_3}}   \\
         %%%%%%%%%%%%%%%%%%%%
    \end{array}
    \right) \, 
    \label{eq:Oi}
\end{equation}
where $\kappa_L$ is defined in Eq.~\eqref{eq:kL} and where we have defined 
\begin{equation}
    N_3 = \kappa_+^2 v_L^2 + \kappa_L^2 v_R^2 \, .
\end{equation}
The elements of the matrix $\widetilde{M}_i^2$ with $\widetilde{M}_{ij}^I = \widetilde{M}_{ji}^I$ read as
\begin{eqnarray}
     \widetilde{M}_{11}^I &=& \frac{1}{2 \kappa_+^2 \kappa_-^2} \Big\{ 4 \sqrt{2} \kappa' \kappa_+^2 \mu_4 v_L v_R - (2 \rho_1 -\rho_2) v_L^2 v_R^2 (3 \kappa'^2 + \kappa^{2}) \nonumber \\
     %%%%%%%%%%%%%
     && \hspace{2mm}+ 4 \kappa_-^2 \kappa_+^4 (-2 \lambda_2 + \lambda_3) + \alpha_3 \kappa_+^4 (v_L^2 + v_R^2) \Big\} \, ,\nonumber \\[5pt]
     %%%%%%%%%%%%%%
     %%%%%%%%%%%%%
     \widetilde{M}_{12}^I &=& \frac{\sqrt{N_3} }{2 \kappa \kappa_+^2} \Big\{ \sqrt{2} \kappa_+^2 \mu_4 - (2 \rho_1 - \rho_2) \kappa' v_L v_R \Big\} \, ,\nonumber \\
     %%%%%%%%%%%%%%%%
     %%%%%%%%%%%%%%%%
     \widetilde{M}_{22}^I &=&  \frac{-2\rho_1 + \rho_2}{2 \kappa_+^2} N_3.
     \label{eq:B4}
\end{eqnarray}
One can further rotate $\widetilde{M}_i^2$ to the physical basis to get the masses of the pseudoscalars. 
%%%%%%%%%%%%%%%%%%%%%5
%%%%%%%%%%%%%%%%%%%%%%%%

Next, we examine the the mass matrix of real scalars in the original basis, $\{ h_1^{0r} , h_2^{0r}, h_3^{0r}, h_4^{0r} \}$. We use the following orthogonal matrix to transform the original basis into an intermediate basis. 
\begin{equation}
    O^r M_r^2 O^{rT} = \widetilde{M}_r^2 \, ,
\end{equation}
where $M_r^2$ is the mass matrix in the $\{ \phi_1^{0r} , \phi_2^{0r}, \chi_L^{0r}, \chi_R^{0r} \}$ basis, $\widetilde{M}_r^2$ is the same matrix in the new basis, and where we have chosen 
\begin{equation}
O^r = \left(
    \begin{array}{cccc}
        \frac{\kappa}{\kappa_L} & \frac{\kappa^\prime}{\kappa_L}  & \frac{v_L}{\kappa_L} & 0  \\
        \frac{\kappa^\prime }{\kappa_+}  & \frac{-\kappa }{\kappa_+} & 0 & 0  \\
         \frac{-\kappa v_L }{\kappa_+  \kappa_L}  &  \frac{-\kappa^\prime v_L }{\kappa_+  \kappa_L}  &  \frac{\kappa_+ }{ \kappa_L} & 0  \\
            0 & 0  & 0 & 1
    \end{array}
    \right).
    \label{eq:Or}
\end{equation}
The Matrix $\widetilde{M}_r^2$ is symmetric, $\widetilde{M}_{ij}^r = \widetilde{M}_{ji}^r$, with its elements given by
\begin{eqnarray}
    \widetilde{M}_{11}^r &=&   \frac{2}{\kappa_L^2} \{  \lambda_1 \kappa_+^4 + (4 \kappa \kappa^\prime \lambda_4 +  \alpha_1 v_L^2) \kappa_+^2 + 4 \kappa^2 \kappa^{\prime 2} \lambda_{23}  \nonumber\\
    %%%%%%%%%%%%%%%%%%%%%%
    && + 4 \kappa^\prime v_L^2  \alpha_2 \kappa + \alpha_3   v_L^2 \kappa_+^2 +  \rho_1 v_L^4) \} \, , \nonumber \\[5pt]
    %%%%%%%%%%%%%%%%%%%%%
    \widetilde{M}_{12}^r &=& \frac{1}{\kappa_+ \kappa_L} \{ -2 \kappa_-^2 (2 \kappa \kappa^\prime \lambda_{23} + \kappa_+^2 \lambda_4) - (  \alpha_3 \kappa \kappa^\prime + 2 \alpha_2 \kappa_-^2)v_L^2 \} \, , \nonumber \\[5pt]
    %%%%%%%%%%%%%%%%%%%%%
     \widetilde{M}_{13}^r &=& \frac{v_L}{\kappa_+ \kappa_L^2} \{ -2\kappa_+^2  (\kappa_+^2 \lambda_1 + 4 \lambda_4 \kappa \kappa^\prime) + 2  (-4 \kappa^2 \kappa^{\prime 2} \lambda_{23} + \kappa_+^2 \rho_1 v_L^2)   \nonumber \\
     %%%%%%
     && \hspace{1mm} + (\alpha_1 \kappa_+^2 + \alpha_3 \kappa^{\prime 2}  + 4 \alpha_2 \kappa \kappa^\prime) (\kappa_+^2 - v_L^2) \} \, , \nonumber
     \\[5pt]
    %%%%%%%%%%%%%%%%%%%%%
     \widetilde{M}_{14}^r &=& \frac{1}{\kappa_L} \{ 4 \alpha_2 \kappa \kappa^\prime + \alpha_3 \kappa^{\prime 2}  + \alpha_1 \kappa_+^2 + 2 \rho_1 v_L^2 \} v_R  \, , \nonumber\\[5pt]
    %%%%%%%%%%%%%%%%%%%%%
     \widetilde{M}_{22}^r &=& \frac{1}{2 \kappa_+^2 \kappa_-^2} \{ 4 \kappa_-^6 \lambda_{23} - 4 \sqrt{2} \kappa' \kappa_+^2 \mu_4 v_L v_R - (3 \kappa'^2 + \kappa^2) \rho_{12} v_L^2 v_R^2  \nonumber \\
     %%%%%%%%%%%%%%%%%%
     && + \alpha_3 \kappa_+^4 (v_L^2 + v_R^2) \} \, , \nonumber\\[5pt]
    %%%%%%%%%%%%%%%%%%%%%
      \widetilde{M}_{23}^r &=& \frac{1}{2 \kappa \kappa_+^2 \kappa_L} \{ -2 \alpha_3 \kappa' \kappa^2 \kappa_+^2 v_L  + 4 \kappa \kappa_-^2 v_L (2 \kappa \kappa' \lambda_{23} + \kappa_+^2 \lambda_4) \nonumber\\
      && \hspace{1mm} +\kappa' v_L \rho_{12} \kappa_L^2 v_R^2 - 4 \alpha_2 \kappa \kappa_+^2 \kappa_-^2 v_L - \sqrt{2} \mu_4 v_R \kappa_+^2 \kappa_L^2 \} \, , \nonumber\\[5pt]
    %%%%%%%%%%%%%%%%%%%%%
       \widetilde{M}_{24}^r &=& \frac{1}{2 \kappa \kappa_+} \{- \sqrt{2} \mu_4 v_L \kappa_+^2  - 2 \alpha_3  \kappa' \kappa^2 v_R + (- 4 \alpha_2 \kappa \kappa_-^2 + \kappa' \rho_{12} v_L^2) v_R  \} \, , \nonumber \\[5pt]
    %%%%%%%%%%%%%%%%%%%%%
      \widetilde{M}_{33}^r &=& \frac{1}{2 \kappa_+^2 \kappa_L^2} \{ 4 v_L^2 (- \alpha_1 \kappa_+^4 + \kappa^\prime (4 \alpha_2 \kappa \kappa_+^2  + 4\kappa^2 \kappa^\prime \lambda_{23}) + \kappa_+^4 \rho_1  \nonumber \\
      &&\hspace{1mm} - \alpha_3 \kappa'^2  \kappa_+^2 - 4 \kappa \kappa^\prime \lambda_4 \kappa_+^2 + \kappa_+^4 \lambda_1) - \rho_{12} \kappa_L^4 v_R^2 \} \, , \nonumber\\[5pt]
    %%%%%%%%%%%%%%%%%%%%%
      \widetilde{M}_{34}^r &=& \frac{1}{2 \kappa_+ \kappa_L} \{ -2( 4 \kappa^\prime  \alpha_2 \kappa  + \alpha_1 \kappa_+^2 + \alpha_3 \kappa^{\prime2}  ) + \kappa_+^2 \rho_{12}^\prime - v_L^2 \rho_{12} \} v_L v_R \, , \nonumber \\[5pt]
    %%%%%%%%%%%%%%%%%%%%%
      \widetilde{M}_{44}^2 &=& \frac{-\rho_{12} v_L^2}{2} + 2 \rho_1 v_R^2 .
      \label{eq:B8}
\end{eqnarray}
Here $\lambda_{23} = 2 \lambda_2 + \lambda_3$, $\rho_{12} = 2 \rho_1 - \rho_2$, and $\rho_{12}^\prime = 2 \rho_1 + \rho_2$. This $4 \times 4$ matrix can be diagonalized numerically to obtain the mass eigenvalues of the scalar fields.  Note that the field $h1^{0r}$ is the SM-like Higgs boson, which has small mixings with the heavier states.

%%%%%%%%%%%%%%%%%%%%%%
%%%%%%%%%%%%%%
\subsection{Different topology for the generation of \texorpdfstring{$\nu_R$}{NRR} Majorana mass}\label{sec:Topology}

\renewcommand{\theequation}{C.\arabic{equation}}
  % redefine the command that creates the equation no.
  \setcounter{equation}{0} 
\begin{figure}[t]
  \subfigure[]{
   \includegraphics[scale=0.34]{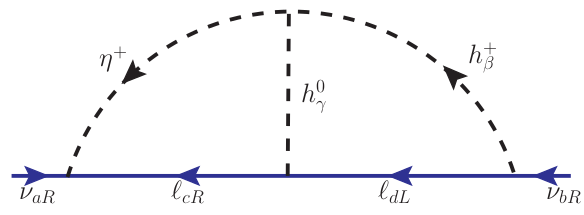}
   }\hspace{5mm}
  \subfigure[]{
    \includegraphics[scale=0.34]{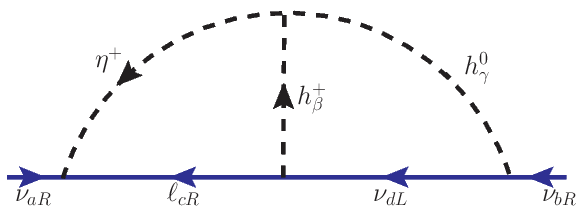}
    } 
 \subfigure[]{
    \hspace{35mm} \includegraphics[scale=0.22]{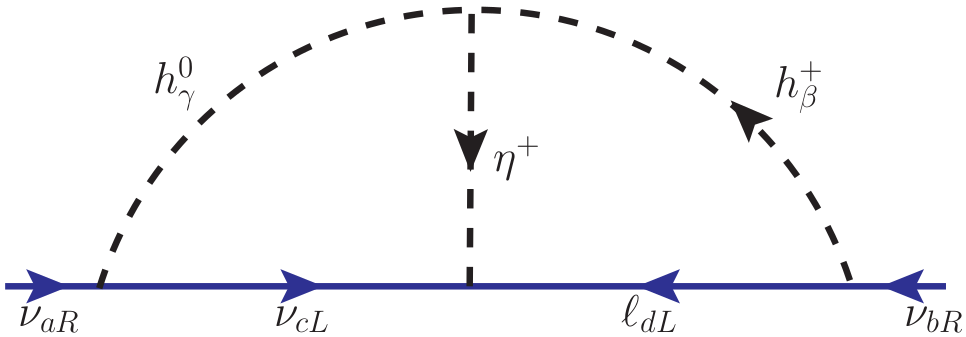}
   } 
  
  \caption{Topology for right-handed Majorana neutrino mass generation with with various arrangement of scalar fields. There are three more diagrams with internal particles replaced by their charge conjugates.}
  \label{fig:Topology}
\end{figure}
%%%%%%%%%
%%%%%%%%%
%%%%%%%%%
\begin{figure}[t]
    \centering
    \includegraphics[scale=0.3]{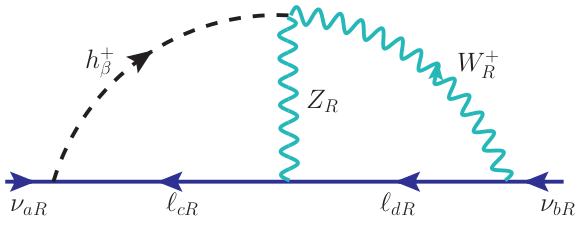}
    \caption{A typical two-loop diagram involving gauge bosons that induce right-handed Majorana neutrino mass in the model.}
    \label{fig:gaugeloop}
\end{figure}
%%%%%%%%%

In this section we show different topology for the right-handed Majorana neutrino mass generation. Unlike in Fig.~\ref{fig:loopdiag}, these are represented in the physical basis for the scalar fields. There are three more diagrams with internal particles replaced by their conjugates. Note that in the electroweak symmetry conserving limit, the neutral and charged scalars have the same mixing matrix and are degenerate. Moreover, scalar and pseudoscalars do not mix, and the $\eta^+$ field will remain a mass eigenstate, the remaining Higgs field mix, and a common $3\times3$ unitary matrix $V$ diagonalizes both charged and neutral scalars. 
%%%%%
There are Goldstone modes associated with various rotations. We work in the Feynman gauge, where the Goldstone bosons are treated just as the physical scalars, with their masses identified as those of $W$ and $Z$ bosons.  Note that there are other topology associated with gauge fields such as the one  in Fig.~\ref{fig:gaugeloop} that needs to be considered.
However, these diagrams are suppressed as they require $\eta^+$ field mixing with $\phi_{1,2}^+$, which is only possible after electroweak symmetry breaking. Thus, we do not include these contribution in our analysis. 

%%%%%%%%%
%%%%%%%%%
\subsection{Evaluation of \texorpdfstring{$I_{45}^{132}$}{Icd} }\label{sec:Icd}
\renewcommand{\theequation}{D.\arabic{equation}}
  % redefine the command that creates the equation no.
  \setcounter{equation}{0} 
%%%%%%%%%%%%%%%%%%%%%%%%%%%%%
Here we evaluate general loop integral $I_{45}^{132}$ given in Eq.~(\ref{eq:MajR}).
After performing a Wick rotation and setting $q \to -q$, Eq.~(\ref{eq:Ir123}) becomes (up to an overall sign):
\begin{equation}
    I_{45}^{132} = \int \int \frac{d^4 p}{(2 \pi)^4} \frac{d^4 q}{(2 \pi)^4} \frac{q.p}{(q^2 + m_{1}^2)(q^2 + m_5^2)(p^2 + m_{2}^2)(p^2 + m_4^2)((p+q)^2 +  m_{3}^2)} \, .
    \label{eq:B2}
\end{equation}
Here $p$ and $q$ are Euclidean four-vectors. $m_4$ and $m_5$ are the charged lepton masses. ($m_{1}$, $m_{2}$) and $m_{3}$ are the masses of the outside and inside scalars. In the limit of keeping only the linear terms in $\epsilon = \frac{\kappa'}{\kappa}$, $\epsilon' = \frac{v_L}{\kappa}$, and $v_R >> v_L, \kappa, \kappa'$, scalar masses are given in Table \ref{tab:Higgs}.
We use some useful relation in evaluating Eq.~(\ref{eq:B2}).
\begin{equation} 
{\frac{1}{\left(q^{2}+m_{1}^{2}\right)\left(q^{2}+m_{2}^{2}\right)}=\frac{1}{\left(m_{1}^{2}-m_{2}^{2}\right)}\left[\frac{1}{q^{2}+m_{2}^{2}}-\frac{1}{q^{2}+m_{1}^{2}}\right]} \, .
\label{eq:B3}
\end{equation}
%%%%%%%%%%%%%%%%%%%%%%
%%%%%%%%%%%%%%%%%%%%%%%%%
\begin{eqnarray}
\frac{p \cdot q}{\left(p^2+m_{1}^{2}\right)\left(q^{2}+m_{2}^{2}\right)((p+q)^{2}+m_{3}^{2})} = 
 %%%%%%%%%%%%%%%%%%%%%%
 \frac{1}{2} \Bigg\{ -\frac{\left(m_{3}^{2}-m_{1}^{2}-m_{2}^{2}\right)}{\left(p^{2}+m_{1}^{2}\right)\left(q^{2}+m_{2}^{2}\right)\left((p+q)^{2}+m_{3}^{2}\right)}  \nonumber\\
 %%%%%%%%%%%%%%%%%%%
 -\frac{1}{\left(p^{2}+m_1^{2}\right)((p+q)^{2}+m_{3}^{2})}  - \frac{1}{\left(q^{2}+m_2^{2}\right)((p+q)^{2}+m_{3}^{2})} + \frac{1}{\left(p^{2}+m_1^{2}\right)\left(q^{2}+m_{2}^{2}\right)} \Bigg\} \, \, . \nonumber\\
 \label{eq:D4}  
\end{eqnarray}
 %%%%%%%%%%%%%%%%%%%%%%%%%%%%%%%%%%%%%%%%%%%%%%%%%%%%%
 In addition, introducing a new notation, \cite{vanderBij:1983bw} one can write the integral in a compact form as: 
\begin{eqnarray}
 \left(m_{1} | m_{2}\right) &=& \int d^{n}p \int d^{n}q \frac{1}{\left(p^{2}+m_{1}^{2}\right)\left(q^{2}+m_{2}^{2}\right)} \, .\label{eq:D5}\\
 %%%%%%%%%%%%%%%%%%%%%%
 \left(m_{1} | m_{2} | m_{3}\right)&=&\int d^{n}p \int d^{n}q \frac{1}{\left(p^{2}+m_{1}^{2}\right)\left(q^{2}+m_{2}^{2}\right)\left((p+q)^{2}+m_{3}^{2}\right)} \, .
 \label{eq:B5}
\end{eqnarray}
%\hrule
%\vspace{3mm}
%%%%%%%%%%%%%%%%%%%%
From Eq.~(\ref{eq:B3}), (\ref{eq:B4}), (\ref{eq:D5}) and (\ref{eq:B5}) (with $n = 4$) we can write Eq.~(\ref{eq:B2}) as follows:
\begin{eqnarray}
    I_{45}^{1 2 3} &=& \frac{1}{2 (2 \pi)^8 (m_{1}^2 - m_5^2) (m_{2}^2 - m_4^2)} \bigg\{ (m_4|m_5) - (m_{2}|m_5) - (m_4|m_{1}) + (m_{2}|m_{1} ) \nonumber \\[3pt]
    %%%%
    && - (m_{3}^2-m_4^2 - m_5^2) (m_4|m_5|m_{3}) + (m_{3}^2-m_{2}^2 -m_5^2) (m_{2}|m_5|m_{3})  \nonumber \\[3pt]
    && + (m_{3}^2-m_4^2 -m_{1}^2) (m_4|m_{1}|m_{3}) - (m_{3}^2-m_{2}^2 -m_{1}^2) (m_{2}|m_{1}|m_{3})
      \bigg\} \, .
    \label{eq:D7}
\end{eqnarray}
%\hrule
%\vspace{3mm}
Solving Eq.~(\ref{eq:D5}) and Eq.~(\ref{eq:B5}), gives the solution to $I_{45}^{132}$. In order to evaluate $(m_1|m_2)$ we use the following identity with $\epsilon = n-4$ :
\begin{multline}
    \int d^{n}p \frac{1}{p^2 + m^2} = i \pi^2 m^2 \bigg[\frac{2}{\epsilon} -1 + \gamma_E + \log{(\pi m^2)} \bigg] + i \pi^2 m^2 \epsilon \Big[ \frac{\pi^2}{24} + \frac{1}{4} \gamma_E^2 - \frac{1}{2} \gamma_E  \\ 
     + \frac{1}{2} + \frac{1}{2} (\gamma_E - 1) \log{(\pi m^2)} + \frac{1}{4} \log^2{\pi m^2} \Big] + \mathcal{O}(\epsilon^2) \, .
    \label{eq:D8}
\end{multline}
%%%%%%
Furthermore, $\left(m_{1} | m_{2} | m_{3}\right)$ can be expanded to 
\begin{equation}
\left(m_{1} | m_{2} | m_{3}\right)=\frac{1}{3-n}\left[m_{1}^{2}\left(m_1 m_{1}\left|m_{2}\right| m_{3}\right)+m_{2}^{2}\left(m_2  m_{2}\left|m_{1}\right| m_{3}\right) \\ {\left.+m_{3}^{2}\left(m_3 m_{3} | m_{1} | m_{2}\right)\right]}\right. \, ,
\label{eq:B7}
\end{equation}
where the term in Eq.~(\ref{eq:B7}) is given by \cite{Ghinculov:1994sd,McDonald:2003zj} 
\begin{eqnarray}
    \left(m_1 m_{1}\left|m_{2}\right| m_{3}\right) &=& \int d^{n}p \int d^{n}q \frac{1}{\left(p^{2}+m_{1}^{2}\right)^2\left(q^{2}+m_{2}^{2}\right)\left((p+q)^{2}+m_{3}^{2}\right)} \nonumber \\[3pt]
%%%%%%%%%%%%%%%%%%%%%%%%%%%
    &=& \frac{-\pi^{4}\left(\pi m_1^{2}\right)^{n-4} \Gamma\left(2-\frac{n}{2} \right)}{\Gamma\left(3-\frac{n}{2}\right)} \int_{0}^{1} d x \int_{0}^{1} d y(x(1-x))^{n/2-2} y(1-y)^{2-n/2}  \nonumber\\[3pt] 
    %%%%%%%%%%%%%%%%%%%%%
    &\times& \left[\Gamma(5-n) \frac{\mu^{2}}{\left(y+\mu^{2}(1-y)\right)^{5-n}}+\frac{n}{2} \Gamma(4-n) \frac{1}{\left(y+\mu^{2}(1-y)\right)}\right]
\end{eqnarray}
where
\begin{equation}
\mu^{2} = \frac{a x+b(1-x)}{x(1-x)}, \quad a = \frac{m_{2}^{2}}{m_1^{2}}, \quad b = \frac{m_{3}^{2}}{m_1^{2}} \, .
\end{equation}
%%%%%%%%%%%%%%%
 Now, letting $\epsilon = n-4$ and expanding in the limit $\epsilon \to 0$, 
 \begin{eqnarray}
     \left(m_1 m_1\left|m_{2}\right| m_{3}\right) &=&\pi^{4}\left[\frac{-2}{\epsilon^{2}}+\frac{1}{\epsilon}\left(1-2 \gamma_{E}-2 \log \left(\pi m_1^{2}\right)\right)\right]+ \pi^{4}\bigg[-\frac{1}{2}-\frac{1}{12} \pi^{2}+ \gamma_E-\gamma_{E}^{2}  \nonumber\\ 
     %%%%%%%%%
     && + \left(1-2 \gamma_{E}\right) \log(\pi m_1^{2}) -\log ^{2}\left(\pi m_1^{2}\right)-f(a,b)\bigg]+ \mathcal{O}(\epsilon) \, ,
     \label{eq:B10}
 \end{eqnarray}
 where the function $f(a,b)$ is given by:
\begin{equation}
    f(a,b)=\int_{0}^{1} d x\left(\operatorname{Li}_{2}\left(1-\mu^{2}\right)-\frac{\mu^{2} \log \mu^{2}}{1-\mu^{2}}\right) \, ,
    \label{eq:B11}
\end{equation}
and the dilogarithm function $\mathrm{Li}_2$ is defined as:
%%%%%%%%%%%%%%%%%%%
\begin{equation}
\mathrm{Li}_{2}(x)=-\int_{0}^{x} \frac{\log (1-y)}{y} d y \, .
\end{equation}
%%%%%%%%%%%%%%%%%%%%%%%%%%%%%%%%%%%%%%%%%%%%%%%%%%%%%
Performing the $y$ integration, $f(a,b)$ in Eq.~(\ref{eq:B11}) becomes \cite{vanderBij:1983bw, McDonald:2003zj}

\begin{align} f(a,b)=&-\frac{1}{2} \log a \log b-\frac{1}{2}\left(\frac{a+b-1}{\sqrt{\Delta}}\right)\Bigg\{ \operatorname{Li}_{2}\left(\frac{-x_{2}}{y_{1}}\right)+\operatorname{Li}_{2}\left(\frac{-y_{2}}{x_{1}}\right)-\operatorname{Li}_{2}\left(\frac{-x_{1}}{y_{2}}\right) \nonumber\\
%%%%%%%%%%%%%%%%%
&-\operatorname{Li}_{2}\left(\frac{-y_{1}}{x_{2}}\right) +\operatorname{Li}_{2}\left(\frac{b-a}{x_{2}}\right)+\operatorname{Li}_{2}\left(\frac{a-b}{y_{2}}\right)-\operatorname{Li}_{2}\left(\frac{b-a}{x_{1}}\right)-\operatorname{Li}_{2}\left(\frac{a-b}{y_{1}}\right)\Bigg\} \, ,
\label{eq:D18}
\end{align} 
%%%%%%%%%%%%%%%%%%%%%%%%%%%%%%%%%%%%%%%%%%%%%%%%
where $\Delta=\left(1-2(a+b)+(a-b)^{2}\right)$ and 
\begin{align}
x_{1}=\frac{1}{2}(1+b-a+\sqrt{\Delta }), &
%%%%%%
\hspace{5mm} 
x_{2}=\frac{1}{2}(1+b-a-\sqrt{\Delta}) \, , \nonumber\\ 
%%%%%%
y_{1}=\frac{1}{2}(1+a-b+\sqrt{\Delta}), & 
%%%%%%%%%%
\hspace{5mm} {y_{2}=\frac{1}{2}(1+a-b-\sqrt{\Delta})} \, .
\end{align}
%%%%%%%%%%%%%%%%%%%%%%%%%%%%%%%%%%%%%%%%%%%%%%%%%%%%%%%
%%%%%%%%%%%%%%%%%%%%%%%%%%%%%%%%%%%%%%%%%%%%%%%%%%%%%%%
By explicit symmetrization between $a$ and $b$ ($a \leftrightarrow b$) and using the relation 
\begin{align}
{\rm Li}_{2}(1-z)&=-\mathrm{Li}_{2}(z)-\log z \log (1-z)+\frac{1}{6} \pi^{2} \, , \nonumber \\ 
%%%%%%%%%%%%%%%%%%%%%%%%%5
{\rm Li}_{2}\left(\frac{1}{z}\right) &=-\mathrm{Li}_{2}(z)-\frac{1}{2} \log ^{2}(-z)-\frac{1}{6} \pi^{2} \, ,
\end{align}
the function simplifies to:
\begin{eqnarray}
    f(a,b)=-\frac{1}{2} \log a \log b &- & \left(\frac{a+b-1}{\sqrt{\Delta }}\right) \bigg\{ \mathrm{Li}_{2}\left(\frac{-x_2}{y_{1}}\right)+\mathrm{Li}_{2}\left(\frac{-y_{2}}{x_{1}}\right)+\frac{1}{4} \log ^{2} \frac{x_{2}}{y_{1}} \nonumber  \\
    %%%%%%%%%%%%%%%%%%%%%%%
 &&+\frac{1}{4} \log ^{2} \frac{y_{2}}{x_{1}} +\frac{1}{4} \log ^{2} \frac{x_{1}}{y_{1}}-\frac{1}{4} \log ^{2} \frac{x_{2}}{y_{2}}+ \frac{\pi^2}{6}\bigg\} \, .
 \label{eq:D16}
\end{eqnarray}
%%%%%%%%%%%%%%%%%%%%%%%%%%%%%%%%%%%%%%%%%%%%%%%%%%%%%%%%%
It is to be noted that function $f(a,b)$ in Eq.~(\ref{eq:D16}) can have non-zero imaginary part \cite{Coleman:1965xm, McDonald:2003zj}. However, the imaginary component of the function $f(a,b)$ cancels out with judicious logarithmic branch choice. The real part of Eq.~(\ref{eq:D16}) is in full agreement with Eq.~\eqref{eq:D18}. By expanding Eq.~(\ref{eq:D7}) as in the relation of Eq.~(\ref{eq:B7}) and making use of expression given in Eq.~(\ref{eq:B10}) together with Eq.~(\ref{eq:D8}), $I_{45}^{132}$ is obtained as 
\clearpage
%%%%%
\begin{align}
    I_{45}^{132}\ =\ &  \frac{1}{(16\pi^2)^2}\ \Bigg[- \frac{1}{\epsilon}+ 2 - \gamma_E - \log{\pi} - \log{\mu^2} + \Big[ \frac{m_1^2}{(m_1^2-m_5^2)}\ \Big\{ \frac{1}{4} \log^2\bigg(\frac{m_1^2}{\mu^2} \bigg)   \nonumber \\[3pt]
    %%%%%%%%%%%
     & 
     -  \frac{1}{2} \log\bigg(\frac{m_1^2}{\mu^2}\bigg)  \Big\} + \frac{m_2^2}{(m_2^2-m_4^2)}\ \left\{ \frac{1}{4} \log^2\bigg(\frac{m_2^2}{\mu^2} \bigg) - \frac{1}{2}  \log\bigg(\frac{m_2^2}{\mu^2}\bigg)  \right\}    \nonumber \\[3pt] 
     %%%%%%%%%
    &
    - \frac{m_4^2}{(m_2^2-m_4^2)}\ \Big\{ \frac{1}{4} \log^2\bigg(\frac{m_4^2}{\mu^2} \bigg) - \frac{1}{2}  \log\bigg(\frac{m_4^2}{\mu^2}\bigg)  \Big\} - \frac{m_5^2}{(m_1^2-m_5^2)}  
     \nonumber\\[3pt]
     %%%%%%%%%
      &
      \times \left\{ \frac{1}{4} \log^2\bigg(\frac{m_5^2}{\mu^2} \bigg) - \frac{1}{2}  \log\bigg(\frac{m_5^2}{\mu^2}\bigg)  \right\} - \frac{1}{2(m_1^2-m_5^2)(m_2^2-m_4^2)}     \nonumber \\[3pt]
      %%%%%%%%%%%%%%%%%
      &
      \times \bigg\{  \bigg( m_2^2\log\bigg(\frac{m_2^2}{\mu^2}\bigg) - m_4^2\log\bigg(\frac{m_4^2}{\mu^2}\bigg) \bigg)   \bigg( m_1^2\log\bigg(\frac{m_1^2}{\mu^2}\bigg) - m_5^2\log\bigg(\frac{m_5^2}{\mu^2}\bigg) \bigg) \bigg\} \Big] \nonumber \\[3pt]
      %%%%%%%%%
      & 
      +  \frac{1}{2(m_1^2-m_5^2)(m_2^2-m_4^2)}\ \Big[ (m_1^2+ m_2^2 - m_3^2)\  (m_2^2 f_2^{13} + m_1^2 f_1^{23} + m_3^2 f_3^{21})   \nonumber \\[3pt]
      %%%%%%%%%
      & 
      - (m_1^2- m_3^2 + m_4^2)\  (m_4^2 f_4^{13} + m_1^2 f_1^{43} + m_3^2 f_3^{41}) -(m_2^2- m_3^2 + m_5^2)  \nonumber\\
    %%%%%%%%%%%%%%
     &
     \times(m_3^2 f_3^{25} + m_5^2 f_5^{23} + m_2^2 f_2^{53}) - (m_3^2- m_4^2 - m_5^2) (m_3^2 f_3^{45} + m_5^2 f_5^{43} + m_4^2 f_4^{53})   \Big] \Bigg] \, ,
    \label{eq:D19}
\end{align}
where $f_i^{jk} \equiv f\bigg( \frac{m_j^2}{m_i^2}, \frac{m_k^2}{m_i^2} \bigg)$, $(i,j,k) = (1,2,3,4,5)$ and $F$ is given by Eq.~(\ref{eq:D16}). In the limit of $m_4 = m_5 = 0$, one can reduce Eq.(\ref{eq:D19}) as follows:
\begin{eqnarray}
    I_{00}^{132} &=& \frac{1}{(16\pi^2)^2} \Bigg[- \frac{1}{\epsilon}+ 2 - \gamma_E - \log{\pi} - \log{\mu^2} - \frac{1}{2} \log\bigg(\frac{m_{1}^2 m_{2}^2 }{\mu^4} \bigg) + \frac{1}{4}\log^2\bigg(\frac{m_{1}^2}{m_{2}^2} \bigg)   \nonumber\\[3pt]
    %%%%%%%%%%%%
     && + \frac{1}{2 m_{1}^2 m_{2}^2}  \bigg[-  \frac{\pi^2}{6}  m_3^4 +  (m_1^2+ m_2^2 - m_3^2) (m_2^2 f_2^{13} + m_1^2 f_1^{2 3} + m_3^2 f_3^{2 1})   \nonumber \\[3pt]
     %%%%%%%%%%%%
     && - (m_1^2- m_3^2) (m_1^2 f_1^{0 3} + m_3^2 f_3^{0 1}) - (m_2^2- m_3^2) (m_3^2 f_3^{2 0} + m_2^2 f_2^{0 3})   \bigg] \, ,
  \label{eq:D20}    
\end{eqnarray}
where $f_i^{0k} \equiv F\bigg( 0 , \frac{m_k^2}{m_i^2} \bigg)$, $f_i^{j0} \equiv F\bigg(\frac{m_j^2}{m_i^2}, 0 \bigg)$, and F[a,0] = F[0,a] = $\mathrm{Li}_{2}(1-a)$.

\subsection{Evaluation of \texorpdfstring{$M_{\nu_R}$}{MvR} }\label{sec:MvR}
\renewcommand{\theequation}{E.\arabic{equation}}
  % redefine the command that creates the equation no.
  \setcounter{equation}{0} 
%%%%%%%%%%%%%%%%%%%%%
To evaluate the neutrino mass, one needs to finally sum over all  possible diagrams. Recognizing outside scalars in the diagram as  $\alpha, \beta$, and inside scalar as $\gamma$, one needs to sum over $\alpha, \beta,$ and $\gamma$. In doing so, all the constants that appear in the integral vanish owing to the unitarity condition. The total contribution to the neutrino mass is given in Eq.~\eqref{eq:2loopMR} and Eq.~\eqref{eq:flavorst} with $\eta$ replaced by $\alpha$. As an illustration we have, 
\begin{equation}
    A_{1ab} = \mathcal{F}_{\alpha \gamma \beta}^{cd}\ I^{\alpha \gamma \beta}_{cd}
\end{equation}
where $\mathcal{F}_{\alpha \gamma \beta}^{cd}$ is the flavor structure with linear combinations of unitary matrices associated with the two-loop neutrino mass matrix in Eq.~\eqref{eq:flavorst}. $\mathcal{F}_{\alpha \gamma \beta} I_{cd}^{\alpha \gamma \beta} $ in the limit of $m_c = m_d = 0$, vanishing charged lepton masses, is given as
\begin{align}
\mathcal{F}_{\alpha \beta \gamma} I^{\alpha \beta \gamma}_{00} =    
     &  \frac{\lambda_{\alpha \beta \gamma}}{ (16 \pi^2)^2} \Bigg[ -\frac{1}{2} \log\bigg(\frac{m_{\alpha}^2 m_{\beta}^2 }{m_{\gamma}^4} \bigg) + \frac{1}{4}\log^2\bigg(\frac{m_{\alpha}^2}{m_{\beta}^2} \bigg) + \frac{1}{2 m_{\alpha}^2 m_{\beta}^2}  \nonumber\\
     %%%%%%%%%%%%%%%%%%%%%%%%%%%%%%%%%%
     & \times \bigg[ (m_\alpha^2+ m_\beta^2 - m_\gamma^2) (m_\beta^2 f_\beta^{\alpha \gamma} + m_\alpha^2 f_\alpha^{\beta \gamma} + m_\gamma^2 f_\gamma^{\beta \alpha}) - (m_\alpha^2- m_\gamma^2)  \nonumber \\
     %%%%%%%%%%%%%%%%%%%%%%%%%%%%%
     & \times (m_\alpha^2 f_\alpha^{0 \gamma} + m_\gamma^2 f_\gamma^{0 \alpha}) - (m_\beta^2- m_\gamma^2) (m_\gamma^2 f_\gamma^{\beta 0} + m_\beta^2 f_\beta^{0 \gamma}) -  \frac{\pi^2}{6}  m_\gamma^4   \bigg] \Bigg] \, ,
\end{align}
%%%%%%%%
where the function $F$ is given by Eq.~(\ref{eq:D16}) with $a$ and $b$ being the ratio of the masses. Moreover, we take $\mu^2 = m_{\gamma}^2$ in Eq.~(\ref{eq:D20}) in getting the above expression.   

%%%%%%%%%%%%%%%%%%%%%%%%%%%%%%%%%%%%%%%%%%%%%%%%%%%%%%%%%55
%%%%%%%%%%%%%%%%%%%%%%%%%%%%%%%%%%%%%%%%%%%%%%%%%%%%%%%%%%%%%%%%%%%
\subsection{Asymptotic behavior of \texorpdfstring{$I_{cd}^{\alpha \gamma \beta}$}{Icd} in evaluating the right-handed Majorana neutrino mass }\label{sec:AIcd}
%%%%%%%%%%%%%%%%%%%%%%%%%%%%%%%%%%%%%%%%%%%
\renewcommand{\theequation}{F.\arabic{equation}}
  % redefine the command that creates the equation no.
  \setcounter{equation}{0} 
  %%%%%%%%%%%%%%%%%%%%%%%%%%%%%%%%%5
We show here the asymptotic behavior for the two cases, $m_{H_\gamma^0} >> m_{h_\alpha^+} = m_{H_\beta^+}$ and $m_{H_\gamma^0} = m_{H_\beta^+} >> m_{h_\alpha^+}$. 
%%%%
We write $I_{cd}^{\alpha \gamma \beta}$ as $I_{cd}^{\eta \gamma \beta}$; identifying $m_\eta \equiv m_{h_\alpha^+}$, $m_{H_\beta^+}  \equiv (m_{h_1^+},m_{h_2^+},m_{h_3^+})$, and $m_{H_\gamma^0} \equiv (m_{h_1^0},m_{h_2^0},m_{h_3^0})$. In evaluating the asymptotic behavior of the neutrino mass, we have to sum over all possibilities in $\beta$ and $\gamma$. Furthermore, to simplify the flavor structure, we consider $\widetilde{y} > y$, and take all the phases zero. 
%%%
%%%%%%
%%%%%%%%%%%5

\subsubsection{\texorpdfstring{$m_{H_\gamma^0} >> m_{\eta} = m_{H_\beta^+} = m_\beta $}{meta} }
Since the masses of leptons are much smaller than Higgs masses, the terms with $m_c^2$ and $m_d^2$ are suppressed and can be ignored in Eq.~(\ref{eq:D7}). Thus, in this limit, we obtain the right-handed neutrino Majorana mass as
\begin{align}
     (M_{\nu_R})_{ab}\ \approx\ & \frac{\ \alpha_4 \, v_R }{(16 \pi^2)^2}\ (f_{ac} \, \widetilde{y}_{cd}^\star\, \widetilde{y}_{db}\ \lambda_{ \beta \gamma} + {\rm Transpose} )\, \bigg\{ - \frac{m_\beta^2}{m_{H_\gamma^0}^2} + \frac{7}{2} \frac{m_\beta^2}{m_{H_\gamma^0}^2} \log\bigg( \frac{m_\beta^2}{m_{H_\gamma^0}^2} \bigg) \bigg\}\, .
     %%%%%%%
\end{align}
%%%%%%%%%%%%%%%%%%%%%%%%%%
%%%%%%%%%%%%%%%%%%%%%%%%%

\subsubsection{\texorpdfstring{$m_{H_\gamma^0} = m_{H_\beta^+} = m_\beta >> m_\eta$}{m}}
Here we take two masses being equal to each other and much heavier than the third. Ignoring the masses of leptons we obtain the RH Majorana mass as
\begin{equation}
     (M_{\nu_R})_{ab} \approx\frac{\ \alpha_4 \, v_R }{2\, (16 \pi^2)^2}\ (f_{ac} \, \widetilde{y}_{cd}^\star\, \widetilde{y}_{db}\  + {\rm Transpose} )\ \lambda_{ \beta \beta}\  \frac{m_\eta^2}{m_\beta^2} \log\bigg(\frac{m_\eta^2}{m_\beta^2} \bigg)  \, .
\end{equation}
%%%%%%%%%%%%%%%%%%%%%%%%%%%%%%%%%%%%%%%%%%%%%%%%%%%%%%%%%%%%%%%%%%%
%%%%%%%%%%%%%%%%%%%%%%%%%%%%%%%%%%%%%%%%%%%%%%%%%%%%%%%%%%%%%%5

\bibliographystyle{utphys}
\bibliography{reference}

\end{document}